\documentclass[twocolumn]{aastex631}

\newcommand{\kms}{\rm km~s\ensuremath{^{-1}\,}}
\newcommand{\msun}{\ensuremath{\rm M_\odot}}
\newcommand{\msunyr}{\ensuremath{\rm M_{\odot}\;{\rm yr}^{-1}}}
\newcommand{\Ha}{\ensuremath{\rm H\alpha}}
\newcommand{\Hb}{\ensuremath{\rm H\beta}}
\newcommand{\lya}{\ensuremath{\rm Ly\alpha}}

\newcommand{\lyb}{Ly$\beta$}

\newcommand{\fluxunits}{\ensuremath{\rm erg~s^{-1}~cm^{-2}}} 
\newcommand{\sbunits}{\ensuremath{\rm erg~s^{-1}~cm^{-2}~arcsec^{-2}}}
\newcommand{\lumunits}{\ensuremath{\rm erg~s^{-1}}}

\newcommand{\secpoint}{\mbox{$''\mskip-7.6mu.\,$}}

\def\ltsima{$\; \buildrel < \over \sim \;$}
\def\simlt{\lower.5ex\hbox{\ltsima}}
\def\gtsima{$\; \buildrel > \over \sim \;$}
\def\simgt{\lower.5ex\hbox{\gtsima}}
\def\arcs{$''~$}

\usepackage[percent]{overpic}
\usepackage{amsmath}
\usepackage{multirow}
\usepackage{booktabs}
\graphicspath{{./}{Figures/}}

\newcommand{\zGOne}{2.4312 ($\pm$ 5.3014 $\rm km~s^{-1}$)}
\newcommand{\linewidthGOne}{81.31 $\pm$ 4.03 $\rm km~s^{-1}$ (H$\alpha$) }
\newcommand{\HaGOne}{5.468 $\pm$ 0.865}
\newcommand{\HbGOne}{1.912 $\pm$ 0.174}
\newcommand{\HgGOne}{0.898 $\pm$ 0.138}
\newcommand{\OIIIOneGOne}{3.554 $\pm$ 0.100}
\newcommand{\OIIITwoGOne}{10.655 $\pm$ 0.300}
\newcommand{\OIIOneGOne}{2.811 $\pm$ 0.244}
\newcommand{\OIITwoGOne}{3.617 $\pm$0.421}
\newcommand{\NIIOneGOne}{0.174 $\pm$ 0.062}
\newcommand{\NIITwoGOne}{0.523 $\pm$ 0.187}
\newcommand{\SIIOneGOne}{0.605 $\pm$ 0.248}
\newcommand{\SIITwoGOne}{0.674 $\pm$ 0.249}
\newcommand{\NeIIIOneGOne}{0.348 $\pm$ 0.171}
\newcommand{\AvGOne}{$\rm 0.21^{+0.73}_{-0.21}$}
\newcommand{\AHalphaGOne}{$\rm 0.17^{+0.43}_{-0.17}$}
\newcommand{\EBVGOne}{$\rm 0.07^{+0.17}_{-0.07}$}
\newcommand{\SFRGOne}{5.82 $\pm$ 0.92}
\newcommand{\SFRbGOne}{4.67}
\newcommand{\HaHbGOne}{3.05 $\pm$ 0.56}
\newcommand{\HgHbGOne}{0.45 $\pm$ 0.08}
\newcommand{\OThreeGOne}{0.75 $\pm$ 0.04}
\newcommand{\OThreeTwoGOne}{0.34 $\pm$ 0.03}
\newcommand{\logUOThreeGOne}{-2.56 $\pm$ 0.08}

\newcommand{\logUOThreeTwoGOne}{-2.68 $\pm$ 0.05}

\newcommand{\RTwoThreeGOne}{1.03 $\pm$ 0.05}

\newcommand{\TwelveOHOThreeTwoRTwoThreeMcGGOne}{8.39 $\pm$ 0.10}
\newcommand{\OHsolarOThreeTwoRTwoThreeMcGGOne}{-0.30}
\newcommand{\NTwoGOne}{-1.02 $\pm$ 0.17}

\newcommand{\logNONTwoGOne}{-1.20 $\pm$ 0.23}
\newcommand{\NOsolarNTwoGOne}{-0.34}
\newcommand{\OThreeNTwoGOne}{1.77 $\pm$ 0.17}
\newcommand{\TwelveOHOThreeNTwoGOne}{8.38 $\pm$ 0.04}
\newcommand{\OHsolarOThreeNTwoGOne}{-0.31}
\newcommand{\NTwoOTwoGOne}{-1.09 $\pm$ 0.16}
\newcommand{\logNONTwoOTwoGOne}{-1.21 $\pm$ 0.13}
\newcommand{\NOsolarNTwoOTwoGOne}{-0.35}
\newcommand{\NTwoSTwoGOne}{-0.39 $\pm$ 0.20}

\newcommand{\NeThreeOTwoGOne}{-1.27 $\pm$ 0.22}
\newcommand{\logUNeThreeOTwoGOne}{-3.03 $\pm$ 0.16}

\newcommand{\zDLALBG}{3.1509 ($\pm$ 4.4409 $\rm km~s^{-1}$)}
\newcommand{\linewidthDLALBG}{50.99 $\pm$ 1.06 $\rm km~s^{-1}$ ([OIII]) }
\newcommand{\HaDLALBG}{\nodata}
\newcommand{\HbDLALBG}{1.758 $\pm$ 0.400}
\newcommand{\HgDLALBG}{\nodata}
\newcommand{\OIIIOneDLALBG}{2.997 $\pm$ 0.178}
\newcommand{\OIIITwoDLALBG}{8.973 $\pm$ 0.519}
\newcommand{\OIIOneDLALBG}{0.536 $\pm$ 0.224}
\newcommand{\OIITwoDLALBG}{0.463 $\pm$0.132}
\newcommand{\NIIOneDLALBG}{\nodata}
\newcommand{\NIITwoDLALBG}{\nodata}
\newcommand{\SIIOneDLALBG}{\nodata}
\newcommand{\SIITwoDLALBG}{\nodata}
\newcommand{\NeIIIOneDLALBG}{0.333 $\pm$ 0.149}
\newcommand{\AvDLALBG}{\nodata}
\newcommand{\AHalphaDLALBG}{\nodata}
\newcommand{\EBVDLALBG}{\nodata}
\newcommand{\SFRDLALBG}{\nodata}
\newcommand{\SFRbDLALBG}{10.01}
\newcommand{\HaHbDLALBG}{\nodata}
\newcommand{\HgHbDLALBG}{\nodata}
\newcommand{\OThreeDLALBG}{0.71 $\pm$ 0.10}
\newcommand{\OThreeTwoDLALBG}{1.08 $\pm$ 0.11}
\newcommand{\logUOThreeDLALBG}{-2.61 $\pm$ 0.15}

\newcommand{\logUOThreeTwoDLALBG}{-2.10 $\pm$ 0.10}

\newcommand{\RTwoThreeDLALBG}{0.87 $\pm$ 0.10}

\newcommand{\TwelveOHOThreeTwoRTwoThreeMcGDLALBG}{7.82 $\pm$ 0.17}
\newcommand{\OHsolarOThreeTwoRTwoThreeMcGDLALBG}{-0.87}
\newcommand{\NTwoDLALBG}{\nodata}

\newcommand{\logNONTwoDLALBG}{\nodata}

\newcommand{\OThreeNTwoDLALBG}{\nodata}
\newcommand{\TwelveOHOThreeNTwoDLALBG}{\nodata}

\newcommand{\NTwoOTwoDLALBG}{\nodata}
\newcommand{\logNONTwoOTwoDLALBG}{\nodata}

\newcommand{\NTwoSTwoDLALBG}{\nodata}

\newcommand{\NeThreeOTwoDLALBG}{-0.48 $\pm$ 0.22}
\newcommand{\logUNeThreeOTwoDLALBG}{-2.53 $\pm$ 0.17}

\newcommand{\MdynGOne}{$\rm 9.85 \pm 0.01 ~ (M_\odot)$}
\newcommand{\MdynDLALBG}{$\rm 9.22 \pm 0.01 ~  (M_\odot)$}

\newcommand{\zLyarGOne}{2.4343 $\pm$ 0.0008}

\newcommand{\vLyarGOne}{271 $\pm$ 72}

\newcommand{\LyarfluxGOne}{0.94 $\pm$ 0.02}

\newcommand{\LyarewGOne}{-6.19 $\pm$ 0.97}
\newcommand{\LyatotfluxGOne}{0.94 $\pm$ 0.02}
\newcommand{\LyatotlumGOne}{41.64 $\pm$ 0.02}
\newcommand{\LyaHalphaGOne}{0.17 $\pm$ 0.16}
\newcommand{\LyaescGOne}{0.020 $\rm ^{+0.028}_{-0.020}$}
\newcommand{\LyaCubexredareaGOne}{5.08}
\newcommand{\LyaCubexredfluxGOne}{2.94 $\pm$ 0.01}
\newcommand{\LyaCubexredlumGOne}{42.14 $\pm$ 0.01}
\newcommand{\LyaCubexredLyaHalphaGOne}{0.54 $\pm$ 0.16}
\newcommand{\LyaCubexredLyaescGOne}{0.06 $\rm ^{+0.02}_{-0.02}$}
\newcommand{\LyaCubexblueareaGOne}{1.12}
\newcommand{\LyaCubexbluefluxGOne}{0.46 $\pm$ 0.03}

\newcommand{\LyaCubexsysareaGOne}{1.40}
\newcommand{\LyaCubexsysfluxGOne}{0.32 $\pm$ 0.01}

\newcommand{\zLyabDLALBG}{3.1450 $\pm$ 0.0008}
\newcommand{\zLyarDLALBG}{3.1533 $\pm$ 0.0008}
\newcommand{\vLyabDLALBG}{-423 $\pm$ 59}
\newcommand{\vLyarDLALBG}{171 $\pm$ 59}
\newcommand{\LyabfluxDLALBG}{0.30 $\pm$ 0.02}
\newcommand{\LyarfluxDLALBG}{4.48 $\pm$ 0.03}
\newcommand{\LyabewDLALBG}{-2.66 $\pm$ 0.59}
\newcommand{\LyarewDLALBG}{-39.45 $\pm$ 0.97}
\newcommand{\LyatotfluxDLALBG}{4.78 $\pm$ 0.05}
\newcommand{\LyatotlumDLALBG}{42.62 $\pm$ 0.01}
\newcommand{\LyaHalphaDLALBG}{0.96 $\pm$ 0.08}
\newcommand{\LyaescDLALBG}{0.110 $\rm ^{+0.014}_{-0.014}$}
\newcommand{\LyaCubexredareaDLALBG}{14.40}
\newcommand{\LyaCubexredfluxDLALBG}{12.50 $\pm$ 0.01}
\newcommand{\LyaCubexredlumDLALBG}{43.04 $\pm$ 0.01}
\newcommand{\LyaCubexredLyaHalphaDLALBG}{2.50 $\pm$ 0.08}
\newcommand{\LyaCubexredLyaescDLALBG}{0.29 $\rm ^{+0.01}_{-0.01}$}
\newcommand{\LyaCubexblueareaDLALBG}{7.11}
\newcommand{\LyaCubexbluefluxDLALBG}{1.15 $\pm$ 0.02}

\newcommand{\LyaCubexsysareaDLALBG}{4.41}
\newcommand{\LyaCubexsysfluxDLALBG}{0.36 $\pm$ 0.01}

\defcitealias{nielsen+2022}{N22}
\defcitealias{rudie+2019}{R19}
\defcitealias{cantalupo+2019}{C19}
\defcitealias{theios+2019}{T19}
\defcitealias{strom+2017}{S17}
\defcitealias{strom+2018}{S18}
\defcitealias{decia+2016}{DC16}
\defcitealias{rudie+2012}{R12}
\begin{document}

\submitjournal{ApJ; Accepted}

\shorttitle{Inner CGM of $z\sim2-3$ Galaxies I}
\shortauthors{Nunez et al.}

\title{KBSS-InCLOSE I: Design and First Results from the Inner CGM of QSO Line Of Sight Emitting Galaxies at $z\sim2-3$\footnote{Some of the data presented herein were obtained at Keck Observatory, which is a private 501(c)3 non-profit organization operated as a scientific partnership among the California Institute of Technology, the University of California, and the National Aeronautics and Space Administration. The Observatory was made possible by the generous financial support of the W. M. Keck Foundation.}}

\correspondingauthor{Evan Haze Nu\~{n}ez}
\email{enunez@astro.caltech.edu}

\author[0000-0001-5595-757X]{Evan Haze Nu\~{n}ez}
\affil{California Institute of Technology, 1200 E. California Blvd., MC 249-17, Pasadena, CA 91125, USA}

\author[0000-0002-4834-7260]{Charles C. Steidel}
\affil{California Institute of Technology, 1200 E. California Blvd., MC 249-17, Pasadena, CA 91125, USA}

\author[0000-0001-6196-5162]{Evan N. Kirby}
\affil{California Institute of Technology, 1200 E. California Blvd., MC 249-17, Pasadena, CA 91125, USA}
\affil{Department of Physics, University of Notre Dame, Notre Dame, IN 46556, USA}

\author[0000-0002-8459-5413]{Gwen C. Rudie}
\affil{The Observatories of the Carnegie Institution for Sciences, 813 Santa Barbara Street, Pasadena, CA, USA}

\author[0000-0001-5847-7934]{Nikolaus Z.\ Prusinski}
\affil{California Institute of Technology, 1200 E. California Blvd., MC 249-17, Pasadena, CA 91125, USA}

\author[0000-0003-4520-5395]{Yuguang Chen}
\affil{University of California, Davis, 1 Shields Ave., Davis, CA 95616, USA}

\author[0000-0002-1945-2299]{Zhuyun Zhuang}
\affil{California Institute of Technology, 1200 E. California Blvd., MC 249-17, Pasadena, CA 91125, USA}

\author[0000-0001-6369-1636]{Allison L. Strom}
\affiliation{Department of Physics and Astronomy, Northwestern University, 2145 Sheridan Road, Evanston, IL 60208, USA}
\affiliation{Center for Interdisciplinary Exploration and Research in Astrophysics (CIERA), Northwestern University, 1800 Sherman Avenue, Evanston, IL 60201, USA}

\author[0000-0001-9714-2758]{Dawn K.\ Erb}
\affil{The Leonard E. Parker Center for Gravitation, Cosmology and Astrophysics, Department of Physics, University of Wisconsin-Milwaukee, 3135 N Maryland Avenue, Milwaukee, WI 53211, USA}

\author[0000-0002-5139-4359]{Max Pettini}
\affil{Institute of Astronomy, Madingley Road Cambridge, CB3 0HA} 
\affil{Kavli Institute for Cosmology, Madingley Road Cambridge, CB3 0H}

\author[0000-0003-3174-7054]{Louise Welsh}
\affiliation{INAF - Osservatorio Astronomico di Trieste, via G. B. Tiepolo 11, I-34143 Trieste, Italy}
\affiliation{IFPU - Institute for Fundamental Physics of the Universe, via Beirut 2, I-34151 Trieste, Italy}

\author[00000-0002-1608-7564]{David S. N. Rupke}
\affiliation{Department of Physics, Rhodes College, 2000 N. Parkway, Memphis, TN 38112, USA}

\author[0000-0001-7653-5827]{Ryan J. Cooke}
\affiliation{Centre for Extragalactic Astronomy, Department of Physics, Durham University, South Road, Durham DH1 3LE, UK}

\begin{abstract}
We present the design and first results of the Inner Circumgalactic Medium (CGM) of QSO Line of Sight Emitting galaxies at $z\sim 2-3$, KBSS-InCLOSE. The survey will connect galaxy properties (e.g., stellar mass $M_*$, interstellar medium ISM metallicity) with the physical conditions of the inner CGM (e.g., kinematics, metallicity) to directly observe the galaxy-scale baryon cycle.
We obtain deep Keck/KCWI optical IFU pointings of Keck Baryonic Structure Survey (KBSS) QSOs to discover new star-forming galaxies at small projected distances $b\lesssim12"$ (98 kpc, $\overline{z}=2.3$), then obtain follow-up Keck/MOSFIRE NIR spectra to confirm their redshifts. We leverage KBSS images and Keck/HIRES QSO spectra to model stellar populations and inner CGM absorption. 
In this paper, we analyze two QSO fields and discover more than 15 new galaxies with KCWI, then use MOSFIRE for two galaxies Q2343-G1 ($z=2.43$; G1) and Q2233-N1 ($z=3.15$; N1), which are both associated with Damped Lyman Alpha absorbers.  
We find that G1 has typical $M_*$,UV/optical emission properties. N1 has lower $M_*$ with very strong nebular emission.
We jointly analyze neutral phase CGM and ionized ISM in N/O (for the first time at this $z$), dust extinction, and high-ionization CGM finding that: G1's CGM is metal poor and less evolved than its ISM, while N1's CGM and ISM abundances are comparable; their CGM shows $\sim1$ dex less dust extinction than the ISM; and G1's CGM has direct evidence of hot, metal-rich galactic outflow ejecta. These findings support that metals and dust are driven into the CGM from outflows, but may also be e.g., stripped ISM gas or satellite enrichment.
The full KBSS-InCLOSE sample will explore these scenarios.
\end{abstract}

\keywords{keywords --- Damped Lyman-alpha systems (349), Galaxy chemical evolution (580), Abundance ratios (11), Circumgalactic medium (1879), Interstellar medium (847), Intergalactic medium (813), Quasar absorption line spectroscopy (1317)}

\section{Introduction} \label{sec:intro}
Connecting galaxy properties and the physical conditions of the inner circumgalactic medium (CGM) of $z\sim2-3$ star-forming galaxies is necessary to understand the evolution of baryons in the galaxy. There have been many statistical analyses of the $z=2-3$ ISM \citep[e.g.,][]{steidel+2003,shapley+2005,shapley+2011,kriek+2015,strom+2017} and CGM absorption properties \citep[e.g.,][]{wolfe+1986,wolfe+2005,simcoe+2006,prochaska_2007,rudie+2012,turner+2014,lehner+2014_KODIAQ,peroux+2020,Lehner+2022_KODIAQZ} but it is difficult to make confident ISM-CGM connections of the same galaxy.

This redshift range lies close to the peak of star formation rate density in the universe (sometimes called ``Cosmic Noon'') when many galaxies were rapidly assembling (i.e., heavily star-forming). 
It has been shown that rapidly star-forming galaxies ubiquitously exhibit strong galaxy-scale outflows with occasional evidence for inflows/accretion \citep[e.g.,][]{steidel+2010,shapley+2011,prusinski+2021}.

Strong outflows suggest that a significant amount of gas is being ejected into the CGM (and/or intergalactic medium; IGM).  Similarly, the presence of inflows may indicate that previously-ejected gas---possibly mixed with primordial gas--- falls back onto the galaxy, providing fuel for further star formation and chemical enrichment \cite[e.g.,][]{angles-alcazar+2017,tumlinson+2017}. The process by which baryons transition through different phases (e.g., $T,\rho$) during a galaxy's star formation history (SFH) is known as the ``baryon cycle.'' Such strong outflows suggest that most of the baryons associated with a galaxy reside in the CGM during the peak epoch of galaxy formation, therefore it \textit{must} be included in our understanding of the galaxy formation and evolution process. 

Even though there is a significant amount of mass in the CGM, the gas is very diffuse and difficult to observe directly in emission due to its intrinsically low surface brightness \citep[e.g.,][]{tumlinson+2017}. We must instead rely on unrelated bright background sources (e.g., QSOs) to provide absorption signatures. A high resolution spectrum of the background QSO provides a detailed, albeit singular, view of the galaxy's CGM at a particular projected distance, or impact parameter ($b$), with the caveat that the pencil beam sightline might not be representative of the entire CGM \citep[e.g.,][]{cooke+2010,rudie+2019}. This suggests that when using galaxy-QSO pairs, one we must build a large statistical sample of galaxies with similar properties (e.g., stellar mass M$_*$, star formation rate SFR, metallicity, etc.)\ with background probes sampling a range of impact parameters.

Galaxy-QSO pairs with small impact parameters (i.e., within the galaxy's viral radius or $R_{\rm vir}$) offer the best chance of seeing (1) CGM gas that is directly associated with the galaxy, and (2) a causal connection between the galaxy's ISM and CGM properties, where ongoing star formation and AGN activity --  and resultant feedback processes -- are likely to be reflected in the physical properties of the CGM \citep[e.g.,][]{turner+2014,prochaska+2017,pratt+2018,zahedy+2021_CUBS3,mintz+2022}.

Work over the past decade using galaxy-QSO pairs has increased our understanding of the CGM at low redshift \citep[$z\sim0.3$; e.g.,][]{tumlinson+2013,werk+2014,tumlinson+2017,prochaska+2017}, intermediate redshift \citep[$z\leq$0.3--1; e.g.,][]{chen+2020_CUBS1,zahedy+2021_CUBS3,qu+2022_CUBSV, qu+2023_CUBSVI, chen+2023CUBS8}, $z\sim2-3$ \citep[e.g.,]{rudie+2012,rakic+2012,turner+2014,turner+2015,rudie+2019,nielsen+2022}, and $z\sim3-4$ \citep[e.g.,][]{lofthouse+2020,lofthouse+2023,galbiati+2023}. 

However, only at low-$z$ and intermediate-$z$ is there a large statistical sample (N$\gtrsim$50) of similar galaxies (e.g., with $L\sim L_*$) with CGM gas probed within the virial radius ($b\leq R_{\rm vir}$).
To date there are about two dozen $z\sim2-3$ galaxies that are thought to be associated with QSO absorption systems of various $N_{\rm HI}$ densities and metal line detections at $b<100$ kpc in the literature \citep[e.g.,][]{weatherley+2005,krogager+2017,rudie+2019}. Of these, less than a dozen have well-characterized ionized ISM and stellar populations \citep[i.e., far ultraviolet FUV and optical nebular emission spectra, FUV--Optical SEDs;][]{rudie+2019}.

\citet{rudie+2019} used a sample of 8 $z\sim2.3$ KBSS galaxy-QSO pairs with known galaxy properties and found that the $b\lesssim R_{\rm vir}$ high-$z$ CGM is multiphase (with singly, doubly, and triply ionized species sharing the same component or ``cloud'' structure), kinematically complex (requiring more than 10 components to model the absorption), contains a significant amount of metals (high covering fraction of metal ions, and a large estimated metal halo mass compared to the ISM), has a high occurrence of gravitationally unbound gas (70\% of the galaxies have absorber components with velocities in excess of the escape velocity $v_{\rm esc}\sim$450-550 \kms), and is thermally supported (thermal broadening dominates over turbulent/non-thermal broadening). These first results must be explored further to understand how and if they correlate with galaxy properties such as stellar mass, SFR, nebular metallicity and ionization, \lya\ halo properties, etc. To accomplish this, we must build a larger sample of $z\sim2-3$ galaxy-QSO pairs probing $b<R_{\rm vir}$.

KBSS-InCLOSE (an extension to KBSS-KCWI; \citealt{chen+2021}) is an ongoing campaign to connect galaxy properties with the physical conditions of the Inner CGM of QSO Line Of Sight Emitting (InCLOSE) galaxies at $z\sim2-3$. InCLOSE will leverage the approach (and ancillary data) employed (gathered) by the original KBSS survey \citep[e.g.,][]{rudie+2012,steidel+2014,turner+2014} to connect galaxies to the exquisite information provided by signal-to-noise ratio $S/N\sim 50-100$ HIRES spectra of QSOs carefully selected to lie just ``behind'' the galaxy survey volume. In spite of years of effort obtaining LRIS and MOSFIRE spectra of more than 3,000 galaxies in the redshift range $2 \le z \le z_{\rm QSO}\sim2.8$, 
the census of galaxies in KBSS was known to be highly incomplete at very small angular scales ($\theta \lesssim 8-10$ arcsec), or impact parameters of $b \lesssim 100$ physical kpc (pkpc) due to the ``glare'' of the PSF of the very bright ($V \sim 16-17$) QSOs. 
As a result, there were only 8 spectroscopically identified galaxies with $b < 100$ pkpc (1 of which with $b < 50$ pkpc, \citealt{rudie+2019}) -- and yet these are the only systems in KBSS where the QSO spectrum is probing gas within the virial radius of the galaxies ($\sim 80-90$ kpc; e.g., \citealt{trainor+2012}), where the response of CGM gas to ongoing star formation and resultant feedback processes will be most evident.

InCLOSE will address the incompleteness of the original KBSS survey at small impact parameters to expand the sample reported in \citet{rudie+2019}. Specifically, Keck Cosmic Web Imager \citep[KCWI;][]{morrissey+2018} optical IFU pointings of all the KBSS QSOs is used to find previously ``missed'' galaxies. Then follow-up NIR spectra using the Multi-Object Spectrometer for InfRared Exploration \citep[MOSFIRE;][]{mclean+2012} will be obtained to spectroscopically confirm and characterize the ionized ISM of newly found galaxies.
Each KCWI pointing focuses on the regions within $10-12$\arcs (82-98 pkpc at $z\sim$2.3) of the QSO sightline, specifically configured to discover previously unseen absorbing galaxies within projected distances of $\le 100$ pkpc. This ``discovery'' phase is critical and relies on KCWI's blue sensitivity for detections of Ly$\alpha$ emission (and FUV continuum) in the redshift range of interest, $z \sim 2.08-2.61$.
Additionally, the data cubes allow for accurate subtraction of the QSO PSF which is required for detecting these near QSO line-of-sight (nLOS) galaxies. 

This redshift range is optimized to allow for observations of a large suite of strong nebular emission lines in the observed-frame near-IR atmospheric bands with MOSFIRE including \Ha, \Hb, [\ion{O}{3}], [\ion{O}{2}], [\ion{S}{2}], and [\ion{N}{2}] 
 at $z\sim2.3$, all with resolving power $R \sim 4000$. 
Rest-frame optical nebular spectra enable the measurement of ionized gas-phase physical conditions (\ion{H}{2} regions) in the inner regions of galaxies, allowing for a direct comparison of the galaxy ionized ISM properties with those measured in the CGM. More explicitly they provide: (1) precise galaxy systemic redshifts (uncertainties $|\delta v| < 20$ km s$^{-1}$) from strong non-resonant emission lines (e.g., [\ion{O}{3}]$\lambda$5008, \Ha), (2) gas-phase oxygen abundance (and, for a subset, abundances of N, and S), (3) measurements of star formation rates (SFRs) derived from the dust-corrected H$\alpha$ luminosity (or limits from \Hb\ luminosity), and (4) dynamical mass measurements, from nebular line widths when combined with \textit{HST} sizes, etc.

Importantly, these critical diagnostics are almost entirely lacking in the literature because the inclusion of the rest-optical spectrum of absorption-bearing nLOS galaxies has been done only a handful of times at this $z$ \citep[e.g., ][]{moller+2002, weatherley+2005,christenson+2004,neeleman+2018,christenson+2019, rudie+2019}.

Each KBSS-InCLOSE galaxy will have (assuming $\bar{z} \rm \sim 2.3$): (1) a rest-FUV spectrum from KCWI blue covering 1060-1665~\AA\ in the rest frame, (2) a rest-optical spectrum from MOSFIRE covering 3650-6800~\AA\ in the rest-frame, (3) deep imaging from ground- and space-based observatories covering 0.1-1.4 $\mu$m in the rest-frame, 
including at least one band observed with \textit{HST}/WFC3-IR, (4) a \lya\ emission and velocity map (from KCWI blue), and (5) a HIRES spectrum of the background QSO with average S/N$\sim 50-100$ covering absorption systems from 970-3045 \AA\ in the rest-frame. Using products (1)--(3) we will characterize the ionized ISM and stellar populations of the galaxies, measuring or inferring properties such as systemic redshift, star formation rate SFR, nebular metallicity, stellar mass, etc., and using product (4) we will characterize the global view of the cold emitting CGM via \lya\ and (5) analyze a highly detailed view (at $b<R_{\rm vir}$) of the CGM via background QSO absorption, measuring or inferring \ion{H}{1} and metal abundance, kinematic properties, and thermal properties. Altogether by combining (1)--(5) for a large sample of galaxies we aim to explicitly connect galaxy and CGM properties of star-forming galaxies at $z\sim2-3$.

In this paper, we focus on two galaxies 
toward two QSOs -- Q2343+1232 ($z_{\rm QSO}=2.573$) and Q2233+1310 ($z_{\rm QSO}=3.295$; which is not part of KBSS) --  
for which we have added extensive new optical IFU data from KCWI\@ \citep{christenson+2004,trainor+2012}. These fields have a rich history in the literature and are known to have at least one nLOS star-forming galaxy per field. One of the galaxies, toward Q2343+1232, was missed from the KBSS survey due to its proximity to the QSO \citep[e..g.,][]{strom+2018} but was found recently by \citealt{nielsen+2022} using KCWI. The other galaxy is towards Q2233+1310. It was discovered using rest-FUV color selection and was found by early optical IFU observations to be very bright in \lya\ \citep[e.g.,][]{steidel+1995,steidel+1996,christensen+2007}.

We selected these two galaxies because their brightness and proximity to QSO sightlines allowed us to develop methods to discover and analyze new galaxies. They should not necessarily be taken as representative of the broader InCLOSE population. Rather, they are systems with high-quality data useful for refining our methodology. These methods will be applied to future observations where we will have little {\it a priori} knowledge about $z\sim2-3$ nLOS galaxies surrounding the KBSS QSOs.
The workflow, techniques, and analysis presented here will be applied to the growing KBSS-InCLOSE sample in future papers.

The main results of this paper can be found in Section \ref{sec:baryoncycle}. The paper is organized as follows. In Section \ref{sec:data}, we discuss our observations and data reduction.
In Section \ref{sec:qso_subtraction}, we show how we remove the bright background QSOs from our datacubes and images. In Section \ref{sec:ism}, we analyze the galaxies' ionized ISM emission via rest-FUV spectra, rest-optical spectra, and rest-FUV to rest-optical SED. In Section \ref{sec:cgm}, we analyze the inner CGM of the galaxies via \ion{H}{1} CGM emission (i.e., Ly$\alpha$ halo, and background QSO absorption. In Section \ref{sec:baryoncycle}, we discuss insights into the galaxy scale baryon cycle by explicitly comparing ISM and CGM properties. In Sections \ref{sec:discussion} and \ref{sec:summary}, we discuss our findings and caveats, then summarize the paper. We adopt solar abundances from \citet[]{asplund+2009}. We adopt the following $\Lambda$CDM cosmological parameters: $\rm H_0 = 70 \; km \; s^{-1} \; Mpc^{-1}, \Omega_m = 0.3, \Omega_\Lambda=0.7$.

\section{Data and Observations} \label{sec:data}

\subsection{Discovering nLOS Galaxies with KCWI} \label{sec:kcwi}
Obtaining KCWI pointings of the KBSS QSO fields is crucial for discovering new nLOS galaxies. Our primary KCWI configuration uses the Medium slicer (slice width$=0\secpoint69$) and large blue (BL) grating to provide an optimal compromise between field of view (FoV$\simeq 16\secpoint5 \times 20\secpoint3$, or $\simeq 135 \times 166$ physical kpc (pkpc)  at z$\sim$2.3), wavelength coverage (3500$-5500$ \AA), and spectral resolution ($2.5$~\AA, or $R=1400- 2240$) across the band. 
 
A summary of the observations are shown in Table~\ref{tab:observations}. 
For Q2233+1310, the data comprise a total of 5 hours integration time taken with the same observational approach, in which each of 15 individual 1200s exposures was obtained with a different position angle of the IFU on the sky, with small ($\simlt 1$\arcs\ moves of the center position of the pointing, in order to achieve reasonable spatial sampling and minimize pixel covariance in the final data cube.

\begin{table*}[htb]
\centering
\caption{New Observations Summary} \label{tab:observations}
    \begin{tabular}{ccccccccc}
    Object          &Type           &Instrument     &RA             &Dec            &$t_{\rm exp}$      &Date           &PI                     &P/ID\\
                    &               &(Config.)      &(J2000.0)      &(J2000.0)      &(hr)               &(Y/M/D)          &                       &\\
    \hline
    \hline
    Q2343+1232      &Optical IFU    &KCWI           &23:46:28.30    &+12:48:57.8    &                   &               &                       & \\
                    &               &Med/BL         &-              &-              &5.2                &2018/11/10       &Steidel                &C305\\    

                    &Optical Images &LRIS           &23:46:28.42    &+12:48:57.4    &                   &               &                       & \\
                    &               &U$_n$,G,R      &-              &-              &1.9,0.7,2.2$^a$   &2022/08/23       &Steidel                &C409\\

                    &NIR Spectra    &MOSFIRE        &23:46:28.42    &+12:48:57.4    &                   &               &                       & \\
                    &               &J,H,K          &-              &-              &1.4,1.0,2.0$^b$    &2022/09/17       &Steidel                &C409\\

    \hline
    Q2233+131       &Optical IFU    &KCWI           &22:36:19.21    &+13:26:19.3    &                   &               &                       & \\
                    &               &Med/BL         &-              &-              &2.0                &2021/09/05     &Steidel              &C249\\
                    &               &Med/BL         &-              &-              &3.0                &2021/08/11     &Erb          &R349\\
                    &NIR Spectra    &MOSFIRE        &22:36:19.27    &+13:26:16.9    &                   &               &                       & \\
                    &               &H,K            &-              &-              &0.9,1.0$^c$        &2022/09/17     &Steidel                &C409\\
    \hline
    \end{tabular}%
    \tablenotetext{a}{U$_n$, G, R band exposure times}
    \tablenotetext{b}{J, H, K band exposure times}
    \tablenotetext{c}{H, K band exposure times}

\end{table*}

For Q2343+1232 we combined our own KBSS-KCWI data set, with a total exposure time of 18700~s (5.2 hours) centered at a position 8\secpoint7 NE of the QSO (such that the QSO falls in the SW corner of the final mosaic, where the net integration time is closer to 17500~s), with additional data retrieved from the Keck Observatory Archive\footnote{\href{https://koa.ipac.caltech.edu/cgi-bin/KOA/nph-KOAlogin}{https://koa.ipac.caltech.edu/}} (KOA) (see \citealt{nielsen+2022}) composed of two partially overlapping footprints taken at the same PA, where the region of overlap is centered on the QSO position with a total exposure time of $\sim 5670$~s (1.6 hours). In the region of full overlap of between the two observations \footnote{This includes the position of galaxy Q2343-G1.}, the total exposure time is $\sim 24400$~s (6.8 hours), but the region SW of the QSO has received significantly less integration time (0.8 hours).

We reduced the KCWI data using a custom version of the publicly available KCWI DRP\footnote{\href{https://github.com/Keck-DataReductionPipelines/KCWI_DRP}{https://github.com/Keck-DataReductionPipelines/KCWI\_DRP}}, with modifications as described by \citet{chen+2021} including a second-pass correction to the background subtraction.

Reduced data cubes from each individual 1200~s exposure were combined into a final mosaic using a custom post-DRP pipeline\footnote{\href{https://github.com/yuguangchen1/KcwiKit.git}{https://github.com/yuguangchen1/KcwiKit.git}} (\citealt{kcwikit}; implementation described in \citealt{chen+2021}) that projects each exposure onto an astrometrically-correct grid with spatial sampling of $0\secpoint 3 \times 0 \secpoint 3$ or $0\secpoint 2 \times 0 \secpoint 2$ and wavelengths sampled with 1~\AA/pixel.

The same procedures were used to re-reduce the archival KCWI data (in the case of the Q2343 field) and incorporate them into the final mosaic.

\subsubsection{Need for QSO PSF Subtraction} \label{sec:qso_subtraction}
Galaxies with small impact parameters will have varying levels of ``contaminant'' emission from the bright background QSOs that must be removed. The QSO contamination affects all derived physical properties of the galaxy deduced from the datacubes, extracted spectra, and images. 

The difficulty of QSO subtraction from the IFU cubes is exacerbated by our goal of detecting the nebular line emission \textit{and} stellar continuum emission of the galaxies in the rest-FUV, e.g., to measure kinematics and flux of spatially extended Ly$\alpha$ emission and interstellar absorption lines. 

In the following sections, we present two methods for subtracting QSOs from the KCWI cubes. The first method recovers only extended line emission from the galaxy 
and is used first to detect new line emitting objects, then for analyzing extended line emission, e.g., \lya\ in the emitting CGM\@. The second method recovers the full rest-FUV spectrum of the galaxy (FUV continnum+Ly$\alpha$ halo) and is used to detect nLOS continuum-emitting sources, then measure line flux, line centroids, and equivalent widths of nebular emission and interstellar absorption lines.


\subsubsection{QSO PSF and Continuum Source Subtraction, and Halo Extraction using CubePSFSub} \label{sec:cxpsf}

CubePSFSub is a routine within the package CubExtractor \citep[][hereafter \citetalias{cantalupo+2019}]{cantalupo+2019} that removes QSO emission using an empirical PSF model constructed from the cube itself. It provides an efficient means to discover new line emitting galaxies and extract physical properties of their \lya\ halos. The package is described in detail in \citetalias{cantalupo+2019} but we summarize it here. The inputs for the program include the position of the \textnormal{QSO}, an inner radius ($R_{\rm in}$) to set the normalization constant for the subtraction, an outer radius ($R_{\rm out}$) within which the subtraction is performed, a spectral width ($\Delta \lambda$) to define the PSF size, 
and number of spectral bins to divide the cube ($N_{\rm bins}$).

CubePSFSub makes pseudo-narrowband images using the \textnormal{mean} flux across the spectral width $\Delta \lambda$ (for each spaxel) using $R_{\rm in}$ (usually a cirular aperture with a radius of 2 pixels; the typical two-dimensional 2D Gaussian full width at half maximum (FWHM) of the QSO $\sim$4 spaxels) to set the peak value of the image. \textnormal{It then subtracts a flux scaled, pseudo-narrowband image from each wavelength slice of the data cube. We ran the routine with no spectral binning 
i.e., a running filter.} We mask wavelength layers where we expect there to be line emission 
to reduce oversubtraction.

\textnormal{We varied the spectral width from 50~\AA\ to 750~\AA\ and found that values between 250--500~\AA\ were preferable. We tested this explicitly by fixing all other parameters ($R_{\rm in}$, $R_{\rm out}$), then extracting the resultant extended emission from Q2233-N1's \lya\ halo using CubEx (assuming spatial and spectral filtering of 1 pixel, and a S/N=1.5). We found that PSF sizes between 250--400 \AA\ resulted in comparably extended emission, while values below (above) this threshold resulted in 6\% (2\%) less detected emission. Additionally, values below this threshold left the cube with significantly increased instances of over-subtraction. Values above the threshold increased the overall residuals.} We used a spectral width of 350~\AA\ for Q2343+1232 and \textnormal{300}~\AA\ for Q2233+1310. 

Finally, after PSF subtraction we removed all continuum sources from the cube using the CubeBKGSub package (in CubExtractor) to ensure that neither QSO continuum nor other continuum sources were left in the cube. We use the same parameters as \citetalias{cantalupo+2019} by setting the median filtering to a bin size of \textnormal{50-100}~\AA\ with a \textnormal{spatial and spectral smoothing of 1 pixel (0.3'', 1\AA\@)}.

We are left with a continuum subtracted datacube that contains only extended line emission, which we use to find new line emitters, then characterize the physical properties of the emitting CGM (e.g., physical size, flux, kinematics) in Section \ref{sec:line_emission}. We refer to these cubes as ``QSO+continuum subtracted.''

Spatially extended line emission maps were extracted from these cubes using a combination of CubExtractor and a custom script to find peaks of Ly$\alpha$ emission per spaxel. CubExtractor was used to make segmentation masks to highlight the voxels (RA, Dec, Wavelength) that are within $\rm \pm3000 \; km \; s^{-1}$ of the emission line of interest (i.e., Ly$\alpha$) and have a signal-to-noise ratio greater than 3.5 after the cube is spatially and spectrally filtered by 1 spatial bin (0.2\arcs--0.3\arcs) and 1 spectral bin (1~\AA), which are typical values \citep[e.g.,][]{cantalupo+2019,langen+2023}. The 3D mask is then flattened to 2D by summing all spaxels to a single value (i.e., collapsing over wavelength) and keeping pixels with values $>$1 (i.e., we ensure that there is a detection in each spaxel; discussed more later). The segmentation mask is then applied to the 3D cube and (1) all of the surface brightness (SB) within each spaxel is summed to create spatial distribution maps, then (2) the peak wavelength of each spaxel is located 
and converted to velocity using the systemic redshift measured from Section \ref{sec:line_emission}. 

We compared our methodology to the maps that can be generated from CubEx and find that they are comparable. We used our custom script due to its flexibility to more easily remove the \textnormal{QSO} residuals from the SB and velocity maps.

\subsubsection{QSO Spectral Subtraction using IFSFIT} \label{sec:ifsfit}
IFSFIT \citep{rupke_ifsfit_2014,rupke+2017} uses a purely spectral approach to subtract QSOs from data cubes. This approach can disentangle the spatially and spectrally varying contributions of the QSO in \textit{each} spaxel so as to remove only the QSO light but preserve nLOS galaxy light. 

\citet{rupke+2017} described the method in detail but here we summarize the main differences in our implementation.
We want to remove all emission related to the QSO including its host galaxy so we modified the method to rely solely on the input QSO spectrum. Our method can be broken down into four steps: 1) QSO continuum extraction, 2) QSO subtraction, 3) QSO halo extraction, and 4) and QSO halo subtraction.

Step (1) is to extract the QSO continuum spectrum from the KCWI data cube. We use a circular aperture centered on the QSO position with a diameter equal to the FHWM (seeing) of the pseudo-broadband cube (2D collapsed by summing from 3500-5500 \AA) i.e., where the QSO dominates the spectrum. The QSO centroid and the seeing were determined from a 2D Gaussian fit performed on the pseudo-broadband image. 

Step (2) is to feed the extracted QSO continuum spectrum into IFSFIT without stellar population synthesis models or emission line modeling (steps 1 and 2 from \citealt{rupke+2017}). Algorithmically, the program extracts the spectrum of each spaxel and models it as two components such that
\begin{equation}
S_{\rm SPAXEL}(x,y)=A(x,y)\times S_{\rm QSO}+S_{\rm Cont}(x,y)
\end{equation}
where $S_{\rm SPAXEL}(x,y)$ is the extracted spectrum from location $x,y$, $A(x,y)$ is a constant that scales the input QSO spectrum according to a series of exponential functions (which keeps the QSO spectrum positive-definite and accounts for changes in seeing and spectral sensitivity) based on the $x,y$ location in the cube,  $S_{\rm QSO}$ is the input \textnormal{QSO} spectrum, and $S_{\rm Cont}(x,y)$ is the featureless additive continuum model. For example, if there is a galaxy present in the spaxels at $(x,y)$ then  $S_{\rm Cont}$ = $S_{\rm Galaxy}$. In all cases, both $A$ and $S$ are fit to minimize the residuals between the data ($S_{\rm SPAXEL}$) and model ($A\times S_{\rm QSO}+S_{\rm Cont}$).

\begin{figure*}
    \centering
    \includegraphics[scale=0.995, trim=0.5cm 0.25cm 0 0]{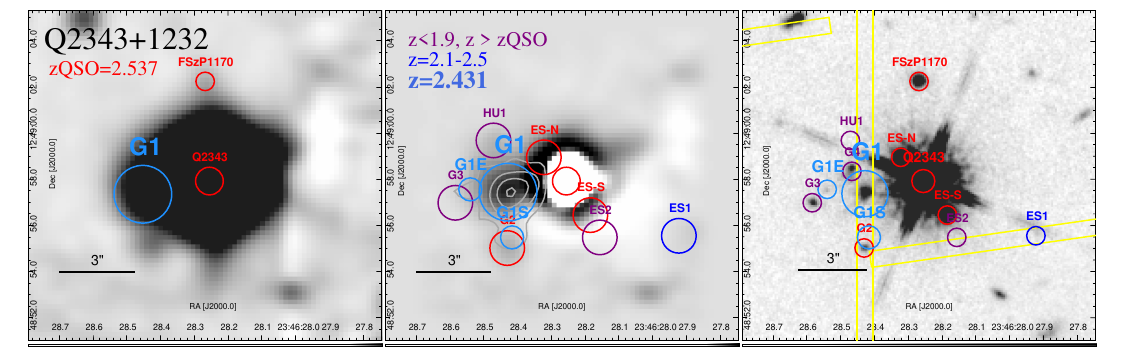}
    \includegraphics[scale=0.995, trim=0.5cm 0.25cm 0 0]{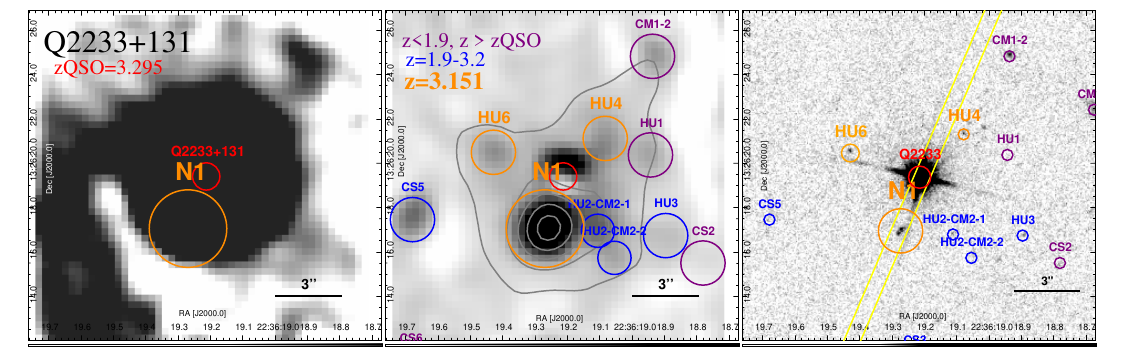}
    \caption{\textnormal{Rest-FUV and rest-Optical images of the QSO sightlines. \textit{Left column:} Keck/KCWI pseudo-broadband (3500--5500 \AA) images before and after (\textit{Middle column}) QSO subtraction; the images are on the same scale and stretch, showcasing the effectiveness of the subtraction (see Section \ref{sec:qso_subtraction}). The grey contours in the post subtracted images show constant \lya\ surface brightness ($\rm v_{\rm \lya} \sim \pm 1000~\kms$; discussed in Section \ref{sec:line_emission}) \textit{Right column:} \textit{HST} images of the QSO sightlines. Objects at a similar redshift have the same color and are summarized in Section \ref{sec:new_objects}. \textit{Top row:} The small red circle shows QSO Q2343, the large blue circle shows galaxy Q2343-G1, the \lya\ contours show 0.35, 0.06, 0.01, 0.0012  $\times10^{-17} \fluxunits~arcsec^{-2}$ for the galaxy, and the \textit{HST}-IR/WFC3 F160W image shows all objects at higher resolution. \textit{Bottom row:} The red circle shows QSO Q2233, the large orange circle shows galaxy Q2233-N1, the \lya\ contours show 0.06, 0.04, 0.02, 0.002 $\times10^{-17} \fluxunits~arcsec^{-2}$ for the galaxy, and the \textit{HST}/WFPC2 F702W image shows all objects at higher resolution. Both galaxies lie at small impact parameter $b\sim~\rm 20 kpc$ (see Table \ref{tab_new_objects}). More than 15 new galaxies were discovered from the KCWI cubes. North is up and east is left. }}
    \label{fig_kcwi_whitelight}
\end{figure*}

The output is a QSO-continuum-subtracted cube, i.e., the difference between the $S_{\rm SPAXEL}$ and $A\times S_{\rm QSO}$, expressed as
\begin{equation}
    S_{\rm cont} = S_{\rm SPAXEL} - A\times S_{\rm QSO}
\end{equation}
In Figure \ref{fig_kcwi_whitelight} we show the results of this procedure before and after being applied to the KCWI cubes towards Q2343+1232 and Q2233+1310. 

Previously known galaxies Q2343-G1 and Q2233-N1 are clearly visible in the subtracted data cube along with more than 7 new galaxies per field.

In order to ensure that the continuum and spectral lines detected in the extracted galaxy spectra were minimally affected by residual contamination from the QSO, we varied the radius of the circular aperture used to define the QSO spectrum and examined the resultant extracted galaxy spectra. 

We found that the galaxy spectra extracted from the QSO-subtracted data cubes were insensitive to the aperture size used to define the QSO ``basis spectrum''; in general, the aperture radius does not affect the strength or centroid of galaxy emission or absorption features or the overall shape of the galaxy continuum (see Appendix \ref{sec:ifsfit_sampling} for more details). We used the aperture that fully \textnormal{sampled} the QSO PSF (i.e., an aperture with a 2-spaxel radius) as our default.

Steps (3) and (4) of the spectral cube subtraction are concerned with extracting and subtracting extended nebular emission associated with the QSO\@ 
because it could affect the extracted spectra of nLOS galaxies, e.g., elevating the continuum surrounding nebular emission, and interstellar absorption lines.
The \lya\ emission is generally more narrow (spectrally) than the emission from the broad-line region meaning it is not completely removed in Step (2). 
To remove the extended narrow-line QSO halo emission, we first visually \textnormal{identified} spaxels that \textnormal{contained} extended \lya\ emission in the QSO continuum-subtracted cube. We \textnormal{extracted} the average \lya\ spectrum ensuring that no spaxels were spatially coincident with emission from other objects in the field (Step 3). In other words, we assume that we can represent the QSO \lya\ halo peak flux and line shape using an average one-dimensional (1D) spectrum\textnormal{,} which should be adequate 
\textnormal{to} remove the most luminous portions of the halo.

We \textnormal{fed} the 1D \lya\ halo spectrum into IFSFIT where it \textnormal{was} run on the QSO continuum-subtracted cube in the same two-component mode described above\footnote{The QSO's \lya\ halo spectrum is featureless (i.e., effectively zero) at all wavelengths besides the QSO peak, because the cube has already been QSO continuum subtracted.}. The resultant cube is QSO continuum+halo subtracted that can be used to extract the spectrum of continuum-detected objects at very small impact parameters with minimal flux from the QSO continuum or \lya\ halo. 

We show the effects of all four steps on the extracted spectra of galaxies Q2343-G1 and Q2233-N1 in Appendix \ref{sec:ifsfit_sampling}.

We refer to the resultant cube as ``QSO-spectrally subtracted.'' With this cube we \textnormal{were} able to find continuum emitting nLOS galaxies and measure their emission/absorption line properties (see Section \ref{sec:fuvcontinuum}). \textnormal{We also used the cubes to verify the extent and flux of \lya\ emission from the galaxies, finding that they are comparable (i.e., Figure \ref{fig_lyamap} is the same regardless of using the QSO-spectrally subtracted cube or the QSO+continuum subtracted cube)}

\textnormal{The 1D rest-FUV spectra of the galaxies were extracted from the QSO-spectrally subtracted KCWI cubes (see Section \ref{sec:ifsfit}) by averaging all of the spaxels that contained continua of the galaxy as seen from the pseudo-broadband images (summed from 3500--5500~\AA\@). We preferred this method over, e.g., weighting the spaxels based on their flux, because the extracted spectrum was otherwise noisier, likely due to to a combination of object faintness and increased noise residuals from the QSO subtraction process.}

\begin{figure}
    \centering
    \includegraphics[scale=0.46]{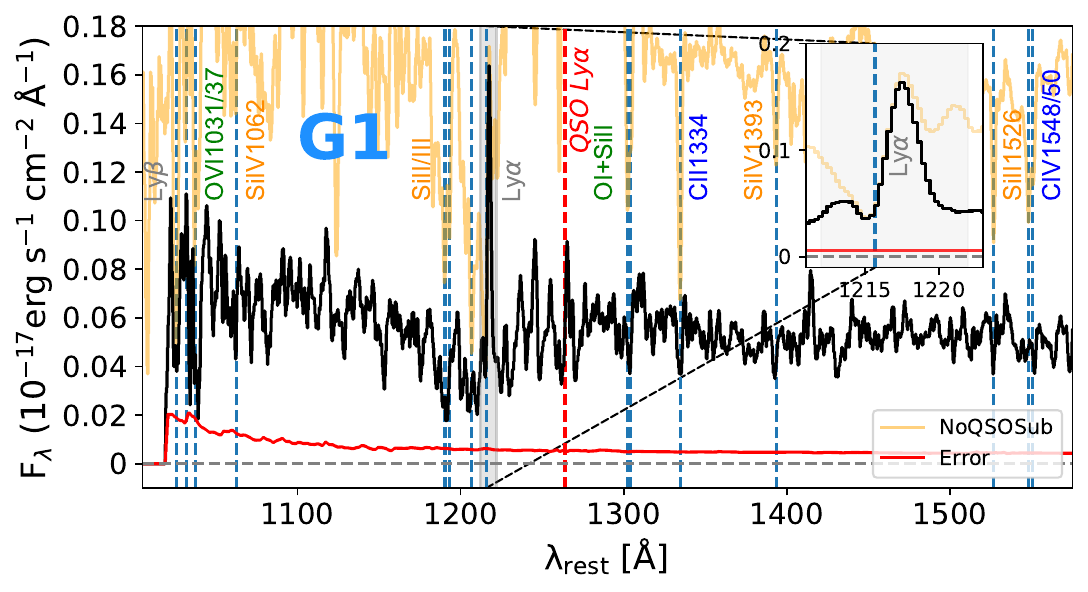}
    \includegraphics[scale=1.03, trim=0.4cm 0 0 0]{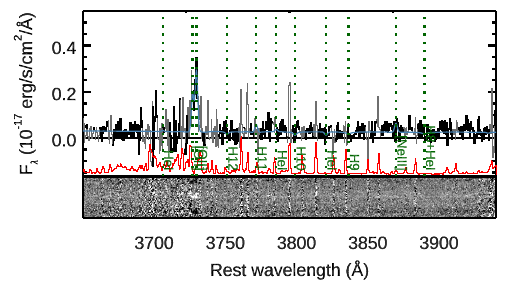} 
    \includegraphics[scale=1.03, trim=0.4cm 0 0 0]{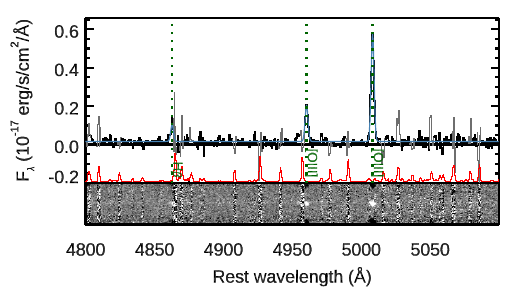}
    \includegraphics[scale=1.03, trim=0.4cm 0 0 0]{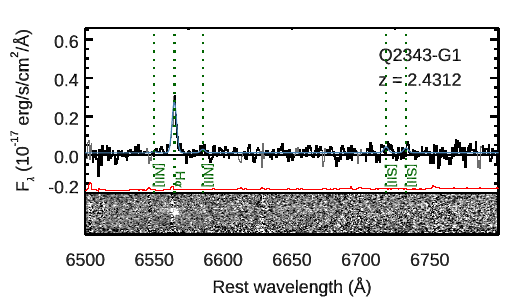}
    \caption{\textnormal{Keck/KCWI and Keck/MOSFIRE spectra of G1. \textit{Top:} Smoothed FUV spectrum from the QSO-spectrally subtracted cube (black), non subtracted cube (yellow-orange), and error spectrum (red). Vertical colored lines show emission and absorption based on G1's systemic redshift. 
    \textit{Bottom panels:} Optical spectrum. The top of each panel shows the 1D spectrum (black), offset error spectrum (red), and modeled emission lines (green dashed). The bottom shows the 2D spectrogram.} 
    }
    \label{fig_kcwi_mosfire_q2343}
\end{figure}

\begin{figure}
    \centering
    \includegraphics[scale=0.46]{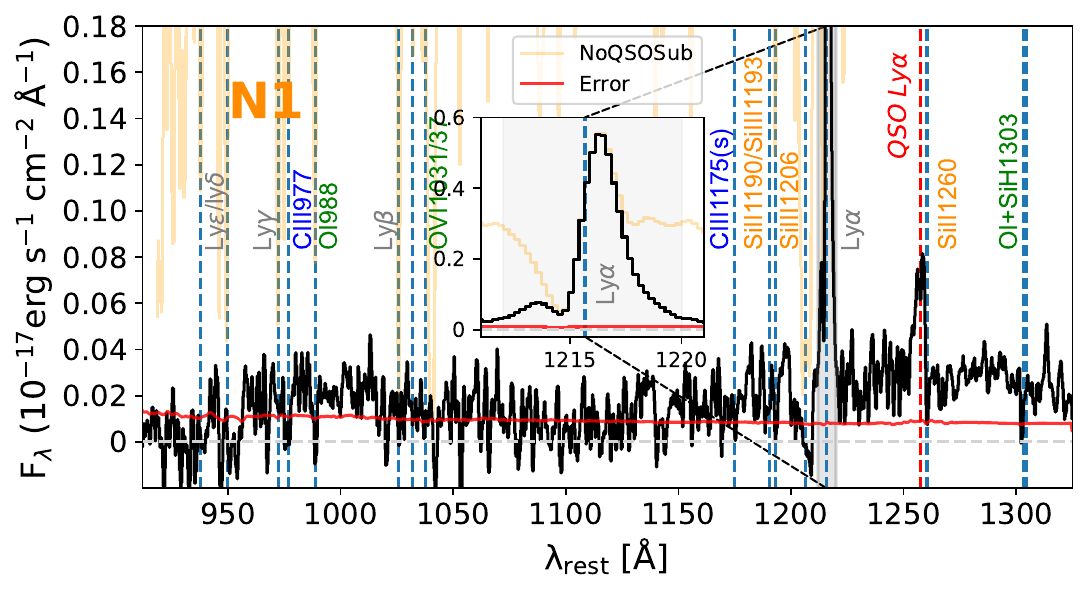}
    \includegraphics[scale=1.04, trim=0.5cm 0 0 0]{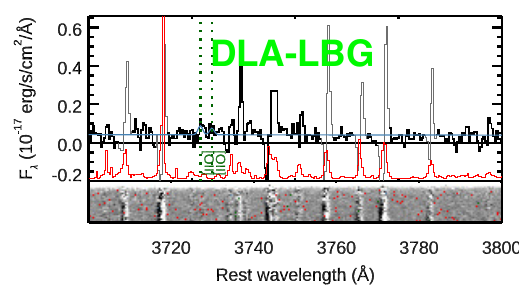} 
    \includegraphics[scale=1.04, trim=0.5cm 0 0 0]{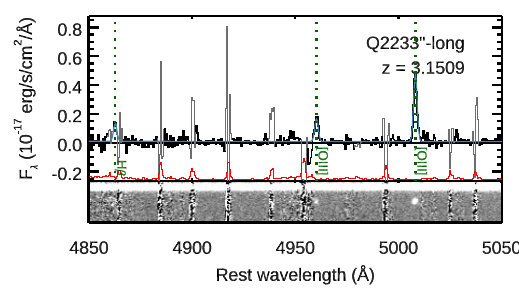}
    \caption{\textnormal{Keck/KCWI and Keck/MOSFIRE spectra of N1. \Ha\ is not accessible from the ground at N1's redshift. Same lines and colors as Figure \ref{fig_kcwi_mosfire_q2343}.}
    }
    \label{fig_kcwi_mosfire_q2233}
\end{figure}

\textnormal{At the top of Figures~\ref{fig_kcwi_mosfire_q2343} and \ref{fig_kcwi_mosfire_q2233} we show the smoothed (3 pixels/\AA; for clarity) rest-FUV spectra of G1 and N1 with IS lines overplotted at the galaxies systemic redshift (Section \ref{sec:mosfire}).
The quality of the KCWI spectrum is such that average IS lines have $S/N\gtrsim2$, which is adequate for checking preliminary redshift; \lya\ has $S/N\geq50$. }

\subsubsection{New Objects Toward Q2343+1232 \& Q2233+131} \label{sec:new_objects}
We \textnormal{identified} previously uncatalogued objects using the deep KCWI cubes. We started classification by making pseudo-broadband images from the non-QSO subtracted cubes where we catalogued new ``Continuum Serendipitous'' sources, which we call CS\#\@; these sources are always found at $b\geq$3--5''. There were sources that would not have been possible to detect without QSO subtraction because they were ``Hidden Under'' the QSO PSF\@.  We call these objects HU\#\@; they are always found at $b\leq$3--5'' and have continuum detections as seen in images made from the QSO-spectrally subtracted cubes (see Section \ref{sec:ifsfit}). Finally, faint line emitting sources were catalogued as ``Emission Serendipitous,'' which we call ES\#\@; these sources are found at all impact parameters by scanning through the QSO+continuum subtracted cubes via narrow bandpasses ($\sim$10--20 \AA)\@; ES\# sources have no continuum detection in the QSO-spectrally subtracted cubes. We catalog the object IDs, their sky position, preliminary FUV redshifts, and some notes in Table \ref{tab_new_objects}. Several objects are omitted from the table and will be discussed in future work.

In field Q2343+1232 (Q2343 hereafter) we discover 7 new objects and \textnormal{recover} 3 objects, the main one being Q2343-G1 (G1 hereafter). G1 was missed by \citet{wang+2015} who searched for H$\alpha$ emission around high metallicity high-\ion{H}{1} absorbers (i.e., Damped Lyman Alpha Absorbers, DLAs) at $z\sim$2.4 due their small footprint of only $\sim$12.5 kpc. \textnormal{G1} was one of three galaxies (G1, G2, G3) \textnormal{discovered} by \citealt{nielsen+2022} (\citetalias{nielsen+2022} hereafter) with KCWI who were searching for host galaxies of strong Mg II absorption at $z\sim2$. Our new cube confirms their main finding that G1 is at the same redshift as the DLA ($z_{\rm DLA}\sim2.431$) but shows that galaxies G2 and G3 are not. They appear as bright continuum emitting objects in the QSO-spectrally subtracted cubes but show no \lya\ or IS lines at the DLA redshift. However, we find two new objects at the same redshift of G1 from their \lya\ emission which we call G1-E (east; towards G3) and G1-S (south; towards G2). It is likely that these are the galaxies that \citetalias{nielsen+2022} originally catalogued. We discuss these new galaxies in detail in Section \ref{sec:galaxy_group?}.

In field Q2233+131 (Q2233 hereafter), we discover 12 new objects and \textnormal{recover} 4 objects, the main one being Q2233-N1 (N1 hereafter). N1 is a well-studied Lyman Break Galaxy \citep[LBG; e.g., ][]{sargent+1989,steidel+1996,moller+2002,christensen+2007,christenson+2019} that was discovered in a search for hosts of optically thick absorbers near QSO sightlines from its UGR continuum photometry \citep[][]{steidel+1995}; the first optical IFU observations were acquired by \citet{christenson+2004,christensen+2007} and the first NIR IFU observations were by \citet{christenson+2019}. 

Recently, \citealt{ogura+2020} acquired ALMA Band 8 (600-800 $\mu$m) observations of the Q2233+1310 sightline and surrounding field to search for \textnormal{$\rm [C II]\lambda 158~\mu$m} emission from N1. No significant emission from N1 was detected, but two other galaxies with significant dust continuum emission were identified, located at angular distances of 4\secpoint7 and 8\secpoint1  from the QSO, respectively. 

We report a significant detection of object Q2233-CS3 which has a previous detection (and photometry) from \citet[][object \#1]{steidel+1995}, and a recent detection in the sub-mm from \citet[][SMG 2]{ogura+2020}. These are all likely the same source because their coordinates are consistent, the F702W and KCWI pseudo-broadband image show a similar morphology, and its colors are consistent with a Lyman break for a galaxy at \lya\ redshift $z_{\lya}=2.077$ \citep[G-R=0.65; U$_n$-G=1.33;][]{steidel+1995}.

We report a significant detection of object Q2233-CS4 which has a previous detection (and photometry) from \citet[][object \#7]{steidel+1995}. CS4 is bright in FUV continuum, and has colors that are consistent with the $z_{\lya}=2.088$ we measure from KCWI \citep[G-R=0.26; U$_n$-G=0.82; ][]{steidel+1995}.

We report a significant detection of Q2233-CS5 and have marginal evidence that it might have been detected recently in the sub-mm by \citet[][SMG 1]{ogura+2020}. The uncertainty arises from its published coordinates and the ensuing $\sim$1\arcs difference that we measure from its position with KCWI. However, we measure $z_{\lya}=2.078$ which is typical of sub-mm galaxies with no continuum detection \textnormal{in} non sub-mm bands \citep[e.g.,][]{chapman+2005,danielson+2017}.

Since Q2233 is not a part of KBSS it lacks crucial ancillary data so we do not include it in the InCLOSE sample.
Nonetheless, it serves as an excellent example of using KCWI to find new nLOS galaxies. We have at least three objects per field that are at redshifts appropriate for rest-optical follow-up. Namely G1, Q2343-G1E, Q2343-G1S, Q2343-ES1 N1, Q2233-HU4, and Q2233-HU6.

\begin{deluxetable*}{lcccccl}
\label{tab_new_objects}
\tabletypesize{\footnotesize}
\tablecaption{New KCWI Objects}

\tablehead{ 
\colhead{KCWI-ID} 	 &\colhead{RA} 	 &\colhead{Dec} 	 &\colhead{$z_{\rm KCWI}$} 	 &\colhead{$b$} 	 &\colhead{$b$} 	 &\colhead{Notes} 	 \\ 
\colhead{} 	&\colhead{(J2000.0)}	 &\colhead{(J2000.0)}	 &\colhead{}	 &\colhead{(")}	 &\colhead{(kpc)}	 &\colhead{}	 \\ 
} 
    
    \startdata 
    \underline{\textnormal{Q2343+1232}} 	                & 	& 	                & 	        		&		& 	        &\\ 
    \underline{\textit{$\rm{1.9} < z < z_{\rm QSO}$}} 	& 	& 	                & 	        		&		& 	        &\\ 
    Q2343-G1		              &23:46:28.4278		&+12:48:57.440		&2.4343			&2.57		&20.8		&\citet{nielsen+2022}\\ 
    Q2343-G1E		             &23:46:28.5397		&+12:48:57.561		&2.4335			&4.17		&33.9		&Same $z_{\rm Lya}$ as G1, towards G3\\ 
    Q2343-G1S		             &23:46:28.4160		&+12:48:55.480		&2.4359			&3.39		&27.5		&Same $z_{\rm Lya}$ as G1, towards G2\\ 
    Q2343-ES1                   &23:46:27.9218		&+12:48:55.550		&2.097			&5.42		&45.1		&\ion{H}{1} but no metals in HIRES\\ 
    \underline{\textit{$z < \rm{1.9}  \;or\;  z \geq z_{\rm QSO}$}}& & 	     	    & 	       		&		    &           &\\ 
    Q2343-G2		              &23:46:28.4303		&+12:48:55.020		&?	        	&3.83       &		    &\citet{nielsen+2022}\\ 
    Q2343-G3		              &23:46:28.5839		&+12:48:56.980		&?		    	&4.84       &			&\citet{nielsen+2022}\\ 
    Q2343-HU1		             &23:46:28.471		&+12:48:59.68		&3.0449			&3.31       &		    &$z_{\rm \lya}$ and Lyb\\ 
    Q2343-ES2		             &23:46:28.156		&+12:48:55.48		&2.661			&2.86       &			&Marginal in f110w,f140w,f160w,No \ion{H}{1} in HIRES\\ 
    Q2343-ESN		             &23:46:28.3218		&+12:48:58.951		&$z_{\rm{qso}}$ &1.41       &		    &Seen in KCWI Lya maps centered at $z_{\rm{qso}}$\\ 
    Q2343-ESS		             &23:46:28.1840		&+12:48:56.448		&$z_{\rm{qso}}$ &1.78       &			&Seen in KCWI Lya maps centered at $z_{\rm{qso}}$\\ 
    \midrule 
    \underline{\textnormal{Q2233+131}} 	                & 	& 	                & 	        & 	        &       &			\\ 
    \underline{\textit{$\rm{1.9} < z < z_{\rm QSO}$}} 	& 	& 	                & 	        & 	        &       &			\\ 
    Q2233+131-N1		        &22:36:19.2733		&+13:26:16.950		&3.1537			&2.55		&19.3		&\citet{steidel+1995}\\ 
    Q2233+131-CS3		        &22:36:19.2296		&+13:26:11.329		&2.0773		    &8.03       &66.9	    &\#1 in \citet{steidel+1995}, SMG2 \citep{ogura+2020}\\
    Q2233+131-CS4		        &22:36:19.5772		&+13:26:06.951		&2.0888		    &13.4       &113 	    &\#7 in \citet{steidel+1995}\\
    Q2233+131-CS5		        &22:36:19.6789		&+13:26:17.458		&2.0781		    &7.02       &58.5	    &May be SMG1 \citep{ogura+2020},Faint in 702w,\ion{H}{1} in HIRES\\ 
    Q2233+131-HU4		        &22:36:19.0827		&+13:26:21.122		&3.1460			&2.61		&19.8		&Similar $z_{\rm Lya}$ as N1\\ 
    Q2233+131-HU6		        &22:36:19.4286		&+13:26:20.497		&3.144			&3.31		&25.1		&Similar $z_{\rm Lya}$ as N1,near multiple sources in F702W\\ 
    \underline{\textit{$z < \rm{1.9}  \;or\;  z \geq z_{\rm QSO}$}}& 	& 	    & 	        & 	        &       &			\\ 
    Q2233+131-CM1-1		        &22:36:18.6267		&+13:26:22.225		&0.4124			&9.01       &			&$z_{\rm OII}$,1/2 in Complex1,Large spiral galaxy\\ 
    Q2233+131-CM1-2		        &22:36:18.9364		&+13:26:24.828		&0.4124			&6.81       &			&$z_{\rm OII}$,2/2 in Complex1,Same $z$ as CM1-1\\ 
    Q2233+131-CS2		        &22:36:18.7803		&+13:26:15.500		&?		    	&7.37       &			&\\  
    Q2233+131-CS6		        &22:36:19.6848		&+13:26:10.131		&?		    	&11.5       &			&\\ 
    Q2233+131-CS7		        &22:36:20.2202		&+13:26:16.619		&0.1466			&14.7       &			&\\ 
    Q2233+131-HU1		        &22:36:18.9428		&+13:26:20.377		&?		    	&4.06       &			&May be part of HU4 complex\\ 
    Q2233+131-HU3		        &22:36:18.8957		&+13:26:16.737		&?		    	&5.29       &			&Faint in 702w,marginal HI and FeII in HIRES\\ 
    \bottomrule 
    \enddata 

\end{deluxetable*}

\subsection{MOSFIRE} \label{sec:mosfire}
All NIR spectra used in this work were obtained using Keck/MOSFIRE \citep{mclean+2012}. 
MOSFIRE uses a cryogenic configurable slit unit (CSU) to form single-slit or multi-slit masks in the telescope focal plane. The integration times and program IDs are summarized in Table \ref{tab:observations}.

In the case of G1 , the object was included as part of a multi-slit mask that also targeted a number of other galaxies of interest in the Q2343 KBSS survey field. The observations were obtained using the same CSU mask configuration in the J, H, and K bands, using the same approach described in previous KBSS work \citep{steidel+2014,strom+2017,strom+2018}. The sky position angle of the instrument field of view was chosen so that G1's 0\secpoint7 $\times$15\secpoint0 slit would also include galaxy Q2343-G2 and Q2343-G4. Total integration times of 1.4, 1.0, and 2.0 hours were obtained in J, H, and K bands, respectively, over the course of the nights of 2022 September 14-16, under variable seeing conditions (0\secpoint5-0\secpoint8~ FWHM)\@. 

In the case of Q2233-N1, spectra in the H and K bands were obtained using MOSFIRE with a 0\secpoint7 longslit oriented to include both galaxy Q2233-N1 and the QSO, using a 2-position nod sequence and a nod offset of 20\secpoint0 instead of the 3\secpoint0 used for multislit masks. 

The data were reduced using the publicly-available MOSFIRE DRP\footnote{\href{https://keck-datareductionpipelines.github.io/MosfireDRP/}{https://keck-datareductionpipelines.github.io/MosfireDRP/}}, which produces background-subtracted, flat-fielded, wavelength-calibrated, telluric-absorption corrected, heliocentric-velocity shifted, rectified, and stacked 2D spectrograms for each slit on the CSU mask. We refer the reader to \cite{steidel+2014} for more details on the data acquisition and reduction, and to \citealt[][]{strom+2017} for details on the flux calibration and slit loss corrections. 

We use MOSPEC \citep{strom+2017}\footnote{\href{https://github.com/allisonstrom/mospec}{https://github.com/allisonstrom/mospec}} to extract 1D spectra from the 2D spectrograms. For Q2343-G1, the spatial profile of the galaxy was modeled as a Gaussian and used to perform optimal extraction of the 1D spectrum for each band. Slit loss corrections were determined separately for the J, H, and K band spectra using a method described by \citet{strom+2017}, which combines information from the spectrum of the calibration star included on the slitmask with comparison to independent measurements (on other KBSS masks) of emission line fluxes of objects that were also observed on the CSU mask. 
The slit loss corrections for G1 are 1.72 $\pm$ 0.05 (J), 1.61 $\pm$ 0.03 (H), 1.33$\pm$0.2 (K). Slit loss corrections for N1 are 1.85$\pm$0.1 (J,H).

We show the MOSFIRE spectra of G1 and N1 in \textnormal{in the bottom panels of Figures \ref{fig_kcwi_mosfire_q2343} and \ref{fig_kcwi_mosfire_q2233}.} G1's rest-optical spectrum (Figure~\ref{fig_kcwi_mosfire_q2343}) shows significant detections (S/N$> 6.5\sigma$) of most of the major diagnostic emission lines including [\ion{O}{2}]$\lambda\lambda 3727,3729$, H$\gamma$, H$\beta$, [\ion{O}{3}]$\lambda\lambda4960,5008$, H$\alpha$, and marginal detections (S/N$>2\sigma$) of [\ion{Ne}{3}]$\lambda 3829$, [\ion{N}{2}]$\lambda\lambda 6549,6585$, [\ion{S}{2}]$\lambda\lambda 6718,6732$. \textnormal{We measure a nebular redshift of $z_{\rm neb}=\rm 2.4312 \pm 0.0005$} and 
adopt this redshift as the systemic redshift for the remainder of the paper (i.e., $z_{\rm sys} = z_{\rm neb}$).

The MOSFIRE spectra of N1 (Figure~\ref{fig_kcwi_mosfire_q2233}) shows significant detections (S/N$>4\sigma$) of [\ion{O}{3}]$\lambda \lambda 4960,5008$ and H$\beta$, and marginal detections (S/N$>2.2\sigma$)) of [\ion{O}{2}]$\lambda \lambda 3727,3729$, [\ion{Ne}{3}]$\lambda 3869$. \Ha\ was not observed because it was shifted out of the NIR atmospheric bands. We measure a nebular redshift $z_{\rm neb}=\rm 3.1509 \pm 0.0005$ 
and adopt it as $z_{\rm sys}$.
We analyze the MOSFIRE spectra in Section \ref{sec:line_emission_optical}.

\subsection{Images \& Photometry} \label{sec:imaging}
We use extant and archival images, and archival photometry to \textnormal{construct}
spectral energy distributions (SED) of the galaxies.
For Q2343, new ground-based UV and optical images ($U_{\rm n}G {\cal R}$; see e.g. \citealt{steidel+2003}) were obtained on August 28, 2022, from \textit{Keck}/LRIS \citep{Oke+1995,steidel+2004}, Table \ref{tab:observations} shows integration times and dates.
Ground-based JH images \citep[published by e.g.,][]{steidel+2014,strom+2017} were taken using \textit{Magellan}/FourStar \citep{persson+2013}. Ground-based K$_S$ images were taken using \textit{Palomar}/WIRC \citep[data published in e.g.,][]{steidel+2004} \citep{wilson+2003}.
Finally, space-based NIR images were taken with \textit{HST}/WFC3IR-F140W (Trainor PID\#14620 2016October05) and \textit{HST}/WFC3IR-F160W (Law PID\#11694 2010June13), and a reduced and science-ready F110W image was pulled from the Hubble Legacy Archive \citep[HLA][]{HLA_lindsay+2010} \textit{HST}/WFC3IR-F110W (Forster Schreiber PID\#12578 2017August21).

For N1, reduced and science-ready images were pulled from the HLA including \textit{HST}/WFPC2IR-702W (Macchetto PID\#6307 19997May09) and \textit{HST}/NICMOS-F160W (Warren PID\#7824 1998Aug07).
UGR photometry is from \citet{steidel+1995}, based on images taken from the William Herschel Telescope (U$n$,R) and \textit{MDM}/Hiltner (G,R), and $K_{\rm s}$ band photometry is from \citet{steidel+1996} based on images taken from \textit{Keck}/NIRC1 \citep{NIRC1_Matthews+1994}.
Our photometry measurements 
are discussed in detail in Section \ref{sec:2d_qsosub}.

\textnormal{In the right panels of Figure \ref{fig_kcwi_whitelight} we show G1, N1, background QSOs Q2343 and Q2233, the newly identified and reobserved KCWI objects (Table \ref{tab_new_objects}), the MOSFIRE slits, and the surrounding foreground and background objects (relative to the QSOs).} 


\subsubsection{Image QSO Subtraction and Photometry} \label{sec:2d_qsosub}
To subtract the QSOs from the ground-/space-based images, we construct effective point spread functions \citep[ePSF][]{anderson+2000,anderson+2016} for each imaging band \textnormal{(U, G, R, J, H, and K)}. 
We find field stars of comparable brightness to the QSOs using IMEXAM \citep{IMEXAM_2022ascl.soft03004S}\footnote{\href{https://github.com/spacetelescope/imexam}{https://github.com/spacetelescope/imexam}}.
Using at least one  field star, we build the ePSF using the EPSFBuilder class in the PHOTUTILS package \citep{bradley_2022}, which is based on the formalism described in detail by \citet{anderson+2000,anderson+2016}. We set all parameters to 1.
Finally, we fit and scale (centroid and peak flux) the ePSF to the QSO 
using a Levenberg-Marquardt fitter, then subtract it. 

We measure the photometry of the galaxies using PHOTUTILS by placing circular apertures at $x$ and $y$ offsets from the centroid of the QSO. 
We take the average median of two separate nearby circular apertures of the same size for background sky subtraction. 
As expected we find very large differences of up to 2.5 mag between QSO-subtracted and non-subtracted images.

We compared the QSO-subtracted ground-based magnitudes to space-based magnitudes in similar bands (e.g., \textit{HST}/F160W and H-band) 
because the higher spatial resolution of HST reduces the contribution of the QSO's PSF wings. We generally found good agreement between the magnitudes \textnormal{0.1--0.3 mag}. 
Additionally, this was useful for placing limits on the photometry error introduced from the QSO subtraction process. We adopt 0.2 magnitude for all photometry measured from QSO subtracted images.  



\subsubsection{SED Fitting} \label{sec:sed_method}
We use a custom SED fitting routine \citep{reddy+2012,theios+2019} that uses pre-computed grids of SED models (Binary Population and Spectral Synthesis models version 2.2, \citealt[][hereafter BPASS]{BPASSv2.2}) 
with a Kroupa initial mass function (IMF \citep[][]{kroupa+2001}), an upper IMF mass of 100 \msun, 
\textnormal{low stellar metallicity $Z_*$=0.002~($Z_*/Z_\odot \sim 0.14$)} 
an SMC-like extinction curve, and a constant SFH. These default parameters were informed by past work on SED fitting of $z\sim \rm 2-3$ galaxies \citep[e.g.,][]{shapley+2005,erb+2006,reddy+2012,reddy+2015,strom+2017}. \textnormal{Specifically, \citet{theios+2019} found that this stellar metallicity was required to reconcile the FUV and optical spectra of a sample of $z\sim2$ KBSS star-forming galaxies; this stellar metallicity also well described a sample of $z\sim3$ LBGs \citep[e.g., Keck Lyman Continuum Survey;][]{steidel+2018,pahl+2023}. Similarly, the resultant nebular metallicities are consistent with previous results \citep[][]{steidel+2016,strom+2018,strom+2022} and recent direct $T_e$-based nebular metallicities from \textit{JWST} \citep[e.g.,][]{rogers+2024,sanders+2024}.}
We run an additional model that forces the stellar ages to be $\geq$50 Myr to more realistically model young stellar populations whose SED ages were younger than the inferred dynamical time estimated from their sizes and nebular line widths \citep[e.g.,][]{reddy+2012}.

\subsection{HIRES} \label{sec:hires}
We use extant and archival optical high-resolution spectra of background QSOs from Keck/HIRES \citep{vogt+1994} to probe CGM absorption from the nLOS galaxies.

Q2343 was observed as part of the original KBSS program \citep[e.g.,][]{rudie+2012,steidel+2014,turner+2014}). We refer the readers to \citet{rudie+2012} for details on the observations and reduction of the spectra but note that the final spectrum includes data obtained from both HIRES and VLT/UVES on UT2 \citep{dekker+2000}. The spectra were reduced, continuum normalized, and coadded, resulting in a spectral resolution of $R \simeq 45,000$ \textnormal{$(FWHM = 6.7~\kms)$}, an \textnormal{average $S/N\sim40-70$ per} resolution element, and \textnormal{spectral range $\lambda \sim3150-10,090~\AA$} \citep{rudie+2012}.

Q2233 was observed a couple of years after HIRES's commissioning (PI: Sargeant, PID: C15H 1996A, Date: 1996-07-21) and was recently included in the KODIAQ survey \citep{omeara+2015_KODIAQ1,moeara+2017_KODIAQ2,omeara+2021_KODIAQ3}. We pulled the fully reduced, continuum-normalized, 1D spectrum from KOA/KODIAQ\@. The spectrum has a resolution of $R\simeq48,000~(FWHM = 6.3 \kms)$, \textnormal{average $S/N\sim15-25$ per} resolution element, and \textnormal{spectral range $\lambda \sim4900-7300 \AA$}.

We analyze the CGM absorption spectra in Section \ref{sec:cgm_abs}.

\section{Ionized ISM and Stellar Population Properties} \label{sec:ism}
In this section we analyze the ionized ISM and stellar population of galaxies G1 and N1 from their rest-FUV \textnormal{spectra}, rest-optical spectra, and SEDs to infer systemic redshift, nebular linewidths, star formation rates, metallicity, ionization, dynamical and stellar mass, and dynamical and stellar age.

\subsection{FUV Continuum} \label{sec:fuvcontinuum}
The FUV spectrum of galaxies is dominated by emission from massive OB stars and contains information about the warm-hot ISM.  
Interstellar absorption (IS) lines seen in the FUV could arise from anywhere between the observer and the ISM\@. Indeed, previous studies have used FUV IS lines to place constrains on metal absorption in the CGM \citep[e.g.,][]{steidel+2010,chen+2020}. 
In our case, the main utility of the rest-FUV spectrum is to measure a preliminary redshift to determine priorities for rest-optical follow-up. Additionally, we measure \lya\ emission flux, \lya\ halo morphology and size, and use IS absorption lines to estimate $z_{\rm{sys}}$ and corroborate $z_{\lya}$.

To measure \lya\ flux and equivalent width we directly integrated over the lines, assuming that the continua on either side of the line was representative of the continuum underlying the emission. We measured the line center by finding the peak flux.
We tabulate the \lya\ measurements for G1 and N1 in Table \ref{tab_fuv_spectra} referring to them as \textnormal{``Continuum Aperture''}.

\textnormal{G1's FUV spectrum (top panel of Figure~\ref{fig_kcwi_mosfire_q2343}) shows strong emission from Ly$\alpha$ and IS absorption from e.g., Ly$\beta$, \ion{O}{6}, \ion{Si}{2}, \ion{Si}{3}, \ion{O}{1}+\ion{Si}{2}, \ion{C}{2}, \ion{Si}{4}, and \ion{C}{4}\@.} 
\textnormal{It has a single red peaked \lya\ emission profile.} 
We measure a \lya\ velocity difference (from $z_{\rm sys}$) of $\Delta v \sim\rm +271~\kms$, 
$F(\lya)_{\rm G1}=(0.94 \pm 0.02)\times10^{-17}~\fluxunits$, using $z_{\rm sys}$ we convert to a luminosity $\log{(L_{\rm \lya}/\rm erg~s^{-1})}=41.64$, and a rest-equivalent width $W_{\rm rest}=(-6.19\pm0.4)$~\AA\@. These values are typical of $z=2-3$ star-forming galaxies of comparable mass to G1 \citep[][]{steidel+2010,steidel+2016,prusinski+2021}.

\textnormal{N1's FUV spectrum (top panel of Figure~\ref{fig_kcwi_mosfire_q2233}) shows strong emission from Ly$\alpha$ and IS absorption from, e.g., higher order Lyman series lines, \ion{C}{2}, \ion{O}{1}, Ly$\beta$, \ion{Si}{2}, \ion{Si}{3}, \ion{O}{1}+\ion{Si}{2}, \ion{Si}{4}, and \ion{C}{4}\@.}
\textnormal{It has a strong, double peaked \lya\ emission profile with a peak separation of $v_{\rm red}-v_{\rm blue}\sim 600~\kms$, blue to red peak ratio of 0.07, and a red peak equivalent width $W_{r}=-39.5$~\AA\@. These values place N1 in the lower quartile of $z\sim3$ LBG continuum-selected galaxies \citep[e.g.,][]{reddy+2009,steidel+2014,trainor+2015}. Interestingly, N1 appears to have properties in between $z\sim2.7$ faint, low mass narrow-band selected LAEs (KBSS-LAEs) and higher-mass $z\sim3$ Lyman Continuum LBGs (LyC-LBGs) analyzed by \citet{trainor+2015}. N1's $EW_{\rm \lya}$ and red-peak velocity offset are consistent with KBSS-LAEs ($EW_{\rm LAE}\sim44$~\AA, $v_{\rm red, LAE}\sim+200~\kms$) while its \lya\ flux and red-blue peak velocity separation are more consistent with LyC-LBGs ($F_{\rm LyC-LBG}\gtrsim 7\times 10^{-17} \fluxunits$, $v_{\rm red-blue, LyC-LBG}\sim+300~ \kms$). Though there are exceptional sources in each sample that have properties that overlap with N1, this comparison suggests that N1 is not necessarily a typical $z\sim3$ LBG or $z\sim2.7$ LAE. We discuss this more in Section \ref{sec:bulk}.}

\begin{table}[htb]
\centering
\caption{\lya~Line Measurements} \label{tab_fuv_spectra}
\resizebox{\columnwidth}{!}{%
    \begin{tabular}{ccc}
    \hline
    \hline
                                        &G1                                 &N1\\
    \hline
    \hline
    \underline{Kinematics$^\dagger$}   &                                   &    \\
    $z_{\rm red}$                      &\zLyarGOne                         &\zLyarDLALBG\\
    $\Delta v_{\rm red}$ (\kms)        &\vLyarGOne                         &\vLyarDLALBG\\
    $z_{\rm blue}$                     &\nodata                            &\zLyabDLALBG\\
    $\Delta v_{\rm blue}$ (\kms)       &\nodata                            &\vLyabDLALBG\\
    \hline
    \underline{Continuum Aperture$^*$}  &                                   &\\
    $F_{\rm red}~(10^{-17}~\fluxunits$)  &\LyarfluxGOne                      &\LyarfluxDLALBG\\
    $F_{\rm blue}~(10^{-17}~\fluxunits$) &\nodata                            &\LyabfluxDLALBG\\
    $F_{\rm tot}~(10^{-17}~\fluxunits$)  &\LyatotfluxGOne                    &\LyatotfluxDLALBG\\
    $\log{(L_{\rm tot}/\lumunits)}$    &\LyatotlumGOne                     &\LyatotlumDLALBG\\
    $W_{\rm red,rest}$ (\AA)            &\LyarewGOne                        &\LyarewDLALBG\\
    $W_{\rm blue,rest}$ (\AA)           &\nodata                            &\LyabewDLALBG\\
    (\lya/\Ha)$_{\rm cont}$             &\LyaHalphaGOne                     &\LyaHalphaDLALBG\\
    $f_{\rm \lya\ esc,cont}$                  &\LyaescGOne                        &\LyaescDLALBG\\
    \hline
    \underline{Halo Aperture$^+$}  &                                   &\\
    $F_{\rm 1}~(10^{-17}~\fluxunits$)     &\LyaCubexredfluxGOne~(G1)$^a$      &\LyaCubexredfluxDLALBG~(N1)$^d$\\
    $\log{(L_{\rm 1}/\lumunits)}$      &\LyaCubexredlumGOne                &\LyaCubexredlumDLALBG\\
    (\lya/\Ha)$_{\rm halo}$             &\LyaCubexredLyaHalphaGOne          &\LyaCubexredLyaHalphaDLALBG\\
    $f_{\rm \lya\ esc,halo}$                  &\LyaCubexredLyaescGOne             &\LyaCubexredLyaescDLALBG\\
    $F_{\rm 2}~(10^{-17}~\fluxunits)$     &\LyaCubexsysfluxGOne~(G1-E)$^b$    &\LyaCubexsysfluxDLALBG~(HU6)$^e$\\
    $F_{\rm 3}~(10^{-17}~\fluxunits)$     &\LyaCubexbluefluxGOne~(G1-S)$^c$   &\LyaCubexbluefluxDLALBG~(HU4)$^f$\\
    \hline
    \end{tabular}
}
    \scriptsize{\tablenotetext{\scriptsize{\dagger}}{\scriptsize{Redshift and velocity have the same error $\pm$ 1 \AA}}}
    \scriptsize{\tablenotetext{\scriptsize{*}}{\scriptsize{Extracted from spaxels showing continuum flux in the QSO-spectrally pseudo-narrowband KCWI images; \textnormal{see Fig.~\ref{fig_kcwi_whitelight}}}}}
    \scriptsize{\tablenotetext{\scriptsize{+}}{\scriptsize{Extracted from spaxels showing extended \lya~flux in the narrowband KCWI images; see Fig.~\ref{fig_lyamap}}}}
    \scriptsize{\tablenotetext{\scriptsize{a}}{\scriptsize{G1: projected area of \LyaCubexredareaGOne $~\rm arcsec^2$}}}
    \scriptsize{\tablenotetext{\scriptsize{b}}{\scriptsize{G1-E: projected area of \LyaCubexsysareaGOne $~\rm arcsec^2$}}}
    \scriptsize{\tablenotetext{\scriptsize{c}}{\scriptsize{G1-S: projected area of \LyaCubexblueareaGOne $~\rm arcsec^2$}}}
    \scriptsize{\tablenotetext{\scriptsize{d}}{\scriptsize{N1: projected area of \LyaCubexredareaDLALBG $~\rm arcsec^2$}}}
    \scriptsize{\tablenotetext{\scriptsize{e}}{\scriptsize{HU6: projected area of \LyaCubexsysareaDLALBG $~\rm arcsec^2$}}}
    \scriptsize{\tablenotetext{\scriptsize{f}}{\scriptsize{HU4: projected area of \LyaCubexblueareaDLALBG $~\rm arcsec^2$}}}
\end{table}

\subsection{Optical Line Emission} \label{sec:line_emission_optical}
We aim to infer the physical quantities associated with the ionized gas (e.g., \ion{H}{2} regions) 
in G1 and N1. 
Rest-optical emission line properties were measured from the MOSFIRE NIR spectra (discussed in Section \ref{sec:mosfire}) using simultaneous 1D Gaussian fits to all lines in a single band using MOSPEC \citep{strom+2017}. 

The first three sections of Table \ref{tab_optical_spectra} tabulate measured emission line properties including redshift, linewidth, inferred dynamical mass, slit loss- (Section \ref{sec:mosfire}) and extinction-corrected (Section \ref{sec:ism_sfr_dust}) line fluxes, and common strong line ratios. The typical S/N for the brightest emission lines are $>$20 ($>$2.5 for marginal detections).

We adopt \textnormal{the $FWHM$ of} the 1D Gaussian fits as the nebular line widths and account for the spectral resolution of the instrument \citep[$\rm \sim30~\kms$][]{mclean+2012}. We measure $\rm \sigma_{\Ha,G1}\sim81~\kms \; and \; \sigma_{[OIII],N1}\sim51~\kms$.

We estimate the dynamical masses of G1 and N1 using their nebular line widths following the same approach as \citet{erb+2006b}:
\begin{equation}
    M_{\rm dyn} = \frac{C\sigma^2r}{G}
\end{equation} 
where C is a constant related to the galaxies' mass distributions, velocity fields, assumed geometries, and inclination angles; $\sigma$ is the line width taken from \Ha\ for G1 and [\ion{O}{3}] for N1 (see Table \ref{tab_optical_spectra}); and $r$ is the radius that we measure from the \textit{HST}/F160W images. More specifically, the radii are the half width at half maximum \textnormal{($HWHM$)} of a 2D Gaussian fit to their continua yielding $r_{\rm G1}=\rm 1.56~kpc$ and $r_{\rm N1}=\rm 1.23~kpc$. The value of C in principle ranges between 1 and 5 (face on to edge on) but we adopt $C \simeq 3.4$ owing to limited information on the galaxies' morphology (average inclination \textnormal{and} velocity correction $\pi/(2v_{\rm obs})$) and not measuring the true circular velocity ($v_{\rm obs}=2.35\sigma/2$). Their resulting dynamical masses are typical for galaxies at their redshift with $\rm \log{(M_{dyn,G1}/\msun)}=9.9~and~\log{(M_{dyn,N1}/\msun)}=9.2$ but N1 is more towards the low-mass end of KBSS-MOSFIRE sample \citep{steidel+2014,strom+2017} \textnormal{and in the lower quartile of the $z\sim3$ LBG mass distribution \citep[KLCS;][]{steidel+2018,pahl+2023}.}

\citet{christenson+2019} acquired adaptive optics-assisted, high spatial resolution NIR IFU data using Keck/OSIRIS to construct [\ion{O}{3}]$+{\rm H\beta}$ emission line maps of N1 in the observed K-band window. They detected [O III]$\lambda5008$ with flux $F(\rm [O~III]\lambda5008) = 2.4\pm 0.5 \times 10^{-17}~\fluxunits$ and marginally detected  [O III]$\lambda4960$ and H$\beta$. The line kinematics and spatial extent were used to infer a dynamical mass of $\sim 3.1 \times 10^9~M_\odot$ and a SFR $\sim {\rm 7.1-13.6~M_\odot yr^{-1}}$ based on their marginal detection of H$\beta$. These values are all consistent with what we have found. 

G1's spectrum appears to be blended with a foreground galaxy at $z_{\rm \Ha}=$1.589. We call the galaxy G1fg (foregound), its trace can be seen a few pixels above G1's emission peaks in the J (top) and H (middle) spectra; its \Ha\ emission is just blueward of G1's [\ion{O}{3}]$\rm \lambda 4960$ emission line (Figure \ref{fig_kcwi_mosfire_q2343}), and there are marginal detections of [\ion{O}{3}] in the J band (top) spectra. We discuss the implication of this on the SEDs in Section \ref{sec:sed_fitting}.

Additionally, the MOSFIRE slit centered on G1 was oriented such that it included G2 and G4 (partially). We detect diffuse emission near G2 at the observed wavelength of the QSO's \Ha\ peak (see the bottom row of Figure \ref{fig_kcwi_mosfire_q2343}) and did not detect optical line emission from G4. This suggests G2 is at the same redshift as the QSO, and that G4 is not at the redshift of G1 and/or that it has weak line emission.


\textnormal{The emission line ratios will be used to infer dust attenuation (\Ha/\Hb), instantaneous SFR (\Ha), ionization parameter (O3, O32, Ne3O2), and nebular metallicity (R23,O32,N2,N2O2) in the following sections.}

\begin{table}[htb]
\centering
\caption{Rest-Optical Measurements} \label{tab_optical_spectra}
\resizebox{\columnwidth}{!}{%
    \begin{tabular}{ccc}
    \hline
    \hline
                                        &G1                                 &N1\\
    \hline
    \hline
    \underline{Line Measurements}       &                                   &\\    
    $z_{\rm [OIII]}$                    &\zGOne                             &\zDLALBG\\
    $\rm Linewidth$                     &\linewidthGOne                     &\linewidthDLALBG\\
    $\log{M_{\rm dyn}}$                 &\MdynGOne                          &\MdynDLALBG\\
    \hline
    \underline{Flux Measurements$^*$}   &$\rm 10^{-17} erg~s^{-1}~cm^{-2}$  &$\rm 10^{-17} erg~s^{-1}~cm^{-2}$\\
    $\rm H\alpha$                       &\HaGOne                            &\HaDLALBG\\
    $\rm H\beta$                        &\HbGOne                            &\HbDLALBG\\
    $\rm H\gamma$                       &\HgGOne                            &\HgDLALBG\\
    $\rm [O~II]\lambda3727$             &\OIIOneGOne                        &\OIIOneDLALBG\\
    $\rm [O~II]\lambda3729$             &\OIITwoGOne                        &\OIITwoDLALBG\\
    $\rm [Ne~III]\lambda3869$           &\NeIIIOneGOne                      &\NeIIIOneDLALBG\\
    $\rm [O~III]\lambda4960$            &\OIIIOneGOne                       &\OIIIOneDLALBG\\
    $\rm [O~III]\lambda5008$            &\OIIITwoGOne                       &\OIIITwoDLALBG\\
    $\rm [N~II]\lambda6549$             &\NIIOneGOne                        &\NIIOneDLALBG\\
    $\rm [N~II]\lambda6585$             &\NIITwoGOne                        &\NIITwoDLALBG\\
    $\rm [S~II]\lambda6718$             &\SIIOneGOne                        &\SIIOneDLALBG\\
    $\rm [S~II]\lambda6732$             &\SIITwoGOne                        &\SIITwoDLALBG\\
    \hline
    \underline{Line Ratios$^\#$}    &                   &\\
    $\rm H\alpha/H\beta ^| $        &\HaHbGOne          &\HaHbDLALBG\\    
    $\rm H\gamma/H\beta ^| $        &\HgHbGOne          &\HgHbDLALBG\\    
    $\rm O3^a$                      &\OThreeGOne        &\OThreeDLALBG\\
    $\rm O32^b$                     &\OThreeTwoGOne     &\OThreeTwoDLALBG\\
    $\rm R23^c$                     &\RTwoThreeGOne     &\RTwoThreeDLALBG\\
    $\rm O3N2^d$                    &\OThreeNTwoGOne    &\OThreeNTwoDLALBG\\
    $\rm N2O2^e$                    &\NTwoOTwoGOne      &\NTwoOTwoDLALBG\\
    $\rm N2^f$                      &\NTwoGOne          &\NTwoDLALBG\\
    $\rm N2S2^g$                    &\NTwoSTwoGOne      &\NTwoSTwoDLALBG\\
    $\rm Ne3O2^h$                   &\NeThreeOTwoGOne   &\NeThreeOTwoDLALBG\\
    \hline
    \underline{Nebular Inferences$^@$}  &                        &\\
    $A_{\rm V}$                         &\AvGOne~mag             &\AvDLALBG\\
    $A_{\rm H\alpha}$                   &\AHalphaGOne~mag        &\AHalphaDLALBG\\
    $E(B-V)_{\rm neb}$                  &\EBVGOne                &\EBVDLALBG\\
    SFR(H$\alpha$)                      &\SFRGOne~\msunyr        &\SFRDLALBG\\
    SFR(H$\beta$)                       &$>$\SFRbGOne~\msunyr    &$>$\SFRbDLALBG~\msunyr\\
    \hline
    \underline{Ionization Parameter$^+$}   &$\log{U}$              &$\log{U}$\\
    $U\rm_{O3}$                            &\logUOThreeGOne        &\logUOThreeDLALBG\\
    $U\rm_{O32}$                           &\logUOThreeTwoGOne     &\logUOThreeTwoDLALBG\\
    $U\rm _{Ne3O2}$                        &\logUNeThreeOTwoGOne   &\logUNeThreeOTwoDLALBG\\
    \hline
    \underline{Strong Line Metallicities$^+$}   &                                   &\\
    \textit{Oxygen Abundance}                   &$\rm 12+\log{(O/H)}$ \; ($\rm [O/H]^{\$}$)                            &$\rm 12+\log{(O/H)}$ \; ($\rm [O/H]^{\$}$)\\
    $\rm O/H_{R23\&O32}$                        &\TwelveOHOThreeTwoRTwoThreeMcGGOne \; (\OHsolarOThreeTwoRTwoThreeMcGGOne)                                                &\TwelveOHOThreeTwoRTwoThreeMcGDLALBG \; (\OHsolarOThreeTwoRTwoThreeMcGDLALBG)\\ 
    $\rm O/H_{O3N2}$                            &\TwelveOHOThreeNTwoGOne \; (\OHsolarOThreeNTwoGOne)                   &\TwelveOHOThreeNTwoDLALBG\\
    \textit{Nitrogen Abundance}                 &$\rm \log{(N/O)}$ \; ($\rm [N/O]^{\$}$)                               &$\rm \log{N/O}$ \; ($\rm [N/O]^{\$}$)\\
    $\rm N/O_{N2}$                              &\logNONTwoGOne \; (\NOsolarNTwoGOne)                                  &\logNONTwoDLALBG\\
    $\rm N/O_{N2O2}$                            &\logNONTwoOTwoGOne \; (\NOsolarNTwoOTwoGOne)                          &\logNONTwoOTwoDLALBG\\
    \hline
    \end{tabular}
}
    \scriptsize{\tablenotetext{\scriptsize{*}}{\scriptsize{Slit loss and extinction corrected \citet{cardelli+1989} , R$_V$=3.1}}}
    \scriptsize{\tablenotetext{\scriptsize{\#}}{\scriptsize{The notation $\lambda\lambda$ refers to the sum of the two lines}}}
        \scriptsize{\tablenotetext{\scriptsize{|}}{\scriptsize{Before extinction correction}}}
    \scriptsize{\tablenotetext{\scriptsize{a}}{\scriptsize{O3=$\rm \log{([O~III]\lambda5008/H\beta)}$}}}
    \scriptsize{\tablenotetext{\scriptsize{b}}{\scriptsize{O32=$\rm \log{([O~III]\lambda\lambda4960,5008/[O~II]\lambda\lambda3727,3729)}$}}}
    \scriptsize{\tablenotetext{\scriptsize{c}}{\scriptsize{R23=$\rm \log{(([O~III]\lambda\lambda4960,5008 +[O~II]\lambda\lambda3727,3729)/H\beta)}$}}}
    \scriptsize{\tablenotetext{\scriptsize{d}}{\scriptsize{O3N2=$\rm \log{([O~III]\lambda5008/H\beta)} -\log{([N~II]\lambda6585/H\alpha)}$}}}
    \scriptsize{\tablenotetext{\scriptsize{e}}{\scriptsize{N2O2=$\rm \log{([N~II]\lambda6585/[O~II]\lambda\lambda3727,3729)}$}}}
    \scriptsize{\tablenotetext{\scriptsize{f}}{\scriptsize{N2=$\rm \log{([N~II]\lambda6585/H\alpha)}$}}}
    \scriptsize{\tablenotetext{\scriptsize{g}}{\scriptsize{N2S2=$\rm \log{([N~II]\lambda6585/[S~II]\lambda\lambda6718,6732)}$}}}
    \scriptsize{\tablenotetext{\scriptsize{h}}{\scriptsize{Ne3O2=$\rm \log{([Ne~III]\lambda3829/[O~II]\lambda\lambda3727,3729)}$}}}
    \scriptsize{\tablenotetext{\scriptsize{@}}{\scriptsize{Calibrations from \citet[]{theios+2019}.}}}
    \scriptsize{\tablenotetext{\scriptsize{+}}{\scriptsize{Calibrations from \citet[]{strom+2018} and \citet{mcgaugh+1991}.}}}
\end{table}

\subsubsection{Dust and Instantaneous SFR} \label{sec:ism_sfr_dust}
We calculate instantaneous SFR (SFR(\Ha)) following the method described in \citealt{theios+2019} (hereafter \citetalias{theios+2019}). Briefly, \citetalias{theios+2019} updated the calibration constant used to convert H$\alpha$ luminosity ($L_{\Ha}$) \textnormal{to $SFR(\Ha)$ by} modeling the SEDs of a representative sample of $z\sim2-3$ KBSS galaxies using stellar population synthesis models that self consistently reproduced the joint rest-FUV and rest-optical spectra of the galaxies. The SFR follows the form 
\begin{equation} \label{eq:sfr}
    \log{(\rm{SFR}_{H\alpha}/M_\odot \rm{yr^{-1}})} = \log{(L_{\rm{H\alpha}}/\rm{erg~s^{-1})}} - C
\end{equation}
where \citetalias{theios+2019} found $C$ to be 41.64, leading to SFRs almost a factor of 2 lower than the canonical values at $z \sim 0$ \citep{kennicutt+1994}. 

To correct $L_{\Ha}$ for dust extinction we use the Balmer decrement ($F$(\Ha\@)/$F$(\Hb\@)) to infer nebular dust attenuation assuming \citet{cardelli+1989} extinction ($R_{\rm V}$=3.1) and case B recombination at 10$^4$ K in the low density limit $\rm 10^2~cm^{-3}$ \citep[$I$(\Ha)/$I$(\Hb\@)=2.86;][]{brocklehurst+1971,osterbrock+1989}. We correct the \Ha\ line flux for dust attenuation, use $z_{\rm sys}$ to convert to luminosity, then convert to SFR(\Ha) using Equation \ref{eq:sfr}.

N1 is at a higher redshift such that \Ha\ is not accessible from the ground so we 
place a lower limit on the instantaneous SFR by assuming no dust and converting \Hb\ to \Ha\ under the same assumptions as above (\Ha\@ = 2.86$\times$ \Hb\@), convert to luminosity then SFR; we call this quantity SFR(\Hb). The lack of a [\ion{C}{2}]$\lambda158\mu m$ nor sub-mm (dust) continuum emission supports that N1 has relatively low dust \textnormal{attenuation} \citep{ogura+2020}.

We tabulate SFR(\Ha), SFR(\Hb), \Ha\ extinction, $A_{\rm V}$ extinction, and nebular reddening in Table \ref{tab_optical_spectra}. 
The SFR(\Ha) and SFR(\Hb) of both galaxies is within a factor of 2 the median SFR of the KBSS-MOSFIRE sample \citep{steidel+2014}. G1's Balmer decrement, extinction, and nebular reddening are also typical. 
We infer a SFR surface density for both galaxies using the effective radius used to calculate dynamical mass in Section \ref{sec:ism}. Assuming a disk geometry at an average inclination angle ($\pi$/4) yields $\Sigma_{SFR(\Ha), G1}=0.48~\msunyr~kpc^{-2}$ and $\Sigma_{SFR(\Hb), N1}\gtrsim1.34~\msunyr~kpc^{-2}$, which are well above the typical threshold \textnormal{found in galaxies that} drive galactic outflows at lower redshift  \citep[$\rm \Sigma_{SFR}\sim0.1$][]{heckman+2015}. 

\subsubsection{Nebular Excitation and Metallicity} \label{sec:ism_metallicity_ionization}
The metal content in the ISM gives insights on recent star formation and feedback sources (e.g., core collapse supernovae CCSNe, asymptotic giant branch AGB stellar winds) present in HII regions. In practice, usually only hydrogen, nitrogen, oxygen, and sulfur abundances are reliable because they \textnormal{are} bright and \textnormal{therefore} often detected at high $z$. It is a non-trivial task to map from emission line flux to abundance (even abundance ratios) due to varying ionizing state, inhomogenities in the gas density and temperature, and dust extinction present in galaxy environments. In the past, authors would use calibrations found at low-$z$ (where most of the degeneracies can be broken) and apply them to high-$z$. Recently, close attention has been paid to the so called ``strong line'' metallicity calibrations specifically for $z\sim2-3$ galaxies, which we discuss and adopt in this paper.

We use strong line calibrations provided by \citealt{strom+2018} (hereafter, \citetalias{strom+2018}) to infer $12+\log{\rm O/H}$, $\log{\rm N/O}$, and $\log{U}$.
\citetalias{strom+2018} developed the calibrations by using photoionization models that account for the chemistry, ionization, and excitation of nebular gas for a representative sample of KBSS-MOSFIRE galaxies.
We tabulate the values in the last two sections of Table \ref{tab_optical_spectra}. Under each of the metallicities we show their logarithmic abundances with respect to solar values.

Both galaxies are in (near) the low-metallicity branch of the O/H vs R23 space based on their R23 ratios ($\rm R23_{N1}=0.87$ and $\rm R23_{G1}=1.03$) but \citetalias{strom+2018} only calibrated $12+\log{\rm O/H}$ for the high-metallicity branch. Therefore we use the calibrations by \citet[][shown explicitly by \citealt{steidel+2016}]{mcgaugh+1991} to map from R23 (and O32) to O metallicity ($\rm O/H_{R23\&O32}$), which we show in the first row of the Oxygen Abundance section of Table~\ref{tab_optical_spectra}.

The inferred $\rm 12+\log{(O/H)}$ values for G1 are internally consistent regardless of ratio used. For G1 we adopt $\rm 12+\log{(O/H)}_{G1,O3N2}=8.39$ ([O/H]$_{\rm G1}=-0.30$).  
This metallicity is typical of KBSS star-forming galaxies. We have only one appropriate O inference for N1, which yields $\rm 12+\log{(O/H)}_{N1,R23\&O32}=7.82$ ([O/H]$_{\rm N1}=-0.87$) making it low-metallicity compared to the KBSS-MOSFIRE sample.

The inferred $\log{(\rm N/O})$ values for G1 are also internally consistent. We adopt $\log{\rm (N/O)}_{\rm G1,N2O2}$=-1.21 ([N/O]$_{\rm G1}=-0.35$).
We do not calculate (N/O) for N1 because [\ion{N}{2}] was not observable from the ground.

\textnormal{The inferred $\log{U}$ for both galaxies are fairly consistent regardless of ratio used.}

\textnormal{Overall, both galaxies ionization parameters are typical of star-forming galaxies at their $z$}. G1's gas-phase nebular metallicity as seen in $\log{\rm (O/H)}$ and $\log{(\rm N/O)}$ is also typical whereas N1's metallicity is \textnormal{low}.

\subsection{SED Properties} \label{sec:sed_fitting}

We want to model and infer physical properties of the galaxies underlying stellar populations.
In Section \ref{sec:imaging} and \ref{sec:2d_qsosub} we constructed QSO-subtracted SEDs. However, bright line emission (e.g., \lya, [\ion{O}{3}], \Ha) can affect the continuum photometry, which can result in inaccurate SED modeling. 

\textnormal{We use the line fluxes measured from the KCWI spectra (from continuum spaxels; see Section \ref{sec:fuvcontinuum}) and slit-loss corrected MOSFIRE spectra (see Section \ref{sec:ism}) to subtract strong line emission from the measured photometry.} The corrections had the \textnormal{form $\rm \Delta M=-2.5\log{((F(Phot~Flux)-F(Line~Flux))/F(Phot~Flux))}$. For} G1 we correct F110W, J, and F140W band photometry for line emission from [\ion{O}{2}]$\lambda\lambda3727,3729$ \textnormal{and} [\ion{Ne}{3}]$\lambda$3869, F140W and F160W band from \Hb and H$\gamma$, and 
K$_S$ band from \Ha, [\ion{N}{2}]$\lambda\lambda6549,6585$, and [\ion{S}{2}]$\lambda\lambda6718,6732$.  The corrections were the largest for J and K$_S$ at 0.1 mag, but small for F110W, F140W, and F160W at 0.03-0.05 mag, smaller than the photometry error. G band requires no line correction because \lya\ falls outside of the bandpass, and U and R bands contain no bright emission lines (as seen in the KCWI rest-FUV spectra).

For N1, we correct G band from \lya,  K$_S$ band from [\ion{O}{3}]$\lambda\lambda4960,5008$, and \Hb, and F160W band from [\ion{O}{2}]$\lambda\lambda3727,3729$ and [\ion{Ne}{3}]$\lambda$3869.  The correction for K$_S$ was very large at 1.2 mag, but small for G and F160W band at 0.1 and 0.08 mag, respectively. U and R bands contain no bright emission lines. 

G1's SED has an additional complication because it is blended with foreground galaxy G1fg ($z_{\Ha}=$1.589; see Section \ref{sec:line_emission_optical}). The foreground galaxy has a detected trace in J and H band, but does not have strong emission lines in any band, and no detected trace in K band (\textnormal{Figure \ref{fig_kcwi_mosfire_q2343})}. The J and H band trace suggests that there \textnormal{is elevated} flux in R, F110W, and F140W mags that will affect the reddening determination from the SED. The lack of a trace in K band suggests that the K$_S$ band photometry is negligibly affected and will therefore not significantly affect the mass determination (which relies strongly on K$_S$ photometry).

We ran each galaxy's QSO-subtracted and emission line-corrected photometry through the SED fitting routine discussed in Section \ref{sec:sed_method}. From the best-fit BPASS v2.2 SED models we infer stellar mass $M_*$, stellar age $t_*$, dust attenuation (continuum color excess) $E(B-V)_{\rm SED}$, and star formation rate $\rm SFR_{SED}$\@. 

\begin{figure}
    \centering
    \begin{overpic}[scale=0.3]{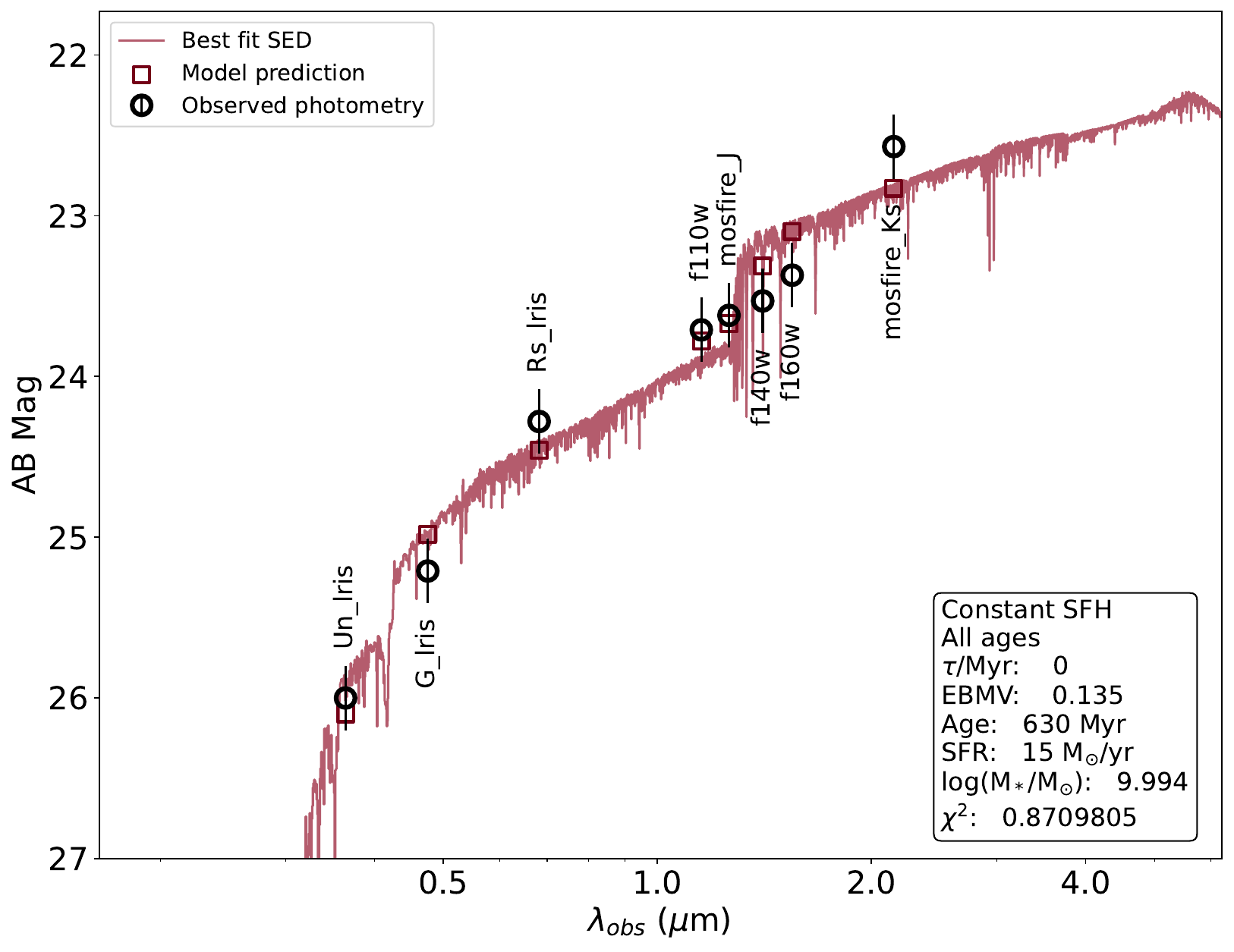} \put(50,68){\color{blue}\large G1}
    \end{overpic}
    \begin{overpic}[scale=0.3]{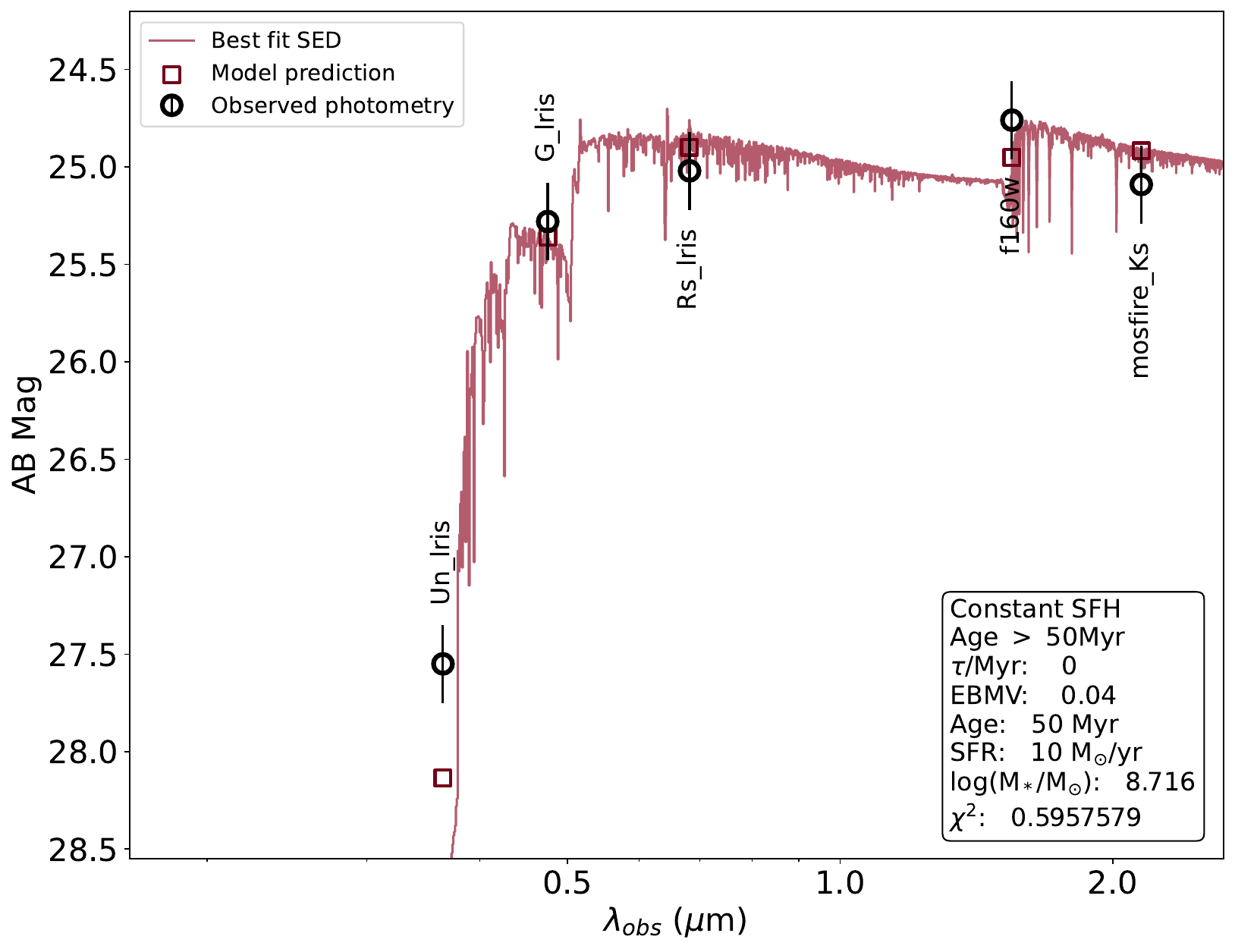} \put(50,68){\color{orange}\large N1}
    \end{overpic}
    \caption{Best fit SEDs from BPASSv2.2 for galaxies G1 (\textit{top}) and N1 (\textit{bottom}). Black points show observed AB magnitudes (after QSO subtraction and bright emission line correction) and red points/curves are from the best fit model. The best fit parameters for the galaxy are shown at the bottom right of plot and are tabulated in Table \ref{tab_sed}. \textit{Top:} G1 and \textit{Bottom:} N1. 
    }
    \label{fig_sed}
\end{figure}

We show the best-fit SED models in Figure \ref{fig_sed} and tabulate the best-fit values in Table \ref{tab_sed}. 
We can see that the QSO- and emission line-corrected photometry (black points) are well fit by their respective models. 

G1 has a stellar mass and age that is consistent with its dynamical time and dynamical mass, its SFR(SED) is within a factor of 2.5 of SFR(\Ha), and it has similar SED and nebular reddening. This agreement shows a consistency between the two independent methods and is evidence that the foreground galaxy G1fg had a small effect on its SED.

For N1, we adopt the model that constrains the age to be $\geq$50 Myr because the best fit BPASS model with no age constraint preferred an age $t_*\sim\rm 10~Myr$ which is about $\sim$20 Myr \textit{younger} than the dynamical time inferred from its size and nebular line width $\tau_{\rm N1,dyn}\sim30$ Myr. With the new age-constrained model, the dynamical mass and SED mass differ by 0.5 dex ($\rm \log{(M_{SED}/\msun)}=8.7 \;and\; \log{(M_{dyn}/\msun)}=9.2$), stellar age and dynamical time are within 0.3 dex, and SFR$_{\rm SED}$ is consistent with the lower limit set by SFR(\Hb), and it has negligible reddening (E(B-V$_{SED}$=0.04; which is consistent with N1's lack of a [\ion{C}{2}]$\lambda158\mu m$ nor sub-mm (dust) continuum emission \citep{ogura+2020}.). \textnormal{These masses are low, but not necessarily atypical, when compared to the $z\sim3$ Keck Lyman Continuum Survey (KLCS) which has a mean stellar mass and standard deviation of $M_*\sim9.6 \pm 0.6$ \citep{steidel+2018,pahl+2023}}.
\textnormal{This all suggests that N1 is} a young, low dust, lower mass, moderately star-forming galaxy.

\begin{table}[htb]
\centering
\caption{BPASSv2.2 Best-Fit SED Parameters} \label{tab_sed}
    \begin{tabular}{ccc}
    \hline
    Parameter                               &G1     &N1$^+$ \\
    \hline
    $\log{(M_*/M_\odot)}^*$                 &9.99   &8.72\\
    $t_*$ (Myr)                             &630    &50\\
    $E(B-V)_{\rm SED}$                      &0.135  &0.04\\
    SFR$_{\rm SED}$ ($\rm M_\odot/yr$)$^*$  &15     &10\\
    \hline
    \end{tabular}%
    \tablenotetext{*}{Typical uncertainty of $\sim$30\%}   
    \tablenotetext{+}{Due to age constraint, values could vary by $\sim\rm50\%$}
\end{table}

\subsection{Comparing ISM and Stellar Population Properties} \label{sec:bulk}
We now combine the ionized ISM and stellar population properties and put them in context with the KBSS-MOSFIRE sample of $z\sim \rm 2-3$ star-forming galaxies and AGN \citep[][; with a $\rm >2\sigma$ measurement of O3 and N2]{steidel+2014,strom+2017}, and low-$z$ galaxies from SDSS-DR8 \citep{aihara+2011} taking emission line measurements and inferred physical properties from the MPA-JHU\footnote{\dataset[https://www.sdss4.org/dr17/spectro/galaxy\_mpajhu/]{https://www.sdss4.org/dr17/spectro/galaxy\_mpajhu/}} catalogs. As noted by \citealt{strom+2017} (\citetalias{strom+2017} hereafter), this particular set of SDSS galaxies was chosen due to their similar detection properties to KBSS-MOSFIRE\@. 
 They have $\rm{0.04}\le z \le \rm{0.1}$, $\rm >50\sigma$ H$\alpha$ detections, $>3\sigma$ detection's of H$\beta$, [\ion{O}{3}], a good redshift measurement and ``reliable'' flag from the MPA-JHU pipeline. The reported $M_*$, SFR, sSFR, are the median estimates from the ``total'' value probability distribution functions.

\begin{figure}
    \centering
    \includegraphics[scale=0.7]{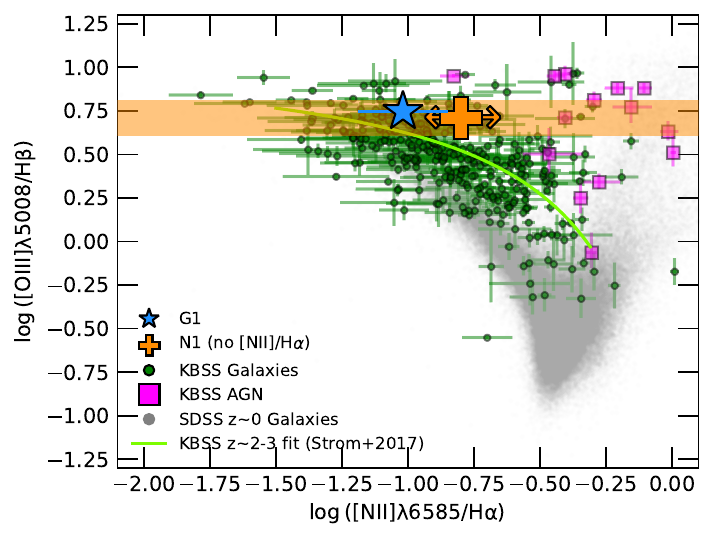}
    \includegraphics[scale=0.7]{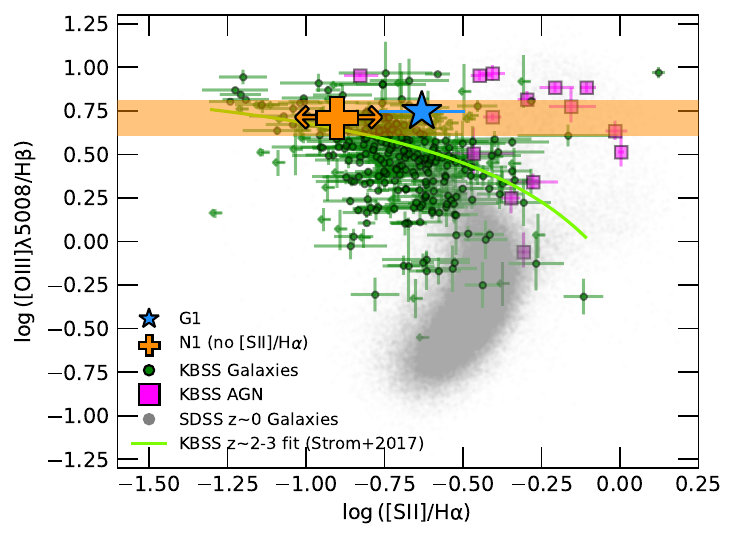}
    \caption{BPT diagrams showing G1 (blue filled star) and N1 (orange filled plus). Green points are $z\sim2-3$ KBSS star forming galaxies, magenta points are KBSS AGN, and grey points are $z\sim0$ SDSS star-forming galaxies. Light green lines are fits to the KBSS galaxies from \citetalias{strom+2017}. Note that N1 does not have an observation of [\ion{N}{2}], \Ha, or \ion{S}{2} so we show it is an unbound point and orange horizontal bar whose width is equal to its O3 error. \textit{Top:} N2-BPT diagram. \textit{Bottom:} S2-BPT diagram.}
    \label{fig_BPT}
\end{figure}

\begin{figure}
    \centering
    \includegraphics[scale=0.7]{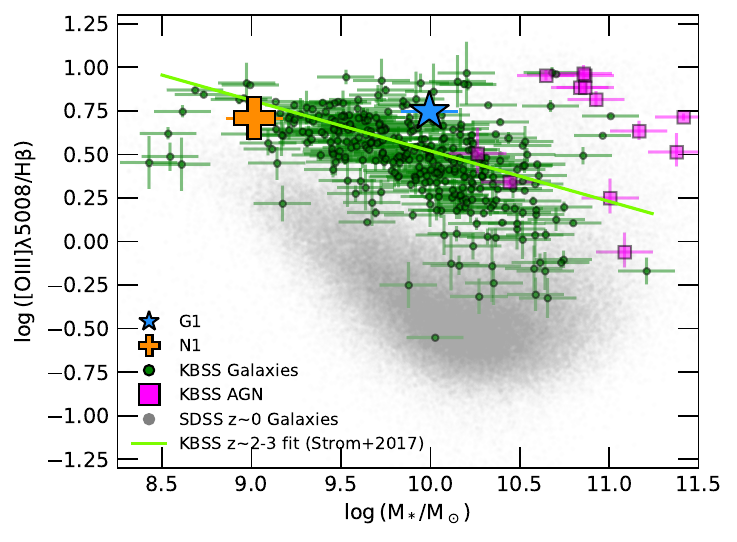}
    \includegraphics[scale=0.7]{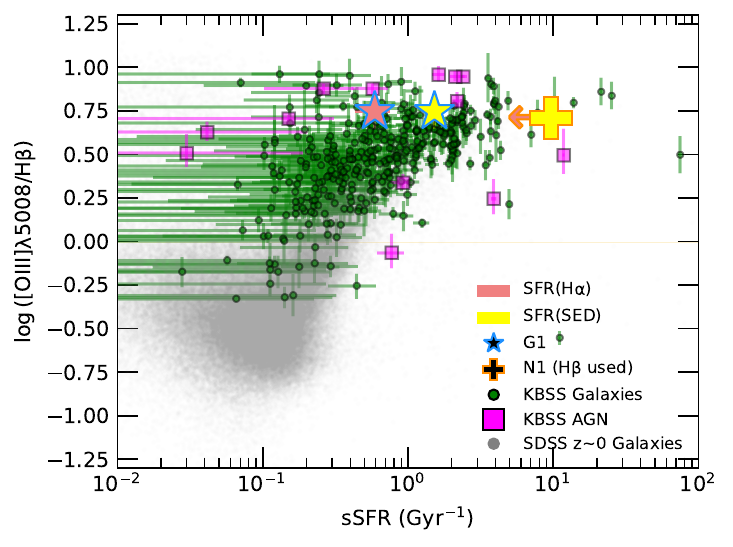}
    \caption{Combined rest-optical and SED properties of G1 (blue outlined star) and N1 (orange outlined cross). The symbols and colored lines are the same as Figure \ref{fig_BPT}.
    \textit{Top:} Stellar Mass--Excitation plot. \textit{Bottom:} sSFR--Excitation plot. Light red filled points show sSFR calculated from SFR(\Ha) (\Hb\ for N1), and yellow filled points are calculated from SFR(SED). N1s \Hb\ SFR(\Hb) (lower limit) and SFR(SED) are roughly the same so appear as one point.}
    \label{fig_MEx_sSFREx}
\end{figure}

Figure \ref{fig_BPT} shows the galaxies on the \textnormal{N2- \citep{baldwin+1981} and S2-BPT \citep{veilleux+1987} diagrams} compared to KBSS-MOSFIRE and SDSS star-forming galaxies. 
Both galaxies have typical O3 ratios compared to KBSS-MOSFIRE that are well above the SDSS galaxies. G1 in particular sits close to both\textnormal{ N2- and S2-BPT} locus fits by \citetalias{strom+2017}. 

The top panel of Figure \ref{fig_MEx_sSFREx} shows the Mass--Excitation ($M_*$--O3; \citep{juneau+2011}) diagram. We use the SED derived mass for G1 (which is within 0.1 dex of the dynamical mass) and use a logarithmic average of the dynamical and SED mass for N1 ($\rm \log{\msun}\sim9.0\pm0.25$). Both galaxies reside within the majority of the KBSS-MOSFIRE sample which itself is vertically offset from $z\sim$0 galaxies indicating higher excitation per unit stellar mass. G1 is in the upper quartile of the distribution towards higher excitation. 
N1 falls \textnormal{on} the low-mass end of the median fit from \citetalias{strom+2017}.

The bottom panel of Figure \ref{fig_MEx_sSFREx} shows the galaxies' specific star formation rate (sSFR) and excitation (O3). We show sSFR calculated from SFR(SED), and SFR(\Ha) (or SFR(\Hb) for N1; see section \ref{sec:ism_sfr_dust}). The \Ha\ and SED SFRs for both galaxies are fairly consistent with one another suggesting a consistent picture between the nebular emission spectra and SED modeling.

\textnormal{Altogether, G1 and N1 have similar O3 ratio and $\log{U}$ but otherwise appear to be on different ends of the $z=2-3$ star-forming galaxy distribution. G1 has a typical stellar mass, sSFR (instantaneous and SED), dust attenuation, \lya\ continuum flux, \lya\ equivalent width, and N2- and S2-BPT diagram location when compared to KBSS-MOSFIRE. N1 on the other-hand is a lower mass (dynamical and SED), high sSFR (SED(SFR) and SFR(\Hb), relatively dust-free, and strong \lya\ and \ion{O}{3} emitting LBG that is in the lower quartile of $z\sim3$ LBG continuum-selected population.}

\section{CGM Properties} \label{sec:cgm}
Most of the baryons in high-redshift galaxies reside in their \textnormal{CGM\@ \citep[e.g.,][]{mitchell+2018}.} Therefore, it is necessary to analyze the CGM to completely study galaxy evolution at the peak epoch of star formation in the universe.

In this section we analyze the CGM properties of the galaxies from their extended \lya\ halo emission in the KCWI cubes and by \ion{H}{1} and metal absorption of background QSOs from HIRES. Specifically, we analyze \lya\ halo morphology, \lya\ velocity maps, and \lya\ physical size, then 
perform component-by-component Voigt profile decomposition of CGM absorption to measure the gas physical properties.

\subsection{Ly$\alpha$ Halo Emission} \label{sec:line_emission}
Extended Ly$\alpha$ halo emission is common in $z\sim$2-3 star-forming galaxies \citep[e.g.,][]{steidel+2010,chen+2021,erb+2023}. Ly$\alpha$ is useful because it is bright, indicative of star-formation and other FUV photon sources, such as AGN, \textnormal{and traces cold ($T\rm \sim10^4~K$) gas in the CGM\@.} Due to the resonant nature of Ly$\alpha$ (i.e., instantaneous absorption and emission that \textnormal{scatters} photons \textit{and} \textnormal{changes} their wavelength), modeling Ly$\alpha$ profiles to extract physical properties is non-trivial. Physical quantities such as \ion{H}{1} column density, temperature, dust content, 3D spatial distribution, intrinsic gas kinematics, etc., are exceedingly difficult to derive unambiguously and independent of model assumptions. Nonetheless, the observational constraints that we compile in this section are vital for context and literature comparisons, and future efforts.

\begin{figure*}
    \centering
    \includegraphics[scale=0.45]{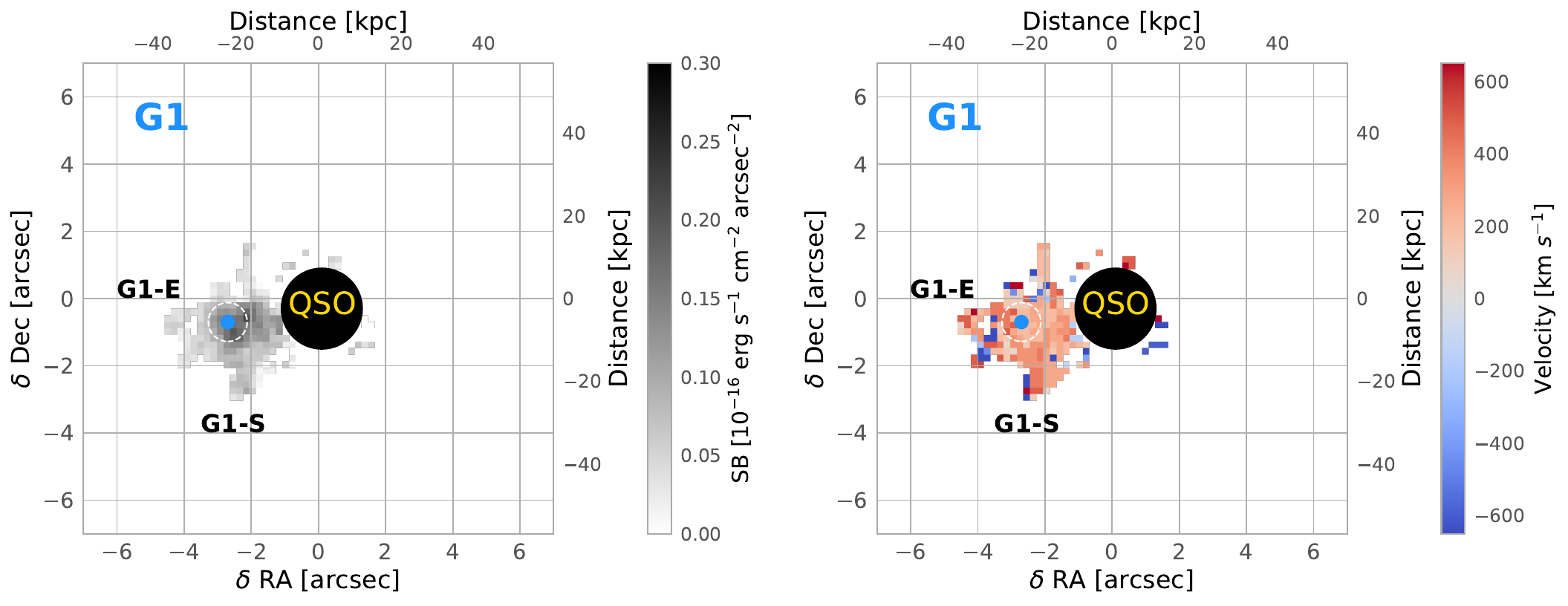}
    \includegraphics[scale=0.45]{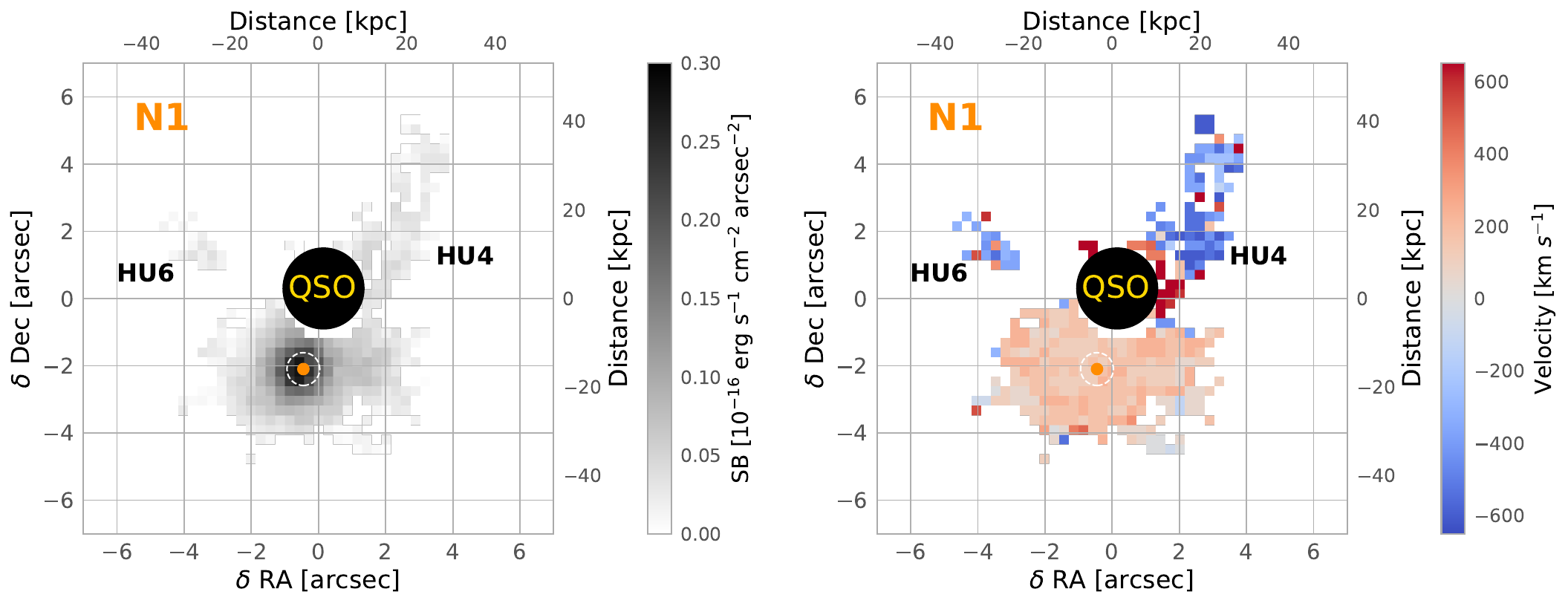}
    \caption{Ly$\alpha$ halos of G1 (\textit{top}) and N1 (\textit{bottom}). The colored dots are the galaxy continuum size and location as measured from \textit{HST}/F160W ($s_{cont,G1}=3.12~kpc$, $\rm s_{cont,N1}=2.26~kpc$), the white dashed circle highlights the continuum location. We show newly discovered objects as black labels. \textit{Left:} Ly$\alpha$ surface brightness spatial map. \textit{Right:} Ly$\alpha$ velocity map showing the peak of Ly$\alpha$ emission in each spaxel in velocity space with respect to the systemic redshifts $z_{\rm G1}$=2.4313 and $z_{\rm N1}$=3.1509. Red is showing redshifted gas and blue shows blueshifted gas.}
    \label{fig_lyamap}
\end{figure*}

In Figure \ref{fig_lyamap} we show \lya\ surface brightness spatial distribution maps and velocity distribution maps for G1 and N1. The surface brightness maps show that both galaxies possess extended \lya\ halos that are at least ten times the extent of their continuum (as measured from \textit{HST}/F160W images; e.g., \textnormal{Figure \ref{fig_kcwi_whitelight}}), and that there is complex (simple) structure in N1's (G1's) velocity distributions.

In Table \ref{tab_fuv_spectra} we show the \lya\ flux for each kinematic component detected and its corresponding area. We calculate the \lya\ fluxes by creating a narrowband image that includes all of the emission of a given component using the QSO+continuum subtracted cubes generated from CubEx (Section \ref{sec:cxpsf}). We 
compute the average SB of the component, then convert to flux using the bandpass and projected area (number of spaxels and spatial scale of the cubes 0.2--0.3\arcs). 

We calculate (or place limits on) the total \lya\ escape fraction by assuming that all of the \lya\ photons originate from the HII regions that we detected in \Ha\ emission (Section \ref{sec:ism}; Table \ref{tab_optical_spectra}), and that \lya/\Ha=8.7 (assuming $T\sim\rm10^4~K$ and $n_e\sim350~{\rm cm}^{-3}$ \citep[][]{osterbrock+1989,langen+2023}). We calculate this for both the total \lya\ from the continuum apertures and from the halo apertures. 

G1's morphology is typical compared to other KBSS galaxies with Ly$\alpha$ halos \citep[e.g.,][]{chen+2021} in that it is circular (perhaps spherical), and is more extended than its continuum with a peak size \textnormal{of $\rm s_{\rm halo}\sim28~kpc~(3.5");~ \rm s_{cont}=3.12~kpc$)}. The two protrusions to the east and south of the halo are the new galaxies G1-E and G1-S. 
G1's velocity map shows a simple kinematic structure with $v\rm \sim +250\rm~km~s^{-1}$. 
We measure a \lya\ halo flux $F_{\rm G1,\lya,halo}=(2.94 \pm 0.01)~\times 10^{-17}\fluxunits$, which gives $(\lya/\Ha)_{\rm G1,halo}\rm=0.54 \pm 0.16$ and $f_{G1,\lya\ esc,halo}=\rm (6\pm2)~\%$. These values are comparable to other star-forming galaxies at $z\sim\rm2-3$, and the low $(\lya/\Ha)$ fraction \textnormal{indicates} that the \lya\ emission mechanism is not dominated by collisional excitation, predicted $(\lya/\Ha)>100$ \citep[e.g.,][]{langen+2023}.

N1's halo morphology is unique in that it has three distinct components that are very extended. The main complex (south) has the highest SB and is spatially coincident with N1; the secondary complex (northwest) has a tapered morphology that widens towards N1 and is spatially coincident with newly discovered galaxy \textnormal{HU4}; and the detached complex (east) is small and coincident with new galaxy HU6. 
The velocities of the complexes are also distinct: $v_{\rm N1}\sim\rm +170~\kms$ for N1, $\rm v_{\rm HU6} \sim -350~\kms$ for HU6, and $v_{\rm HU4} \rm \sim-500~\kms$ for HU4, all with respect to the $z_{\rm sys,N1}$.

N1's \lya\ halo \textnormal{(main complex)} has \textnormal{both} a red and blue component that is 
is spatially extended ($\rm s_{halo}=46~kpc$) compared to $\rm s_{cont}=2.46~kpc$.
We measure a total \lya\ halo flux $F_{\rm N1,\lya,halo}=(12.50 \pm 0.01)\times 10^{-17}\fluxunits$, which gives $(\lya/\Ha)_{\rm N1,halo}\rm=2.50 \pm 0.08$ and $f_{N1,\lya\ esc,halo}=\rm (29\pm1)~\%$ (assuming negligible dust, which is consistent with its SED)\@. As seen before, N1's small $(\lya/\Ha)$ fraction \textnormal{indicates that  collisional excitation is not the dominant emission mechanism for its \lya\ halo. Its large \lya\ equivalent width, moderate red peak velocity, high \lya\ escape fraction, and large $\rm F_{\lya}/F_{\Ha}$ are similar to low mass KBSS-LAEs, while its bright $F_{\rm \lya}$ and large red-blue peak velocity separation are similar to LyC-LBGs. Additionally, the combination of its lower-mass, continuum colors, nebular excitation, low dust extinction, strong \lya\ properties (flux, equivalent width), and its complex and very extended \lya\ halo morphology ($\rm \sim100~kpc$) make it reminiscent of $z\sim 2$  \lya\-emitting extreme emission line galaxies (EELGs) \citep[e.g.,][]{erb+2016,erb+2023}. The sample analyzed by \citet{erb+2023} all showed double-peaked \lya\ profiles, extended halos ($\rm \gtrsim50~kpc$), typical peak separations $v\sim600~\kms$, and median \lya\ escape fraction $f_{\rm esc}\sim0.3$.}

As a summary, both galaxies show extended \lya\ emission in their CGM.  
Their \lya\ is kinematically dynamic (with respect to $z_{\rm sys}$), spatially extended (more than 10x their continuum size), bright (\textnormal{$F_{\rm \lya}>3\times 10^{-17}\fluxunits$}, \textnormal{and is not dominated} by collisional excitation. G1 shows a simple velocity distribution ($v_{\rm peak}\sim$+250 \kms), whereas N1 shows three distinct velocity components that span a large velocity range $v\sim-$600 - +250 \kms (discussed more in Section \ref{sec:galaxy_group?}). \textnormal{G1} has a typical \lya\ escape fraction whereas N1's is very large. This suggests that G1 is a typical \lya\ emitting star-forming galaxy at its $z$\textnormal{. N1 on the other-hand is in the tail distribution of $z\sim3$ LBG continuum-selected galaxies, and analogous to $z\sim2$ EELGs.}

\subsection{QSO Absorption} \label{sec:cgm_abs}
Using an unrelated background source (e.g., QSO) to probe the CGM is an ideal method to obtain a very high quality \textit{but} highly localized view. It is difficult to draw galaxy-wide conclusions from these localized views given the patchy/clumpy nature of the CGM but at small impact parameters, we might be more likely to see trends with galaxy properties (e.g., $M_*$, SFR, $Z_{\rm ISM}$)\@ \citep[e.g.,][]{rudie+2012,werk+2014,tumlinson+2017,rudie+2019}. 

We use HIRES spectra of the KBSS \textnormal{QSOs} to search for CGM absorption at the systemic redshifts of G1 and N1. As noted in the literature, both galaxies are at the same redshifts as DLAs and show detections of metals of varying ionization states. The DLA towards Q2343 was first identified by \citet{sargent+1988}, and has been analyzed using high-resolution VLT/UVES and Keck/HIRES spectra by several groups over the past 20 years \citep{dodorico+2002,desauges-zavadsky+2004,wang+2015,nielsen+2022}. Similar attention has been paid to the sub-DLA towards Q2233 which was analyzed using Keck/HIRES by many groups \citep{sargent+1989,steidel+1995,lu+1997,lu+1998}.

\begin{figure*}
    \centering
    \includegraphics[scale=0.78]{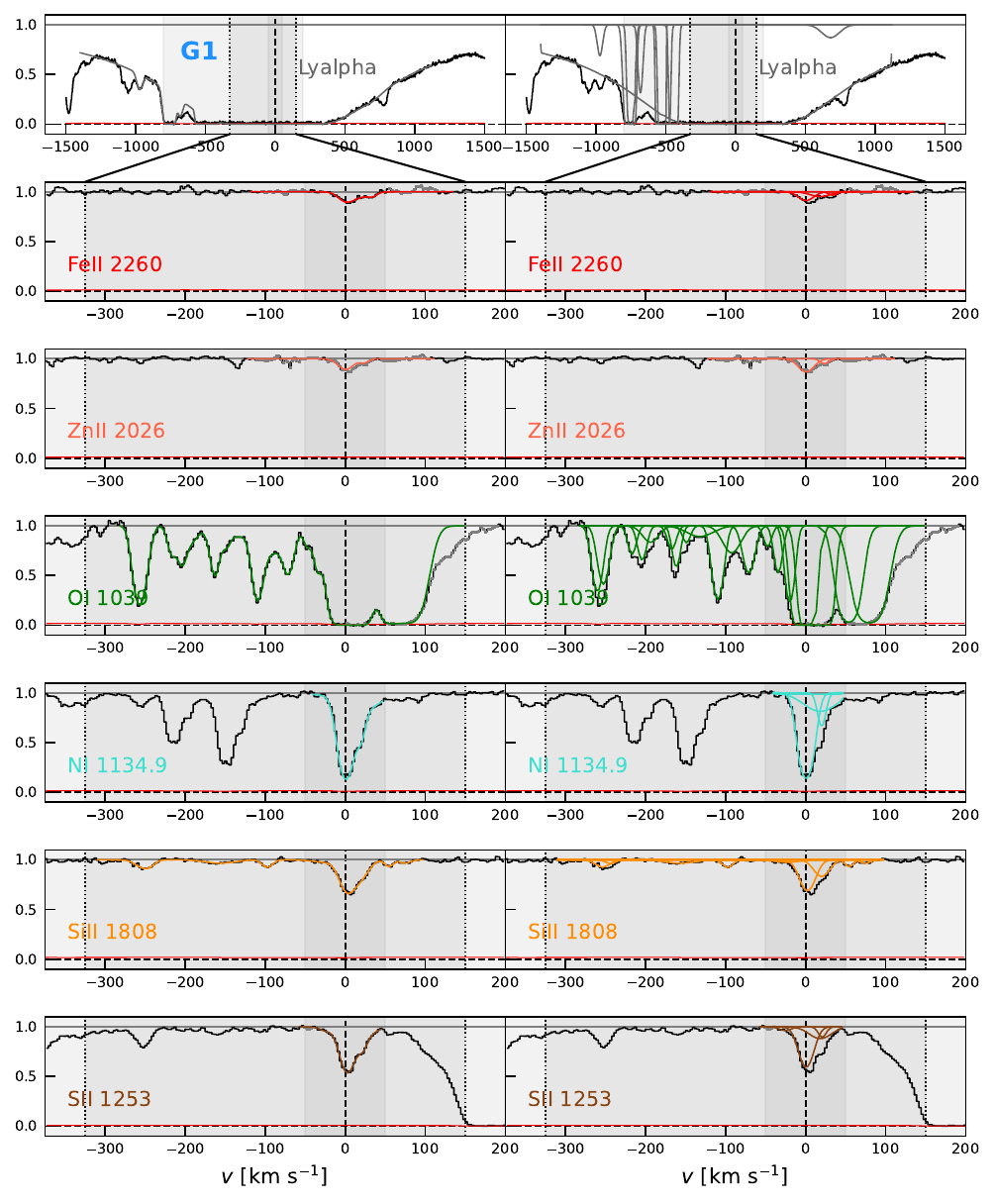}
    \caption{G1's CGM absorption and best fits centered at $z_{\rm G1}$=2.4312. Best fit models are shown as colored lines, and contamination from intervening absorption is shown as light gray lines. Note that \ion{H}{1} (top) is on a different velocity scale. The darkest gray region shows the velocity range associated with the DLA which is defined by the Fe-peak elements velocity range, the lighter gray shaded region shows the full extant of the low-ionization metal absorption, and the lightest gray regions shows the velocity of the intermediate- and high-ionization absorption \textnormal{(discussed in Section \ref{sec:baryoncycle_unboundgas})}. \textit{Left panel:} The product of the individual Voigt profiles. \textit{Right panel:} \textnormal{The} decomposed Voigt profiles. The error spectrum is shown at the bottom as a red solid line. }
    \label{fig_cgmfit_q2343rep}
\end{figure*}

\begin{figure*}
    \centering
    \includegraphics[scale=0.78]{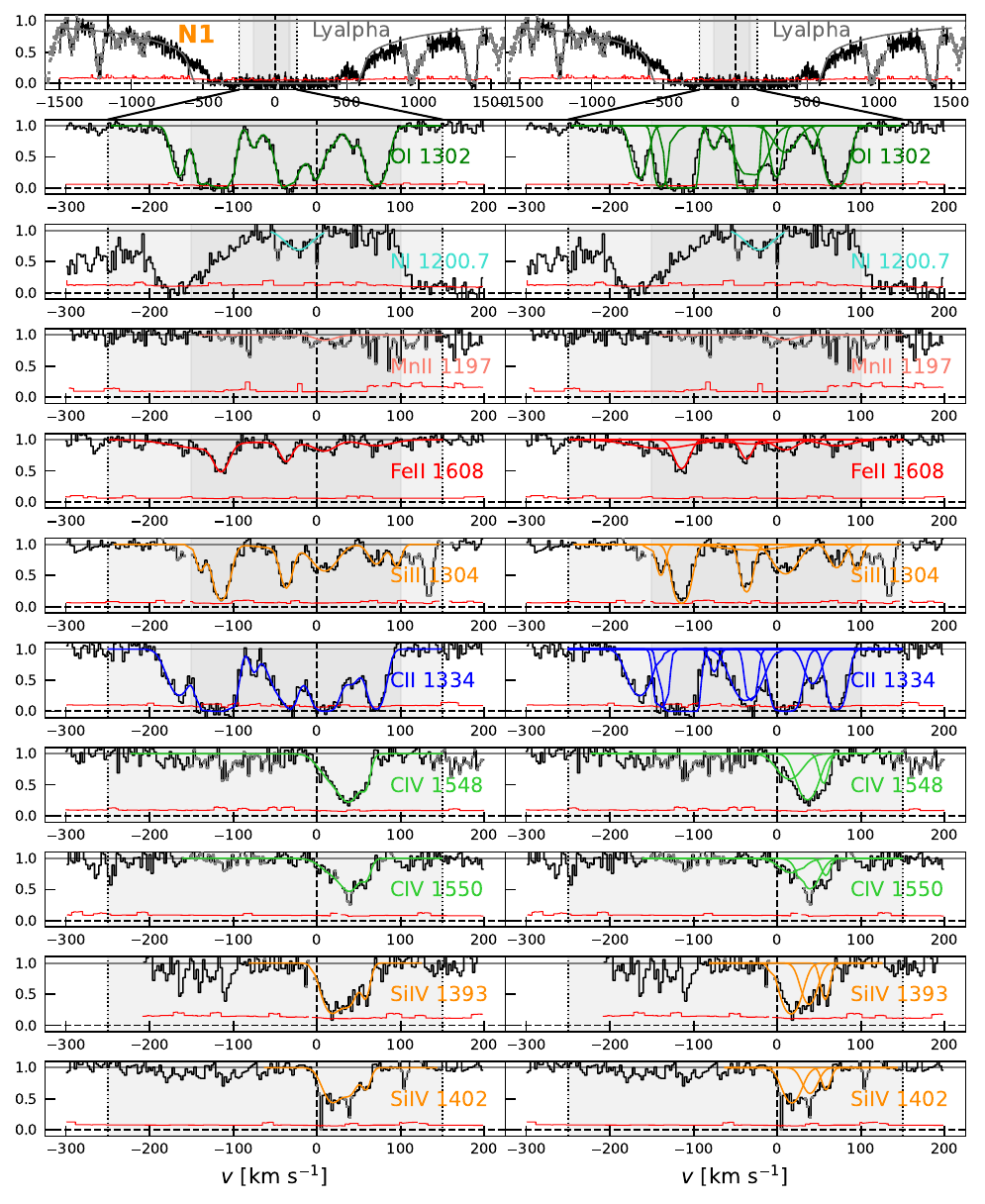}
    \caption{N1's CGM absorption centered at its systemic redshift z$_{\rm N1}$=3.1509. The lines and colors are the same as Figure~\ref{fig_cgmfit_q2343low}.
    }
    \label{fig_cgmfit_q2233}
\end{figure*}

In Figures~\ref{fig_cgmfit_q2343rep} and \ref{fig_cgmfit_q2233} we show representative neutral (e.g., \ion{H}{1}, \ion{N}{1}, \ion{O}{1}), low-ionization (e.g., \ion{C}{2}, \ion{Si}{2}, \ion{S}{2}, \ion{Fe}{2}), \textnormal{intermediate-ionization (\ion{C}{4}, \ion{Si}{4}) and metal transitions}.
The metals extend more than $400~\rm km~s^{-1}$, and for G1 extend \textnormal{from $\rm v\sim-800-+200~\kms$!} 
 The darkest grey shaded region shows the metal components that are self-shielded by the DLA; this was determined by the velocity spread of the low-ionization Fe-peak elements (e.g., \ion{Mn}{2}, \ion{Fe}{2}, \ion{Zn}{2}) which require dense, self-shielded gas for detection due to their low ionization potentials. 

\subsubsection{Voigt Profile Fitting} \label{sec:voigt}
We inferred physical properties of inner CGM gas absorption via Voigt profile decomposition. To fit the complexes, we used a series of individual Voigt profiles of a single redshift $z$, Doppler parameter $b$ (i.e., single temperature), and column density $N_{\rm Z}$. We use two Voigt profile fitting programs for this task: VoigtFit \citep{voigtfit} for quick interactive initial guesses, and ALIS \citep[Absorption LIne fitting Software\footnote{\dataset[https://github.com/rcooke-ast/ALIS]{https://github.com/rcooke-ast/ALIS}};][]{ALIS} for the main fitting.

The HIRES spectra are of high enough quality that we identified absorption components by visual inspection. We started by assigning a component to each absorption trough using the interactive version of VoigtFit, assuming the smallest Doppler parameter that we could realistically resolve $b\sim5~\rm km~s^{-1}$ (given the spectral resolution of $7$ \kms). 

We fit the first guesses with ALIS, then visually inspected the residuals between the best-fit model and data. We used locations of peaks and troughs in the residuals to add or remove components. We continued this process iteratively until the residuals showed no clear peaks and troughs (i.e., residual deviations were below that of the error spectrum).
Finally, once the components showed no strong residual deviations we checked that $\chi^2/\rm DOF\sim 1$. There were instances where fits achieved this but multiple components gave large errors that were \textnormal{comparable to or greater} than the best-fit value. In these cases, we removed the component, then refit, and usually the fit statistic stayed \textnormal{the same}.

We assumed that ions of similar ionization state arise from the same gas so we tie their redshifts (velocities) to one another, and allow column density and Doppler width to vary. 
We fit one metal transition at a time (e.g., all \ion{Fe}{2} transitions) then copy the best-fit component structure to to the next transition (e.g., copied \ion{Fe}{2}, to \ion{Si}{2}), run the fit, check the residuals, \textit{then} add and remove components as needed. 
This procedure resulted in $\sim90\%$ of the best-fit components being tied with at least one component from another metal transition with the same ionization state. We did not tie any components to \ion{H}{1} (or vice versa) due to the difficulty of separating single from blended components.

G1 and N1's low ionization transitions (e.g., \ion{O}{1}, \ion{Si}{2}, \ion{Fe}{2}) always had a similar structure that was distinct from their higher ionization transitions (e.g., \ion{C}{4}, \ion{Si}{4}, \ion{O}{6}) which is typical in DLAs due to self-shielding. We discuss the intermediate- and high-ionization transitions in \textnormal{Section} \ref{sec:baryoncycle_unboundgas}.  In practice, all of the low ions had their $z$ tied to strong non-saturated transitions (e.g., \ion{Si}{2}, \ion{Al}{2}, \ion{Fe}{2}), and the higher ions were tied to each other (e.g., \ion{C}{4}, \ion{Si}{4}; \ion{O}{4}).

To derive column densities of saturated transitions (e.g., \ion{C}{2}, \ion{O}{1}, \ion{Al}{2}) we froze $b$ in the strongest (saturated) components to that of the $b$ of an unsaturated ion of similar mass (usually Si) to place a lower limit on the column density. For G1 this included \ion{O}{1} and \ion{Al}{2}, and for N1 this included \ion{C}{2} and \ion{O}{1}\@. There are other more heavily saturated transitions we did not fit (e.g., \ion{C}{2}, \ion{Mg}{2} for G1).

After determining the tied component parameters, we refit the spectrum by shifting the components together in $z$ while allowing $\log{N}$ and $b$ to vary (unless the absorption was saturated). Once again, we checked the errors and residuals to ensure a good fit a fit statistic $\chi^2/\rm DOF\sim 1$.

We use the $\log{N_{\rm H~I}}$ catalog from \citet[][hereafter \citetalias{rudie+2012}]{rudie+2012} to determine the \ion{H}{1} component structure and total \ion{H}{1} column for systems that were covered in their work, which includes G1. R12 was interested in the \ion{H}{1} absorption systems toward the 15 KBSS QSO sightlines whose redshifts were (1) not proximate to the QSO redshift ($\lesssim 1000~\rm~km~s^{-1}$), and (2) allowed for Ly$\beta$ to fall on the 
detector.  
In practice, we froze
\ion{H}{1} to \citetalias{rudie+2012}'s best fit $z_{\rm{R12}}$, $b_{\rm{R12}}$, and $\log{N_{\rm{R12}}}$ for G1, and we used literature values for N1. 

\citetalias{rudie+2012}'s \textnormal{\ion{H}{1}} catalogs were also useful in fitting transitions that were contaminated with \lya/\lyb\ absorption such as \ion{O}{6}$\lambda\lambda1031,1037$. The \ion{H}{1} catalog allowed us to remove Ly$\beta$ absorption from higher redshift systems. 
Specifically, we divided out the best-fit \citetalias{rudie+2012} absorption components from the entire spectrum, saved it, and used it to fit the contaminated species. 
It is still possible that there is Ly$\alpha$ contamination and unresolved components so we quote lower limits on these column densities $\log{N}$. We did not use the \ion{H}{1}-removed spectrum for all systems because the large residual spikes that surrounded well-fit \ion{H}{1} systems were difficult to work around for some transitions \footnote{Most absorbers with $\log{N_{\rm H~I}}\gtrsim14.5$ are well fit.}. More importantly, the \ion{H}{1}-removed spectrum would have made our results dependent on R12's modeling. \textnormal{This} would add an unnecessary source of uncertainty because \textnormal{most transitions} have little to no contamination.

In Figures \ref{fig_cgmfit_q2343rep} and \ref{fig_cgmfit_q2233} we show the results of the best-model fits and \ion{H}{1} fits from \citetalias{rudie+2012} for some representative transitions of G1 and N1 . We show all of the fits in Appendix \ref{sec:full_voigt}.

The top three panels of Figure \ref{fig_cgmfit_q2343rep} show representative \ion{H}{1} and Fe-peak metal transitions for G1. The HIRES spectral coverage allows measurements of Ly$\alpha$ through Ly$\zeta$ giving strong constraints on $b$ and $\log{N_{\rm H~I}}$. The Fe-peak transitions have a simple structure that is fit well with two or three components. The systems only span $\pm \rm 50~km~s^{-1}$.  
We use these elements to estimate dust depletion in Section \ref{sec:cgm_metallicity}.

The bottom four panels of Figure \ref{fig_cgmfit_q2343rep} show representative neutral and low-ionization metal transitions for G1.
The component structure for each of the ions is very similar and well fit, and almost every component is tied with at one other metal transition. The strongest components (\ion{H}{1} and metals) are all near the systemic redshift of G1 ($\Delta v\sim \rm-8~km~s^{-1}$), but there is still weak metal absorption out to velocities exceeding $\sim\rm-500~km~s^{-1}$.  \textnormal{This high-velocity absorption has the same structure as intermediate- and high-ionization ions and is discussed in Section \ref{sec:baryoncycle_unboundgas} (Figure \ref{fig_cgmfit_q2343medhigh}).}

In Figure \ref{fig_cgmfit_q2233}, we show all of the best-fit \ion{H}{1} and metal transitions for N1. All of the neutral and low ionization metal transitions share a similar component structure, \textnormal{more than 85\% of their components are tied to another}. The dominant \ion{H}{1} component is offset \textnormal{from $z_{\rm sys,N1}$ by $\sim\rm7~km~s^{-1}$,} but the dominant metal absorption complex appears to be blueshifted by $\sim\rm -100$~\kms. The low ions span a velocity range of $\sim\rm-200 - +100~km~s^{-1}$ and might all be associated with the main DLA \textnormal{(\ion{H}{1})} component. The intermediate-ionization ions (\ion{C}{4} and \ion{Si}{4}) have a simpler kinematic structure and smaller velocity range compared to the low ions with an average velocity offset of $\sim 50$~\kms\ spread over $\sim 100$~\kms. \ion{C}{4} and \ion{Si}{4} share no common components with the low ions.

\subsubsection{Column Densities and Kinematics} \label{sec:cgm_column_kinematics}
In this section, we analyze the total column density and component structure of the best-fit model parameters found in Section \ref{sec:voigt}.

In Table \ref{tab_cgm}, we report the total column summed over $\pm 1000~{\rm km~s}^{-1}$ from the systemic velocity for the given ions. 

Due to the difficulty of associating extended metal components with a parent \ion{H}{1} absorber, we report the total column densities, not total metallicities. Indeed, even in ideal scenarios where there are strong constraints on the location of \ion{H}{1} components, large ionization corrections are necessary for absorbers with low $\log{N_{\rm HI}}$ gas making metallicity determination non-trivial \citep[e.g.,][]{zahedy+2021_CUBS3}\@. 

\begin{table*}[htb] 
\centering
\tiny
\caption{CGM Absorption Column Density} \label{tab_cgm} 
    \begin{tabular}{ccccccccccccccccccc} 
    \hline 
    \multirow{2}{*}{Galaxy} 	&\multirow{2}{*}{$z_{\rm sys}$} 	&\multirow{2}{*}{$D_{\rm tran}$} 	&\multicolumn{15}{c}{$\log{(\Sigma N_{\rm X})} [\rm cm^{-2}]^{a,b}$} \\ 
    \cmidrule(lr){4-19} 
     	& 	&(pkpc) 	&HI 	&CII 	&NI 	&OI 	&AlII 	&SiII 	&SII 	&CrII 	&MnII 	&FeII 	&NiII 	&ZnII 	&AlIII 	&CIV 	&SiIV 	&OVI \\ 
    \hline 
    G1 	 &2.4313 	 &20.8 	&20.40 	&\nodata 	&14.70 	&$>$16.72 	&$>$13.91 	&15.31 	&14.73 	&12.82 	&12.26 	&14.45 	&13.25 	&12.07 	&13.26 	&14.74 	&14.03 	&$>$15.39 	\\ 
    N1 	 &3.1509 	 &19.3 	&20.00 	&$>$15.55 	&$>$13.99 	&$>$16.05 	&13.33 	&14.56 	&\nodata 	&\nodata 	&$>$12.57 	&14.28 	&\nodata 	&\nodata 	&\nodata 	&13.90 	&13.59 	&\nodata 	\\ 

    \hline 
    \end{tabular}%
    \tablenotetext{a}{Column density summed over $\rm \pm~1000~km~s^{-1}$} 
    \tablenotetext{b}{Typical uncertainty of 0.1--0.2 dex} 
\end{table*} 

The total column densities we measured are comparable or larger than the highest column densities reported by \citetalias{rudie+2019} for ions we have in common (e.g., \ion{Si}{2}, \ion{C}{3}, \ion{C}{4}, \ion{O}{6}. For some saturated ions e.g., \ion{O}{6}\textnormal{,} we report lower limits on the column density \textnormal{meaning} they could be even larger.

Table \ref{tab_metals_components} shows the number of components required to fit each ion. Note that \textnormal{the} \ion{Si}{2} structure for G1 is slightly misleading because we report only the weak non-saturated transition (\ion{Si}{2}~$\lambda$1808).  There are many components for \textnormal{stronger}, saturated \textnormal{\ion{Si}{2}} transitions. The metal absorption for N1 requires $\gtrsim5$ components, and up to 31 components and for G1. This complicated kinematic structure for $z\sim$2-3 galaxies has been noted before by \citet{rudie+2019}.

\begin{table}[htb] 
\centering 
\caption{Component Structure} \label{tab_metals_components} 
\tiny
    \begin{tabular}{lcccccccccc} 
    \hline 
    \hline 
    \multirow{2}{*}{Galaxy} 	&\multicolumn{9}{c}{$\rm \#^{a,b}$} 	\\ 
     	&HI 	&CII 	&OI 	&AlII 	&SiII 	&FeII 	&AlIII 	&CIV 	&SiIV 	&OVI 	\\ 
    \hline 
    G1 	 &9 	&\nodata 	&19 	&29 	&9$^c$ 	&20$^c$ 	&10 	&31 	&23 	&29 	\\ 
    N1 	 &1 	&9 	&10 	&6 	&\nodata 	&6 	&\nodata 	&3 	&3 	&\nodata 	\\ 
    \hline 
    \end{tabular}%
    \tablenotetext{a}{Total number of single Voigt components required for the composite fit} 
    \tablenotetext{b}{\ion{N}{1}, \ion{S}{2}, \ion{Cr}{2}, \ion{Ni}{2}, and \ion{Zn}{2} have $\leq3$ components and are omitted from this table} 
    \tablenotetext{C}{Does not include saturated components} 
\end{table} 

The total column density and kinematic structure of the components add more evidence that the $z\sim$2--3 CGM (within $R_{\rm vir}$) is kinematically complex requiring at minimum 10 components (for neutral and low ions) with some transitions requiring up to 31 components (for intermediate and high ions), and column densities exceeding the largest found by R19 seen in \ion{C}{2}, \ion{Si}{2}, \ion{C}{4}, \ion{Si}{4}, and \ion{O}{6} suggestive of a high covering fraction of metals within $R_{\rm vir}$ \textnormal{at} $z\sim2-3$ CGM.

\subsubsection{CGM Metallicity} \label{sec:cgm_metallicity}

We  derive CGM metallicity using metal components that are self-shielded by the DLA gas, which reduces ionization effects. For G1, we use the velocity range of the Fe-peak components $v \pm\rm50~km~s^{-1}$ \textnormal{to estimate the extent of the DLA self-shielding}. For N1, \textnormal{we the} location of the strongest \ion{Fe}{2} component ($v\sim$ -150 \kms) and \ion{Mn}{2} (50 \kms) $\rm-150-50~km~s^{-1}$ \textnormal{to estimate the extent of the DLA self-shielding}. 
Additionally, all of the absorption components that we sum over have \ion{O}{1} and \ion{N}{1} in common suggesting that the gas is mostly neutral based on their low ionization potentials \citep{field+1971,steigman+1971}. 
We sum the metal columns, normalize by the \ion{H}{1} column, then normalize by solar values (\citet{asplund+2009}), and report the values in Table \ref{tab_cgm_metals}. For ions where the main components are saturated we quote lower limits on the metallicity. 

The $\alpha$-element metallicities are consistent with one another for G1 (O, S, Si; within 0.1 dex) and N1 (O, S; within 0.2 dex). The Fe-peak element abundances for both galaxies are fairly consistent with one another (within 0.3 dex) with the largest discrepancy found in \ion{Mn}{2}.

\begin{table*}[htb] 
\centering 
\caption{CGM Metallicity$^{a}$} \label{tab_cgm_metals} 
    \begin{tabular}{lccccccccccc}
    \hline 
    \hline 
    Galaxy 	&[C/H] 	&[N/H] 	&[O/H] 	&[Al/H] 	&[Si/H] 	&[S/H] 	&[Cr/H] 	&[Mn/H] 	&[Fe/H] 	&[Ni/H] 	&[Zn/H] 	\\ 
     	&($N_{\rm CII}$) 	&($N_{\rm NI}$) 	&($N_{\rm OI}$) 	&($N_{\rm AlII}$) 	&($N_{\rm SiII}$) 	&($N_{\rm SII}$) 	&($N_{\rm CrII}$) 	&($N_{\rm MnII}$) 	&($N_{\rm FeII}$) 	&($N_{\rm NiII}$) 	&($N_{\rm ZnII}$) 	\\ 
    \hline 
    G1$^a$ 	&\nodata 	&-1.49 	&$>$-0.85 	&$>$-1.11 	&-0.73 	&-0.79 	&-1.22 	&-1.50 	&-1.39 	&-1.35 	&-0.89 	\\ 
    	&(\nodata )	&(14.74) 	&(16.24) 	&(13.74) 	&(15.18) 	&(14.73) 	&(12.82) 	&(12.33) 	&(14.51) 	&(13.27) 	&(12.07) 	\\ 
    N1$^b$ 	&$>$-0.92 	&$>$-1.84 	&$>$-0.72 	&-1.12 	&-0.92 	&\nodata 	&\nodata 	&$<$-0.86 	&-1.22 	&\nodata 	&\nodata 	\\ 
    	&(15.51) 	&(13.99) 	&(15.97) 	&(13.33) 	&(14.59) 	&(\nodata )	&(\nodata )	&(12.57) 	&(14.28) 	&(\nodata )	&(\nodata )	\\ 
    \hline 
    \end{tabular}%
    \tablenotetext{a}{Column density of the gas self-shielded by the G1 DLA: $\rm \pm50~km~s^{-1}$}
    \tablenotetext{b}{Column density of the gas self-shielded by the N1 DLA: $\rm \ -150-100~km~s^{-1}$}
\end{table*} 

\begin{figure*}
    \centering
    \includegraphics[scale=0.55]{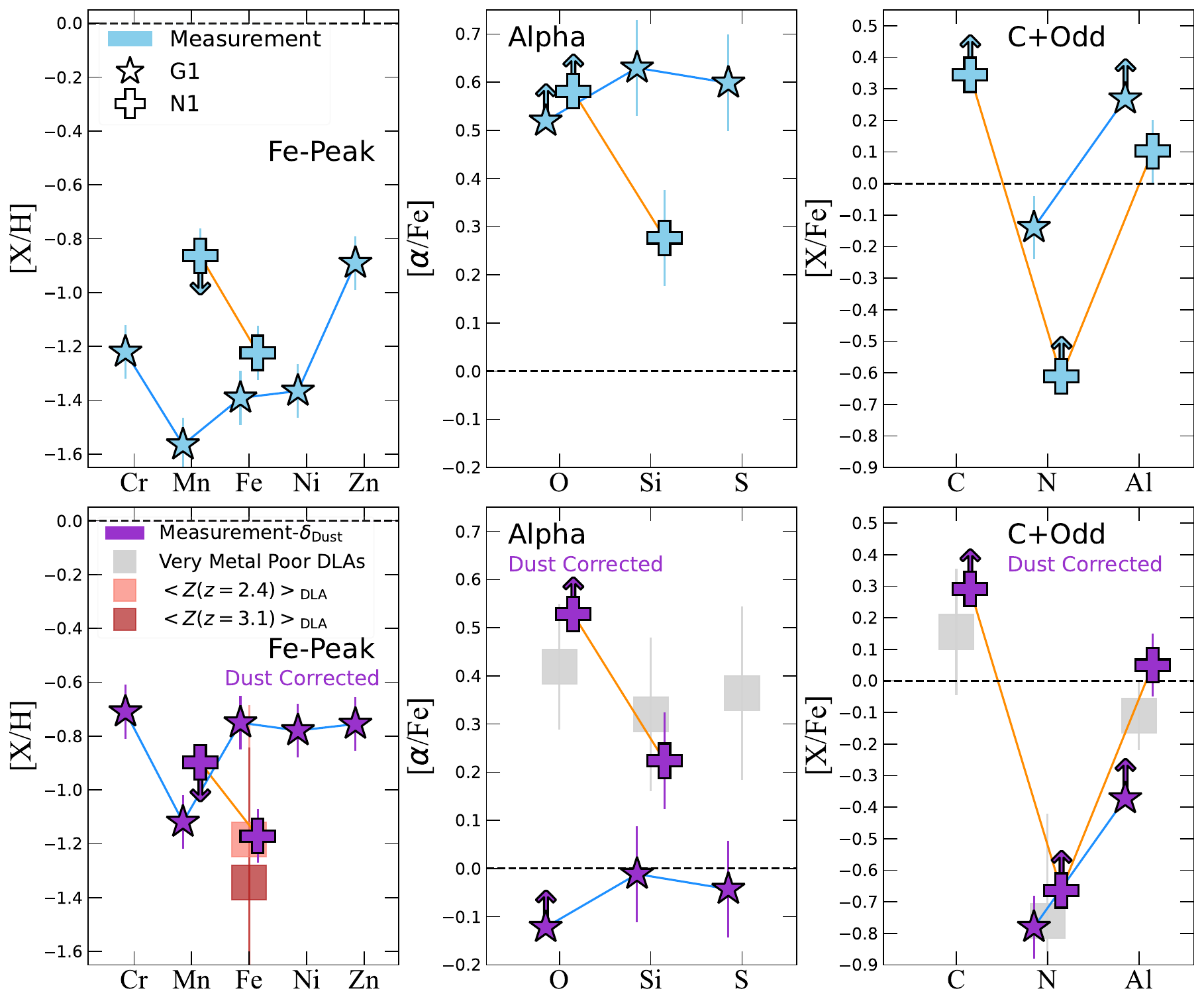}
    \caption{\textnormal{CGM abundance ratios for G1 (star) and N1 (plus). \textit{Top row:} \textit{Blue filled} symbols are direct non-dust corrected metallicity measurements. \textit{Bottom row:} \textit{Purple filled} symbols have been dust-corrected following empirical prescriptions from \citealt{decia+2016}. The \textit{left column} shows Fe-peak abundance; the \textit{middle column} panel shows [$\alpha$/Fe] ratios; and the \textit{right column} show Carbon and odd number elements. The light red and dark red squares show the DLA mean metallicity (and typical dispersion) from a compilation/analysis by \citealt{rafelski+2012} (references therein), the gray squares show the median abundance ratios for very metal poor DLAs compiled by \citealt{nunez+2022}.}
    }
    \label{fig_cgm_metallicity}
\end{figure*}

The top row of Figure \ref{fig_cgm_metallicity} shows a summary of the abundance ratios of G1 and N1. The Fe-peak elements (Cr, Mn, Fe, Ni; top left) are consistent with one another for both galaxies. \textnormal{Mn appears to be underabundant for G1, which is commonly found in DLAs and related to its nucleosynthetic origin \citep[e.g., SNeIa; ][]{nomoto+1997,vladilo+1998,pettini+2000,konstantopoulou+2022}. We place an upper limit on Mn for N1 due to the weakness of the line and difficulty locating the continuum.}
\textnormal{The [$\alpha$/Fe] ratios show $\alpha$-enhancement for both galaxies.}
\textnormal{Carbon and the odd elements show similar deviations from solar. Additionally for G1, there is a clear offset in the Zn abundance compared to the rest of the Fe-peak elements that we discuss in the next section.}

\subsubsection{Dust Depletion} \label{sec:cgm_dust}
It is well known within the literature that DLAs experience dust depletion analogous to that seen in the MW ISM \citep[e.g.,][]{calura+2003,vladilo+2006,khare+2007,jenkins+2009,decia+2016,decia+2018,peroux+2020,konstantopoulou+2022,ramburuth-hurt+2023}. We are interested in the total metallicity of the DLAs (CGM gas), so we must account for metals in all phases, i.e., gas and dust. 
We must rely on ratios of refractory elements (easily condense onto dust grains) and volatile elements (easy to keep in the gas phase), to gauge the amount of dust depletion in DLAs with the important condition that the elements track each other nucleosynthetically. 
The dust depletion correction $\delta_d$ can be used to correct metallicity and place limits on the amount of dust (e.g., $A_{\rm V}$ or $E(B-V)$) in the CGM\@. 

G1 shows detection's of the volatile element Zn, and refractory elements Cr, Mn, Fe, and Ni, which are all near the Fe-peak. For N1, there are no detection's of volatile elements but a few detection's of refractory elements Mn, Fe, and Si. 
We can use Si to place limits on dust depletion because it is not as easily depleted as Mn and Fe.

We estimated the dust depletion corrections using a couple of methods. The simplest approach used the difference (linear ratio) between the Zn and Fe-peak abundances (see Section \ref{sec:cgm_metallicity}). In practice, this meant setting all Fe peak elements abundances to match that of [Zn/H]\@. We then computed the dust depleted ratios [X/Fe]$_d$ = [X/(Fe-$\delta_{\rm Zn}$)] where $\delta_{\rm Zn}$ is the difference between [Zn/H] and [Fe/H]\@. The equivalent expression is [X/Fe]$_d$=[X/Zn]. The advantage of this method is that it is straightforward, completely empirical, and specific to this absorber. Importantly, it does not account for potential depletion effects of Zn and cannot be used in systems that do not have detections of Zn.

The other method uses the relations derived by \citet[][hereafter \citetalias{decia+2016}]{decia+2016,decia+2018} and later improved by \citet{konstantopoulou+2022}, who used an ensemble of abundances measured from DLAs, Galactic absorbers, and SMC and LMC absorbers to predict the average dust depletion of a system given an [Zn/Fe]. The unknown intrinsic abundances of an individual absorber make it difficult to calibrate from one system, but with a large enough sample one could find bulk trends that can be used to find an average dust depletion correction. The critical assumption that the authors made was that the depletion correction $\delta_d=$0 at [Zn/Fe]=0.  
We refer the reader to Section 3 of \citetalias{decia+2016} for a complete description of the method but note that the updated relations published by \citet{decia+2018} allowed dust depletion to be predicted  
with either [Zn/Fe], [Si/Fe], or [S/Fe]. They also computed relations for elements besides Fe (e.g., N, O, S, Al, N) but we correct only the Fe-peak elements because (1) they likely dominate the depletion (they have much higher condensation temperatures than e.g., O, N \citep{lodders_2003}), (2) the other relations rely on the predicted curves from the [Zn/Fe] abundances i.e., a model based upon another model, and (3) the other relations have a significant amount of scatter and much fewer measurements than that used for [Zn/Fe] and [Si/Fe].

Both methods result in dust corrected Fe abundances that were consistent with one another for G1 $\rm [Fe/H]_{Zn}=-0.89\pm 0.1,~[Fe/H]_{DC16}=-0.76\pm 0.15$. We use the depletion corrections from \citetalias{decia+2016} because its predicted values were consistent with our simpler empirical approach and because it allows us to estimate depletion for N1, which has no detection of Zn. \textnormal{\citet{konstanatopoulou+2024} recently analyzed cosmic dust evolution using a large sample of DLAs (and MW, SMC, and LMC absorbers) that included G1.  We find comparable depletion.}

\textnormal{We use the [Si/Fe] ratio to infer N1's dust depletion correction, which is small ($\delta_d=\rm0.05~dex$).} 

The bottom row of Figure \ref{fig_cgm_metallicity} show the dust depletion-corrected abundances for G1 and N1\textnormal{; red and pink squares show the DLA mean metallicity from a sample of \ion{H}{1}-selected DLAs at high-z (analyzed by \citealt{rafelski+2012}) at G1's and N1's redshift; and grey squares show the median abundance ratios of one of the largest samples of very metal-poor (VMP; $\rm [Fe/H] \leq 2$) DLAs compiled by \citet[][]{nunez+2022}.
The CGM metallicity for both G1 and N1 are comparable to that of the DLA mean metallicity at their redshifts (we discuss this more in Section \ref{sec:literature_comparison}). 
The Fe-peak elements for both galaxies are consistent after applying the dust depletion correction, and comparable to what we inferred empirically (i.e., [Zn/H] in the upper plot or $\delta_d=0.6$ dex for G1).}
The [$\alpha$/Fe] abundance patterns (lower middle plot) show a striking difference: G1's CGM [$\alpha$/Fe] is solar and consistent across the three $\alpha$ tracers, and N1's CGM shows $\alpha$-enhancement similar to VMP DLAs.
Interestingly, both of their [C,N,Al/Fe] abundance patterns (right bottom plot) are consistent with the VMP DLAs. There could be a few reasons for this consistency that are related to the nucleosynthetic origin of the elements. 
Nitrogen, which is produced abundantly by AGB stars, might increase in lockstep with Fe, suggesting that the VMP DLA ratios include some enrichment from delayed events.


Altogether, \textnormal{G1's CGM has a moderate-high metallicity} with a moderate amount of dust depletion reminiscent of a chemically evolved absorber, but has a C and odd-element abundance pattern similar to less chemically evolved absorbers. Interestingly, G1 has not been recently enriched, as evidenced by its solar $\rm [\alpha/Fe]\sim0$ ratio. This \textnormal{suggests} that the abundance patterns for individual absorbers are hard to predict and caution must be used if assuming any abundance pattern for analysis (e.g., photoionization modeling).

\textnormal{N1's CGM has a low-moderate metallicity with C, odd-element, and [$\alpha$/Fe] abundance ratios consistent with VMP DLAs. This suggests that it was recently enriched by CCSNe.}

\subsubsection{Comparison with the DLA Literature} \label{sec:literature_comparison}
\textnormal{In this section we compare our measured/inferred DLA properties with those in the literature in terms of \ion{H}{1} column density and metallicity (we assume [Fe/H]$_d$=[M/H]). We discuss dust depletion in Section \ref{sec:baryoncycle_dust}. We refer to the DLA associated with G1 at $z=2.41$ as G1DLAz2, and the DLA associated with N1 at $z=3.15$ as N1DLAz3.}

\textnormal{N1DLAz3} has previously been found to have $\log{N_{\rm H~I}}\rm =20.00 \pm 0.1$ at z$_{\rm abs}$=3.153 ($\delta v_{\rm ISM}<\rm 60~kms^{-1}$), ${\rm  [Fe/H] = -1.4 \pm 0.1}$ \citep{sargent+1989,lu+1997,lu+1998}. Our \ion{H}{1} is the same as the authors and Fe abundance is within 0.2 dex (regardless of dust correction).

\textnormal{We compare the metallicity of both DLAs to the cosmic DLA mean metallicity (and dispersion) at their redshift using the linear relation computed by \citet{rafelski+2012}, who analyzed a large sample of DLAs across a wide redshift range (z$\sim$1.5-5) with little metallicity bias (\ion{H}{1} selected sample). Using the relation at N1's redshift yields $Z(z=3.1509)=-1.34\pm0.5$ making N1DLAz3 comparable to the mean metallicity at its redshift ([Fe/H]$_d\sim$-1.1$\pm$0.1). Even though the mean relation is not a physical model and the scatter is large, it is interesting that N1DLAz3 is not obviously metal-poor because its abundance ratios are similar to that seen in VMP DLAs, which are systematically below the DLA cosmic mean by 1 dex between $z\sim2-3$ (i.e., [M/H]=[Fe/H]$<$-2 by definition).}

\textnormal{The \ion{H}{1} column density of N1DLAz3 ($\log{N_{\rm H~I}}=20.0$) is at the boundary between sub-DLAs ($19\leq \log{N_{\rm H~I}}<20.3$; also known as Super Lyman Limit Systems or SLLS) and DLAs ($\log{N_{\rm H~I}}\geq20.3$). It would therefore be useful to compare it to a survey of sub-DLAs. The KODIAQ-Z \citep{Lehner+2022_KODIAQZ} and HD-LLS surveys \citep{prochaska+2015,fumagalli+2016} analyzed the metallicity distribution of a large number of \ion{H}{1} absorbers with $\log{N_{\rm H~I}}=$14.6--20.3. Among their many findings was an increase in metallicity with $N_{\rm H~I}$ from sub-DLA to DLA, which is consistent with our findings for N1DLAz3 and G1DLAz2, though there is large scatter in these relations ($\Delta[M/H]>0.5$). Additionally, the authors found that the median/mean (standard deviation) metallicity of sub-DLAs is $<Z>_{\rm SLLS}=$-1.90/-1.93$\pm 0.89$ putting N1DLAz3 well above the median/mean. As mentioned before, this is puzzling given that its abundance ratios are similar to that of VMP DLAs.} 

\textnormal{\citet{berg+2015} analyzed a large sample of metal-rich DLAs that we will compare with G1DLAz2. Almost half of their sample had $\log{N_{\rm H~I}}>21$ meaning G1DLAz2 is on the low end of the distribution ($\log{N_{\rm H~I}}=20.4$). The median metallicity of their sample is [M/H]$\sim$0.7 which is comparable to the metallicity of G1DLAz2 ([Fe/H]$_d$=-0.75). Finally, the DLA mean metallicity at this redshift is $Z(z=2.4312)=-1.18\pm0.5$ placing G1DLAz2 well above the mean. This all suggests that it is a relatively low \ion{H}{1}, metal-rich DLA.}

G1DLAz2 was recently studied in detail by \citetalias{nielsen+2022}. Their absorption analysis was based on a component by component (``cloud-by-cloud'') multiphase Bayesian modeling scheme that extracted the kinematic structure and physical conditions for each of the components (``clouds'') independently but self consistently (each component per ion had its own model that folded in multiphase components when necessary) \citep{sameer+2021}. Photoionization grids were used to infer the physical conditions of the gas associated with low $\log{N_{\rm HI}}$ gas ($\log{(N_{\rm HI}/cm^{-2})}<19$ typically; the majority of the metals components). Importantly, the grids assume a solar abundance pattern, which we found is likely not representative of the true intrinsic abundance pattern of the DLA (Section \ref{sec:cgm_metallicity}). Nonetheless, their analyses presents an opportunity for us to compare our flexible ``by hand'' method with their statistically robust method.

One of the most straight-forward comparisons that we can make is the overall quality of fit and total column density, since neither of these should be significantly affected by model assumptions. 
A visual comparison of both our fits to Ly$\alpha/\beta,/\gamma$, \ion{Fe}{2}$\lambda$2344, and \ion{C}{4}$\lambda\lambda$1548,1550 show that the fits are comparable in terms of fit statistics and the kinematic structure (see Fig.~4 in \citetalias{nielsen+2022}). However, a visual comparison of \ion{Si}{4}$\lambda\lambda$1393,1402, show that we were able to obtain a better fit. When comparing the derived total column densities of \ion{Si}{4} though, we obtain similar columns: $\log{(N_{\rm SiIV}/cm^{-2})}_{\rm This~work}=14.04\pm0.05,~\log{(N_{\rm SiIV}/cm^{-2})}_{\rm N22}=14.12\pm0.01$. Indeed, for almost all of the column densities that we have in common (\ion{H}{1}, \ion{Si}{2}, \ion{S}{2}, \ion{N}{1}) our measurements are within 0.2 dex. 

One exception to our agreement is the lower limit that we measure for \ion{O}{1} being 1.5 dex larger than their reported values. \citetalias{nielsen+2022} separately measure \ion{O}{1} from their main fit and achieve a similar column as ours. They acknowledge that their main fit is unable to reproduce this \ion{O}{1} column and might be explained as either the assumed abundance pattern (solar scale from \citet{grevesse+2010}) or additional unseen neutral components superimposed on the \ion{O}{1} complex.

Finally, they find a total $N_{\rm H~I}$ weighted metallicity of $\rm log{(Z/Z)}_\odot=$-0.68, and find a wide distribution of metallicties per cloud from effectively pristine gas ($\rm \log{Z/Z}_\odot<-2$) to super solar metallicity gas ($\rm \log{Z/Z}_\odot>0$). We infer a similar dust-corrected metallicity [Fe/H]$_d$=-0.75 and find components that may have high metallicities (see Section \ref{sec:baryoncycle_unboundgas}). 

This comparison with \citetalias{nielsen+2022} has shown that the flexible fitting method we use is quantitatively and qualitatively comparable to their statistical method.

\section{Insights into the Galaxy-Scale Baryon Cycle at z$\sim$2-3} \label{sec:baryoncycle}

Here we compare the CGM and ISM analyses presented in Sections \ref{sec:ism} and \ref{sec:cgm} to place constraints on the galaxy-scale baryon cycle of galaxies G1 and N1. We will summarize what we have learned about each of the galaxies thus far.

\textnormal{\textit{G1}} is a typical\textnormal{, \lya\ emitting }star-forming galaxy at $z_{\rm sys}=2.4312$. From the KCWI cube we found it possesses a \lya\ halo that extends to more than five times its continuum size (28 kpc), has a simple velocity distribution that peaks \textnormal{at} $v_{\lya}=+271~\kms$, and appears to have small objects that are connected to its \lya\ halo, G1-E and G1-S. From the MOSFIRE spectra and SEDs we found that its stellar mass ($\log{(M_*/\msun)}=9.9$), gas-phase oxygen abundance ($\rm 12+\log{(O/H)}=8.39$), gas-phase N/O ($\rm \log{(N/O)}=-1.51$), star formation rate (SFR=6-15 \msun), ionization parameter ($\log{U}=-3.0-- 2.5$), dust extinction ($A_{\rm V}$=0.21) are all typical of $z\sim2.3$ star-forming galaxies. From the HIRES spectrum we \textnormal{found} that G1's CGM (as probed by \textnormal{DLA} absorption) has a high occurence of metal absorption across many ionization \textnormal{states} with large column densities, its CGM is kinematically complex (requiring more than 10 components to fit the absorption; has metal absorption \textnormal{across $\Delta v\sim 1000~\kms$)}, has a moderate\textnormal{--high} metallicity ([Fe/H]=-0.76, dust-corrected) with an abundance pattern that seems to deviate from solar, and has moderate dust depletion ($\delta_d$=0.6 dex) \textnormal{inferred from line-of-sight extinction}.

\textnormal{\textit{N1}} is a lower-mass star-forming galaxy at $z_{\rm sys}=3.1509$ \textnormal{with properties in the tail distribution of $z\sim3$ LBG population and analogous to $z\sim2$ EELGs}. From the KCWI cube we found it posses a complex \lya\ halo with three components that extends to more than fifty times its continuum size (100 kpc); the main components is double peaked with velocities at $v_{\lya,b}=-439~\kms$ and $v_{\lya,r}=+171~\kms$, and appears to include small objects that are connected to its \lya\ halo, HU4 and HU6. From the MOSFIRE spectra and SEDs we found that N1 has a stellar mass that places it in the bottom quartile of UV-selected galaxies in spectroscopic samples  ($\log{(M_*/\msun)}=8.7-9.2$), low gas-phase oxygen abundance ($\rm 12+\log{(O/H)}=7.82$), moderate star formation rate (SFR$>$10 \msun), strong \lya\ ($\rm F_{\lya}=12.5\times10^{-17}\fluxunits$) and strong \textnormal{[\ion{O}{3}] ($\rm F_{\rm [O~III]}=14.1\times10^{-17}\fluxunits$)}, and a large $f_{\rm esc,\lya}=29\%$. From the HIRES spectrum we \textnormal{found} that N1's CGM has metallic absorption with high column densities across multiple stages of ionization\textnormal{, it} is kinematically complex (requiring more than 6 components to fit the absorption), has metal absorption across hundreds of \kms, has a \textnormal{moderate} metallicity ([Fe/H]=-1.1, dust-corrected) with an abundance pattern that \textnormal{is similar to} chemically young \textnormal{absorbers}, and has low CGM dust depletion \textnormal{($\delta_d$=0.05 dex).}

\subsection{Potential Satellites of G1 and N1} \label{sec:galaxy_group?}
We define a galaxy group as two or more galaxies of comparable mass ($\Delta M \sim$50\%) with projected distances smaller than the $z\sim 2-3$ galaxy autocorrelation length $r^{GG}_{0} = (6.0 \pm 0.5)$ Mpc \citep{peebles+1980}. \textnormal{The virial radii of the two galaxies are informed by previous KBSS studies and ranges from $R_{vir, \rm G1}\sim$80--90 kpc for G1 (based on the clustering of KBSS $L_*$ galaxies with $M_h\sim\rm 10^{12}~\msun$, assuming an NFW dark matter halo profile \citep{trainor+2012,rudie+2019}), and $R_{vir, \rm N1}\sim$60--70 kpc based on the virial radius determined for the median stellar mass $M_*\sim10^9~\msun$ (implied $M_h\sim 10^{11.5}$) of $z\sim2$ low-mass EELGs \citep{erb+2023}}.
\textnormal{The potential satellites are well within the virial radii of G1 and N1; the projected distances between the objects are at most 40 kpc which is about the same distance as the SMC and LMC from the MW\@.} 

\citetalias{nielsen+2022} suggested that the origin of the DLA towards Q2343+1232 is the intragroup medium of a galaxy group at the same redshift. 
We found from our deep KCWI cube, which includes \citetalias{nielsen+2022} exposures, 
that only G1 is at the same redshift as the DLA. 
However, we find evidence \textnormal{for} two satellites of G1: G1-E (east), and G1-S (south). Both sources are visible in the \textit{HST} image of the Q2343 sightline (\textnormal{Figure \ref{fig_kcwi_whitelight}}) but only G1-E appears to be associated with a continuum source. Both objects appear in \lya\ narrow-band images (irrespective of QSO subtraction technique) centered around $z\sim z_{\rm \lya,G1}$ (Fig.~\ref{fig_lyamap}). 
We argue that they are lower mass than G1 based on their photometry and \lya\ fluxes. Specifically, their rest-frame optical photometry (\textit{HST-IR}/F160W; QSO-subtracted) is small (compared to G1 $m_{\rm G1, F160W}=23.37 \pm 0.2$)\textnormal{,} $m_{\rm G1-E, F160W}=25.90 \pm 0.2$, and $m_{\rm G1-S, F160W} > 25.96$, and their \lya\ fluxes are also small compared to G1, $\rm F_{G1-E}/F_{G1}\sim0.11$ and $\rm F_{G1-S}/F_{G1}\sim0.15$ (see Table \ref{tab_fuv_spectra}). Both galaxies may be resonantly scattering the \lya\ from G1's halo because their spaxels connect/overlap in the pseudo-narrow band images. Additionally, G1-S has no optical emission lines \textnormal{detected in} the MOSFIRE spectra. Therefore, G1 is likely not part of a galaxy group but instead is a massive galaxy with two detected satellites. 

\begin{figure}
    \centering
    \includegraphics[scale=0.70]{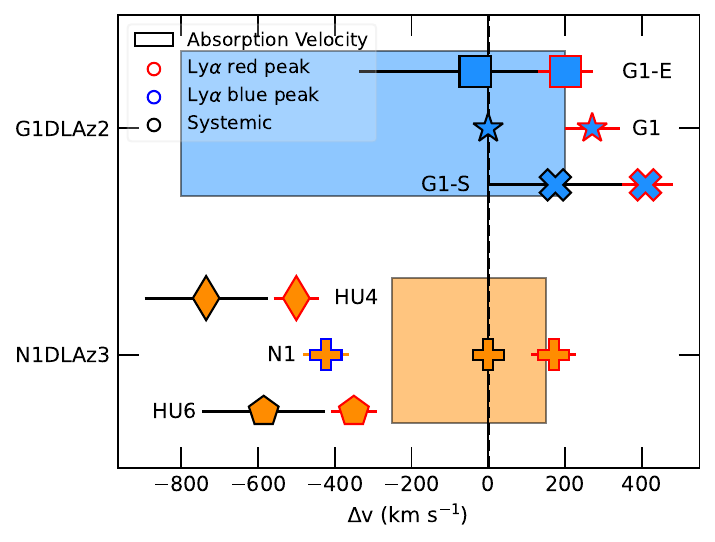}
    \caption{\lya\ emission kinematics (filled points) compared to CGM absorption kinematics (shaded regions) \textnormal{for the DLA associated with G1 (\textit{top}; blue shaded region) and the DLA associated with N1 (\textit{bottom}; orange shaded region).} Red outlined markers show the \lya\ red peak velocity (Section \ref{sec:line_emission},\ref{sec:fuvcontinuum}), black outlined markers are the systemic velocity (Section \ref{sec:mosfire},\ref{sec:line_emission_optical}), and blue outlined markers show blue peak velocity, all with respect to $z_{\rm sys}$ (nebular emission) of G1 and N1. Each marker is labeled in the center of the figure.}
    \label{fig_velocity}
\end{figure}

In Figure \ref{fig_velocity} we compare the kinematics of G1-E and G1-S to explore their potential effects on the dynamics (and perhaps abundances) measured in the DLA\@. We measure $z_{\rm \lya, G1-E} \rm = 2.4335 \pm 0.0008$ ($\Delta v_{\rm \lya,G1-E }=201\pm72~\kms$) and $z_{\rm G1-S,\lya} \rm = 2.4359 \pm 0.0008$ ($\Delta v_{\rm sys}=410\pm72~\kms$). Non-resonant transitions would provide more reliable redshifts but applying the typical $z_{\rm \lya}-z_{\rm sys}$ velocity offset to the galaxies ($\Delta v_{\lya}=-235\pm 101~\kms$ \citep{trainor+2015,steidel+2018}) gives $\Delta v_{\rm G1-E, sys}=-34\pm72~\kms$ and $\Delta v_{\rm G1-S, sys}=175-\pm72~\kms$. Their estimated systemic velocities coincide with metal absorption in the HIRES spectrum between -100 -- +200 \kms\  making it possible that they could be contributing to the absorption.

N1's \lya\ halo appears to have three distinct kinematic components (see Fig.\ref{fig_lyamap}). Interestingly, each component is coincident with continuum source(s) in the \textit{HST}/702W image (\textnormal{Figure \ref{fig_kcwi_whitelight}}); the redshifted component with N1; the intermediate component with HU6 (east of N1); and the blueshifted component with HU4 (and perhaps even more uncatalogued continuum sources; northwest of N1.)

In Figure \ref{fig_velocity} we compare the \lya\ kinematics of N1, HU4, HU6, and the DLA at the bottom of the figure. The 
components 
have velocities that are similar to N1's blue peak (extracted from spaxels that contain the continuum of the galaxy i.e., there is little chance of contamination from the other sources).

The estimated systemic velocities are fairly large when compared to the escape velocity of $z\sim$2 halo with virial mass $\log{(M_{\rm vir}/\msun)}=11.7$ at 20 kpc \citep{trainor+2012,rudie+2019}: $v_{\rm esc}\sim$450 \kms (when assuming a NFW profile). This velocity is an upper limit for N1 because it is low stellar mass ($\log{M_*}\sim8.7-9.2$) compared to more massive KBSS star-forming galaxies. Due to the complexities of \lya\ radiative transfer, it is difficult to determine whether these \lya\ velocities are dominated by bulk gas motion or resonant scattering, but the lack of a velocity gradient in all of the objects suggests that there is no gas acceleration\textnormal{/rotation}. Indeed, all of the \textnormal{objects} have a simple velocity structure with a small velocity dispersion of $\sigma\rm\sim50-100~\kms$. Additionally, we showed in previous sections that the dominant emission mechanism of \lya\ in N1 is \textnormal{not} collisional excitation given its low $\rm \lya/\Ha$. 

\textnormal{Both HU4 and HU6s} velocities do not overlap with the DLA absorption suggesting that they do not significantly affect the CGM absorption seen in the DLA.  
Though, it is still possible that they could have interacted with N1 in the past and/or ejected metals from a previous starburst to the DLA.

\textnormal{The objects} are likely lower mass than N1, considering their small \lya\ fluxes ($F_{\rm HU4,\lya}/F_{\rm N1,\lya}\rm=(2.88\pm0.1)\%$), $F_{HU6,\lya}/F_{N1,\lya}\rm=(9.2\pm0.1)\%$, and fainter F702W magnitudes: $m_{\rm HU6}=25.65\pm0.3$ and $m_{\rm HU4}=25.85\pm0.3$, compared to $m_{\rm N1}=24.45\pm0.3$.

Altogether, the combination of the \textit{HST} images, KCWI, MOSFIRE, and HIRES spectra suggest that the new objects discovered towards G1 and N1 are likely satellites. The \textit{HST} images showed that the objects are at projected distances well within the virial radius of G1 and N1, and that they are lower mass based on their photometry. The KCWI spatial and velocity maps showed that the objects have small \lya\ fluxes \textnormal{($F<0.1F_{\lya}$)} and velocities consistent with G1 and N1, though if a systemic velocity correction is applied to N1 only one of the objects is at a consistent velocity. The KCWI spatial maps and MOSFIRE spectra suggest that the objects near G1 may be scattering photons from its \lya\ halo based on their low \lya/\Ha\ and their \lya\ emission appearing spatially connected to the halo. The KCWI velocity maps and HIRES spectra show: for N1, that the satellites' \textnormal{presence} may not significantly affect the DLA absorption because their \lya\ emission velocities are much larger than the absorption velocity range; 
for G1 the satellites have similar velocities to strong metal absorption in its CGM and may affect the kinematics and abundances. 

 \subsection{Chemical Enrichment} \label{sec:baryoncycle_metallicity}
One of the most straightforward methods to trace the galaxy-scale baryon cycle is to compare metallicity differences between the ISM and CGM\@. Metals are formed via nucleosynthesis and are liberated via supernovae (CCSNe and Type Ia SNe) and other less energetic processes, e.g., the winds from AGB stars. Metals that are found in the CGM therefore likely originated from the ISM.
In Section \ref{sec:ism_metallicity_ionization} we \textnormal{calculated} ISM bulk metallicity using strong line calibrations, and in Section \ref{sec:cgm_metallicity} we inferred the metallicity of self-shielded, neutral phase CGM gas (where ionization effects are minimal) using Voigt profile decomposition. Since different elements can have different channels of production (e.g., hydrostatic vs.\ explosive nuclesynthesis, $\alpha$\-production, thermonuclear explosions) it is useful to compare similar metals. In this case, O and N can be reliably measured from both the ISM and CGM\@. 

\begin{figure*}
    \centering
    \includegraphics[scale=0.77]{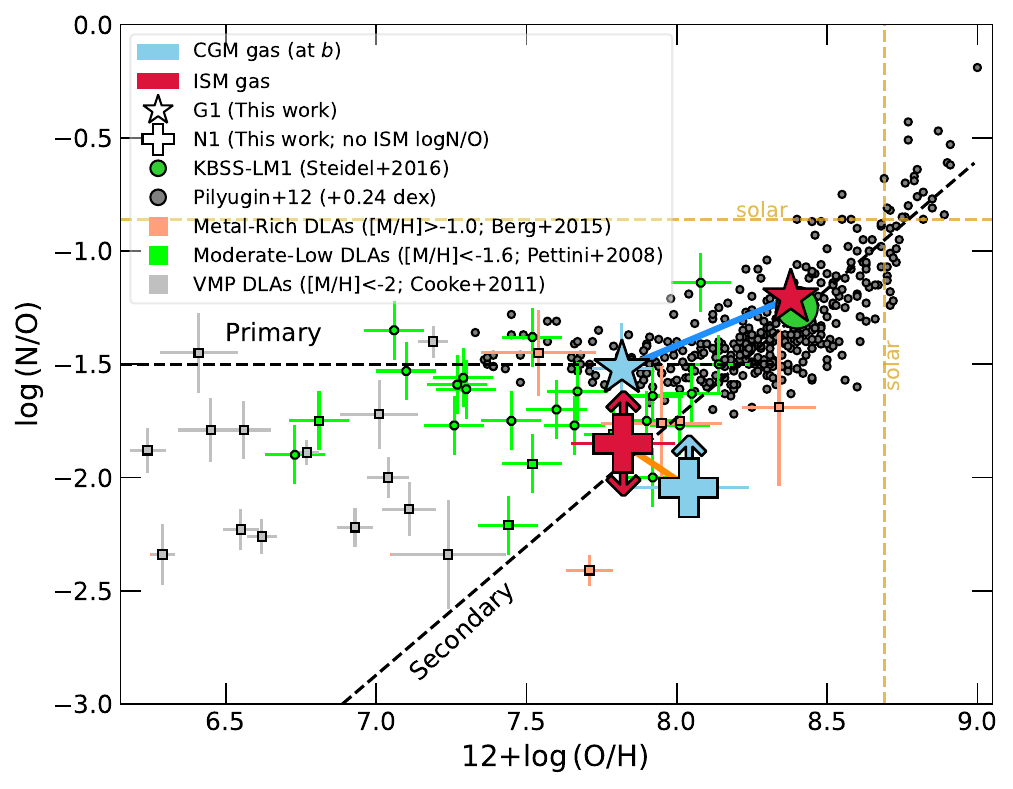}
    \caption{\textnormal{Comparison of gas-phase N/O as a function of oxygen abundance in G1's (star) and N1's (plus) CGM (light blue filled; Section \ref{sec:cgm_metallicity}) and ISM (red filled; Section \ref{sec:ism_metallicity_ionization}) gas. This is the first time this explicit comparison has been made at this $z$ (to our knowledge) and shows that G1's CGM is less chemically evolved than its ISM, and that N1's ISM and CGM are comparable in oxygen abundance. An effective average of the KBSS-MOSFIRE sample is plotted as a green circle \citep[KBSS-LM1;][]{steidel+2016}; pink squares show metal-rich DLAs compiled by \citealt{berg+2015}; light green squares are moderate--low metallicity DLAs compiled by \citealt{pettini+2008} where the pentagons show points where S was converted to O assuming $\rm (O/S)_\odot$=1.57 \citet{asplund+2009}); gray squares show very metal poor DLAs compiled by \citealt{cooke+2011_sources}; grey points show nearby HII regions in SDSS galaxies from \citet{pilyugin+2012}; dashed black lines show the approximate locations of the primary N plateau and the secondary N rise with similar locations and slopes as \citealt{pettini+2008}; and dashed yellow lines show solar values from \citealt{asplund+2009}. 
    REFERENCES. Metal-rich compilation by \citealt{berg+2015}: \citet{lopez+1999,prochaska+1999,prochaska+2001,prochaska+2001a,dessauges-zavadsky+2006,dessauges-zavadsky+2007,berg+2013,berg+2015a,lopez+2002,centurion+2003,petitjean+2008,prochaska+2002PASP,dutta+2014,ledoux+1998,erni+2006,pettini+2008,lopez+2003,srianand+2005,noterdaeme+2012,srianand+2012,henry+2007,prochaska+2003c,prochaska+2007,ledoux+2006,noterdaeme+2008,lu+1996,zafar+2014,levshakov+2002,dessauges-zavadsky+2001}. Moderate--low metallicity compilation by \citealt{pettini+2008}: \citet{desauges-zavadsky+2004,ellison+2001,petitjean+2008,lopez+2003,centurion+2003,henry+2007,dessauges-zavadsky+2006,lu+1998,pettini+2002,ledoux+2006,lopez+2002,dodorico+2002}; Very metal poor compilation by \citealt{cooke+2011_sources}: \citet{cooke+2011_carbon,penprase+2010,pettini+2008,molaro+2000,petitjean+2008,ellison+2010,dessauges-zavadsky+2001,srianand+2010,omeara+2006,prochaska+2002}}.}
    \label{fig_NO_OH}
\end{figure*}

For the first time at this $z$ (to our knowledge), Figure \ref{fig_NO_OH} shows the N/O as a function of O/H for G1 and N1.\textnormal{ We color coded the symbols to show metals from the CGM (blue) and from the ISM (red) compared to nearby \ion{H}{2} regions analyzed by \citet{pilyugin+2012}, a sample of metal-rich ($\rm [M/H]>-1.0$) high-$z$ DLAs compiled by \citet{berg+2015}, moderate-low metallicity ($\rm [M/H]<-1.6$) high-$z$ DLAs compiled by \citet{pettini+2008}, and very metal poor ($\rm [M/H]<-2$) DLAs compiled by \citet{cooke+2011_sources}. Unfortunately, [\ion{N}{2}] was not accessible for N1's ISM, so we show an unbound point (in N/O) at the O/H abundance that we were able to measure, which places it 
in the low-metallicity regime. Although the O CGM absorption lines are saturated, we nonetheless represent the [O/H] as a point and not a limit because both galaxies $\alpha$-element abundances were within 0.2 dex; we adopt this difference as its error.}

\textnormal{The CGM gas lies in the ``primary nitrogen'' plateau where the majority of the N is liberated from CCSNe, which has been observed before in high redshift DLAs \citep[e.g.,][]{pettini+1995,centurion+2003,pettini+2002,pettini+2008,zafar+2014c}. Indeed, we can see that neither the high-, moderate-, or low-metallicity DLA samples populate the secondary nitrogen rise.} G1's ISM gas is in the ``secondary nitrogen'' rise where more delayed enriching events have commenced \citep[e.g., winds from AGB stars,][]{marigo+2001}. This shows that the ISM of G1 is more chemcially evolved, similar to the KBSS-LM1 galaxy stack (an effective average of the KBSS-MOSFIRE sample). More explicitly, this result suggests that the \textnormal{CGM metals (probed from this sightline)} were ejected earlier in G1's SFH when it was in the primary nitrogen plateau or that the CGM has been enriched by SN winds with minimal entrainment of AGB-enriched gas. In the latter scenario, the high-velocity (high-energy; CCSNe and SNeIa) SN winds escaped the galaxy and enriched the CGM whereas the low-velocity (low-energy) winds stayed within the ISM in the recent past.

This positive metallicity gradient ($Z_{\rm{ISM}}$-$Z_{\rm{CGM}}>0$) seen in G1 is common in other galaxy-QSO studies \citep[e.g.,][]{chen+2005,peroux+2011,battisti+2012,prochaska_2007,berg+2023} and aligns well with the picture of the ISM being the main driver of metals into the CGM \textit{and} \textnormal{the presence of inefficient mixing of outflow ejecta with ambient gas.} More specifically, if the ISM is the origin of the bulk of metals seen in the CGM with non-instantaneous mixing, one would expect most sightlines to show $Z_{\rm{ISM}} \gg Z_{\rm{CGM}}$ with some sightlines (or components) being metal rich. 

N1 shows a flat metallicity gradient ($Z_{\rm{ISM}}$-$Z_{\rm{CGM}}\sim0$), i.e., it has similar O abundance in both its CGM and ISM\@. This is not common \citep[but see e.g.,][]{peroux+2011,schady+2024}. Unfortunately, we do not have any constraints on the N/O ratio in the ISM and only a lower limit in the CGM, so it is difficult to make inferences on its enrichment history. But its O abundance suggests that both the ISM and CGM lie in the primary nitrogen \textnormal{plateau} which is consistent with the findings thus far that it has a young stellar population. Indeed, N1's stellar mass is small, its SED is blue, and its sSFR is high, all of which point to a young stellar population with an ongoing/recent starburst (Figs.~\ref{fig_sed} and \ref{fig_MEx_sSFREx}). 

There are a few ways that one could explain N1's flat metallicity gradient: (1) The QSO is intersecting a particularly metal-rich sightline. This would be consistent with the patchy nature of the CGM. (2) We are seeing stripped ISM from HU6. This could be consistent with the \lya\ kinematics ($v_{\rm HU5}\sim$-500 \kms) and tapered morphology that narrows  away from near N1, but is inconsistent with the DLA absorption velocities ($v_{\rm DLA}\sim$-200 - +150 \kms) and lack  of velocity gradient. (3) The satellites of N1 (HU4 and HU6) are driving galactic outflows so that the DLA metallicity (and all other properties) is a combination of two or more of the galaxies. The kinematics of the \lya\ from the satellites do not fully support this interpretation because neither of the satellites are at velocities with DLA absorption ($v_{\rm DLA}\sim$-200 - +150 \kms). \textnormal{(4) N1 ejected very metal-rich (super-solar) gas that efficiently mixed into the CGM such that the metallicity is now similar to the ISM\@. This is the least likely scenario as the CGM is not well mixed \citep[e.g.,][]{faucher-giguere+2023}, but it is possible drive such high metallicity winds \citep[][]{martin+2002,strickland+2009,chisholm+2018}}

Altogether we were able to see that the ISM-CGM metalliciity between the galaxies were very different\textnormal{. G1's CGM was more metal poor and less chemical evolved than its ISM while N1 CGM and ISM had comparable metallicity.} Expanding the number of galaxies that we can map onto this space will help answer questions about the metal enrichment of the CGM.

\subsection{Dust Abundance and Depletion} \label{sec:baryoncycle_dust}
Explicit comparisons between the dust content in the CGM and ISM of individual galaxies are almost entirely lacking in the literature, especially at $z\sim2$. There is work describing cosmic dust content \citep[e.g.,][]{pontzen+2009,ledoux+2015,decia+2018,peroux+2020,konstantopoulou+2023} using DLAs, but the explicit connection to singular galaxies is not common \citep[but see, e.g.;][]{rudie+2017,boettcher+2021}. 

We quantify the amount of dust in the ISM in Section \ref{sec:ism_sfr_dust}) using the Balmer decrement ($F(H\alpha)/F(H\beta)$), then again in Section \ref{sec:sed_fitting}) from the best-fit SED parameters. 
We calculated CGM dust depletion corrections $\delta_d$ in Section \ref{sec:cgm_dust} 
using the empirical scaling relations from \citetalias{decia+2016}.
We convert the CGM depletion corrections to extinction by scaling the galactic $A_{\rm V}$ to $N_{\rm H~I}$ conversion by metallicity (equation 8 from \citetalias{decia+2016}): 
\begin{equation} \label{eq:av_cgm}
    A_{\rm V} = DTM \times \left( \frac{A_{\rm V}}{N_{\rm{HI}}} \right)_{\rm Gal} \times N_{\rm{HI}} \times 10^{[M/\rm{H}]}
\end{equation}
where $DTM$ is the dust to metal ratio equal to $(1-10^{\delta_d})/dtm(\rm{Gal})$ where $dtm(\rm{Gal})$=0.98 \citep[\textnormal{the dust-to-metal ratio of the Milky Way;}][]{decia+2013} and $\delta_d$ is the depletion correction derived in Section \ref{sec:cgm_metallicity}; $\left( \frac{A_{\rm V}}{N(\rm{H~I})} \right)_{\rm Gal} = \rm 0.45\times10^{-21}~mag~cm^{-2}$ \citep{watson+2011} is the galactic conversion from \ion{H}{1} column to extinction; $N(\rm{H~I})$ is the linear neutral hydrogen column density; and [$M$/H] is the metallicity, which we equate to the dust corrected iron abundance $\rm [Fe/H]_d$ (see Section \ref{sec:cgm_metallicity}).

\begin{figure}
    \centering
    \includegraphics[scale=0.7]{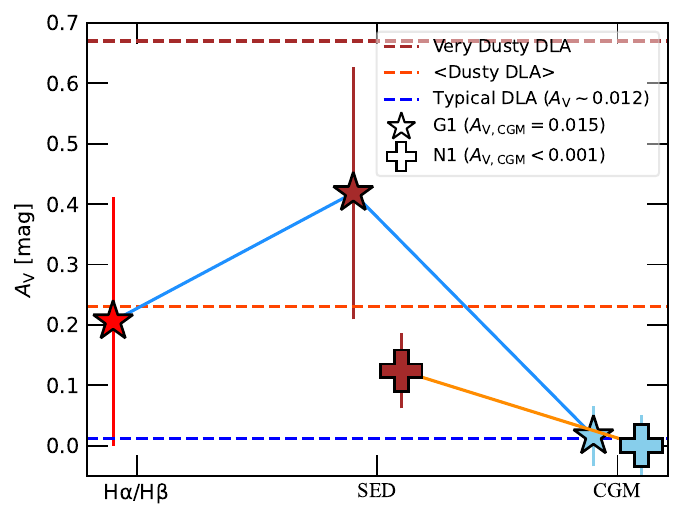}
    \caption{\textnormal{Dust extinction of G1 (blue outlined star) and N1 (orange outlined plus) in the ISM (red fill) and CGM (light blue fill). Extinction measurements using (1) the Balmer decrement are denoted as ``$\rm H\alpha/H\beta$,'' (see Section \ref{sec:ism_sfr_dust}), (2) the best-fit SED models as ``SED,'' (see Section \ref{sec:sed_fitting}) and (3) from the CGM as ``CGM\@'' (see Section \ref{sec:cgm_dust}). We show typical DLA dust extinction as a blue horizontal dashed line from \citep{decia+2016,konstantopoulou+2023}, average dusty DLA extinction \citep{heintz+2018}, and a very dusty DLA that was analyzed by \citealt{konstanatopoulou+2024}.
    We use $R_{\rm V}$=3.1 \citep{cardelli+1989}.}}
    \label{fig_dust}
\end{figure}

In Figure \ref{fig_dust} we show the dust content of G1 and N1 expressed as extinction from \textnormal{\ion{H}{2}} regions (Balmer decrement), continuum photometry (SED), and CGM (using Equation \ref{eq:av_cgm}). We do not have a nebular extinction for N1 because \Ha\ was accessible from the ground. \textnormal{We compare the extinction to that of a typical DLA from the large samples analyzed by \citet{decia+2016,ramburuth-hurt+2023,konstantopoulou+2023}, the average extinction of a sample of dusty DLAs \citep[$A_{\rm V}\gtrsim$0.2;][]{heintz+2018}, and a very dusty DLA \citep[J1056+1208;][]{konstanatopoulou+2024}.  Regardless of the method employed to determine the dust in the ISM, there is at least an order of magnitude less inferred line-of-sight extinction in the CGM (light blue filled symbols).} 

It is possible that we are underestimating the depletion of N1 because we could only use the [Si/Fe] ratio to determine its dust depletion correction. We note that if we were to use the empirical CGM depletion correction for G1 instead of the \citetalias{decia+2016} relations, the 
$A_{\rm V}$ would not change significantly. 

This trend is consistent with dust being driven from the ISM to the CGM.
It seems likely that the dust was formed in the ISM and was expelled via SN winds, just like the metals. \textnormal{Along with dust creation and expulsion, it is also possible that dust was destroyed during this process \citep[e.g.,][]{otsuki+2024}.} There could be a few ways in which the depletion patterned formed.  Either the depletion pattern was always there, or it took time for refractory elements to deplete onto dust grains. 

Regardless of the origin of the dust depletion pattern, \textnormal{we have shown that the inferred line-of-sight extinction is more than ten times smaller} in the CGM as compared to the galactic \ion{H}{2} regions.

\subsection{Unbound Gas in the CGM} \label{sec:baryoncycle_unboundgas}
\begin{figure*}
    \centering
    \includegraphics[scale=0.78]{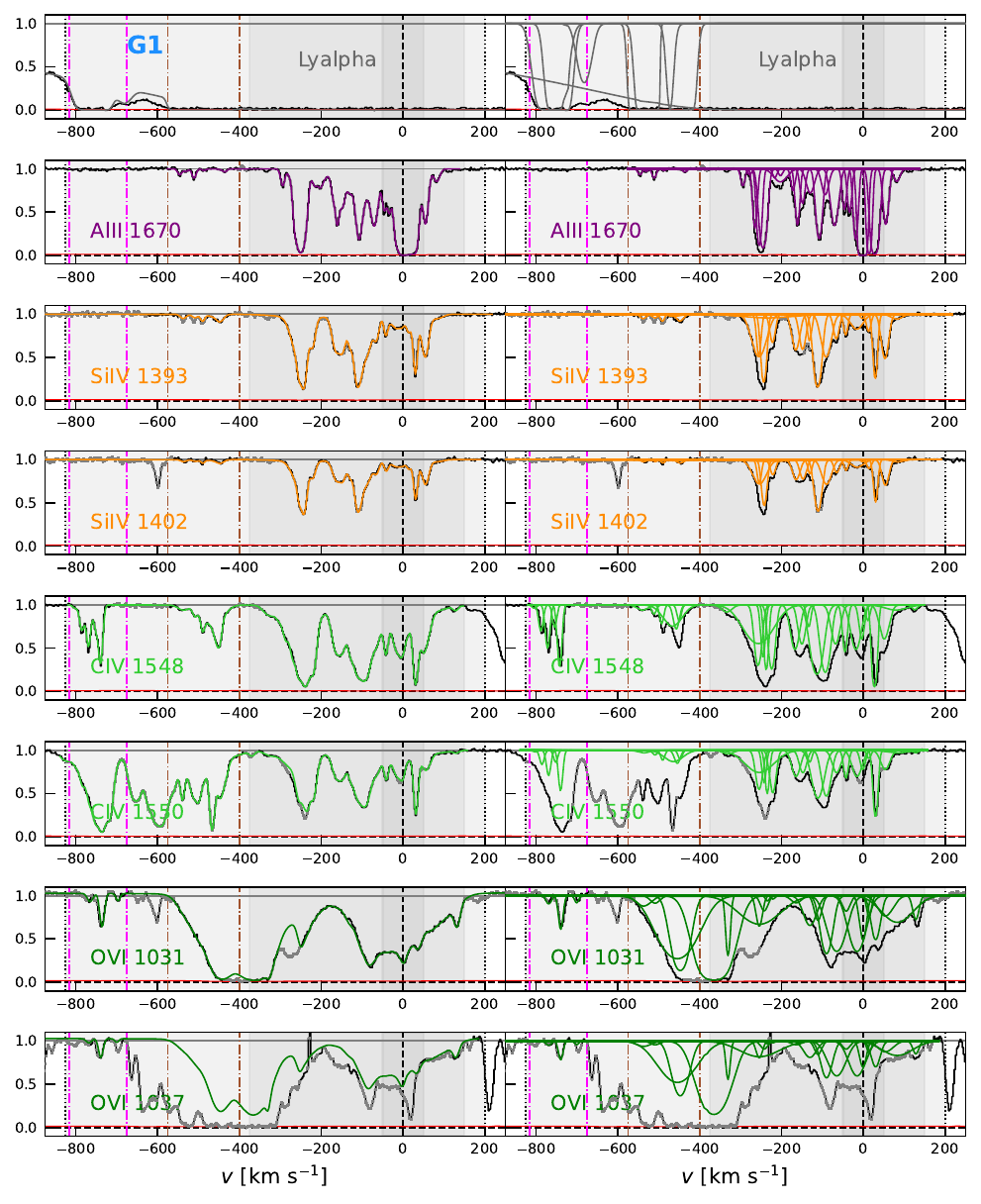}
    \caption{G1's intermediate- and high-ionization metal CGM absorption and and best-fit Voigt profiles centered at $z_{\rm G1}$=2.4312. The magenta dashdot lines shows a metal absorption complex at $v\sim-750~\kms$ and brown dashdot lines show metal absorption at $v\sim-500~\kms$. The lines and colors are \textnormal{the} same as Figure \ref{fig_cgmfit_q2343rep}. 
    }
    \label{fig_cgmfit_q2343repmedhigh}
\end{figure*}

In Figure \ref{fig_cgmfit_q2343repmedhigh}, we show all best-fit intermediate- and high-ionization metal transitions, a representative low-ionization transition, and the best-fit \ion{H}{1} from \citetalias{rudie+2012} for G1. 
The intermediate- and high-ionization transitions (\ion{C}{4}, \ion{Si}{4}, \ion{O}{6}) have a very similar structure, and share $\sim \rm 90\%$ of \textnormal{their} components. 
Their structure is different from that of the low ionization metals which is typical in DLAs due to the self-shielding. 

\textnormal{Some} of the absorbers in G1's CGM have radial velocities $v_r\gtrsim$500 \kms\ relative to the systemic redshift. These velocities are lower limits on the true (three dimensional) velocity of the gas. Absorbers at smaller velocities might also be unbound, but we cannot tell from the current data.  At these large velocities, the absorbers are \textit{unambiguously} gravitationally unbound to the galaxy, given the escape velocity of a typical KBSS $L_*$ star-forming galaxy $v_{\rm esc}\sim$500 \kms at $b\sim$20 kpc and $M_h\sim\rm 10^{12}~\msun$, assuming an NFW dark matter halo profile \citep{trainor+2012,rudie+2019}. Additionally, the impact parameter is a lower limit on the distance from the galaxy, so some absorbers near (or below) the escape velocity may actually be unbound since a farther distance would reduce the escape velocity. \textnormal{N1 might also have unbound gas in its CGM but its absorbers' radial velocities range between -200 -- +100 \kms\ } \textnormal{which is well below its likely escape velocity $v_{\rm esc, N1} \sim 400 ~ \kms\ ~ (M_h \sim 10^{11.5}~ \msun$), so we can conclude only that there is no \textit{unambiguously} unbound gas and do not discuss it in this section \citep[][]{rudie+2019,erb+2023}}.
\textnormal{The sightline to Q2343 has two absorption complexes that, if associated with G1, would have velocities equal to or exceeding its escape velocity}: one near $v\sim$-500 \kms which we refer to as \textnormal{``Complex500''}, and the other near $v\sim$-750 \kms which we refer to as \textnormal{``Complex750''}.

Complex500 (magenta dash-dot lines) is seen across all ionization states observed in the HIRES spectra, including neutral (\ion{H}{1}, \ion{O}{1}, \ion{N}{1}), low (\ion{C}{2}, \ion{Al}{2}, \ion{Si}{2}), low-intermediate (\ion{C}{3}, \ion{Al}{3}, \ion{Si}{3}), intermediate (\ion{C}{4}, \ion{Si}{4}), and high (\ion{O}{6}) ionization metal species. The ions share the same kinematic structure even though they differ significantly in their column densities. For example, $\log{(N_{\rm H~I}/{\rm cm}^{-2})}=15.86\pm0.1$, $\log{(N_{\rm Al~II}/{\rm cm}^{-2})}=11.65\pm0.1$, $\log{(N_{\rm Si~IV}/{\rm cm}^{-2})}=12.43\pm0.2$, $\log{(N_{\rm C~IV}/{\rm cm}^{-2})}=13.58\pm0.2$, and $\log{(N_{\rm O~VI}/{\rm cm}^{-2})}=14.82\pm0.2$ (found by summing over $\sim$-550 -- -450 \kms). \textnormal{We can see that the low-ion metal columns (e.g., \ion{Al}{3}) are small compared to the intermediate- and high-ionization metals (e.g., \ion{C}{4}).} Complex500 is reminiscent of the CGM absorbers analyzed by \citetalias{rudie+2019}, who found that most absorption complexes (within $R_{\rm vir}$ at $z\sim$2--3) had low- to intermediate-ionization (and some high-ionization) ions with the same kinematic structure\textnormal{, with many of the absorbers at velocities that exceeded the galaxies gravitation potential. This suggests that Complex500 is more typical of the $z\sim2$ CGM compared to the $v\sim0~\kms$ absorbers that are self-shielded by the DLA that allow for much larger columns of low-ionization gas.}

\textnormal{Complex750 (brown dashdot lines) is seen only in \ion{H}{1} (two components), \ion{C}{4} (five components), and \ion{O}{6} (three components) spread between -800 -- -700 \kms.} They are the fastest moving absorbers detected in G1's CGM.

Their column densities are more comparable \textnormal{to each another} (in contrast to Complex500). Specifically, $\log{(N_{\rm H~I}/{\rm cm}^{-2})}=14.89\pm0.1$, $\log{(N_{\rm C~IV}/{\rm cm}^{-2})}=13.61\pm0.1$, and $\log{N_{\rm O~VI}/{\rm cm}^{-2}}=13.40\pm0.1$. Their ratios are $\log{(N_{\rm C~IV}/N_{\rm H~I})}=-1.28 \pm 0.2$, $\log{(N_{\rm O~VI}/N_{\rm H~I})}=-1.49\pm0.2$, and $\log{(N_{\rm C~IV}/N_{\rm O~VI})}=0.21\pm0.2$. 
\textnormal{The strongest component of Complex750 is found at $v\sim766~\kms$.}

These two complexes are associated with low $\log{N_{\rm H~I}}$ gas and relatively high columns of metals, suggesting that the gas is either highly ionized, metal-rich, or both. The high velocities associated with the two complexes, especially that of Complex750, point to an energetic origin. The combination of the column densities and velocity suggest that the absorbers were ejected via energetic galactic outflows. Indeed, \citetalias{nielsen+2022} found that the metallicities of the clouds in Complex500 are super-solar, possibly indicating that these absorbers are the products of undiluted CCSNe. This is likely also true for Complex750 given its column density ratios.

These results are similar to previous studies where enhancement of \ion{H}{1}, \ion{C}{4}, and \ion{O}{6} was found within 180 kpc of KBSS galaxies, at column densities and velocities that would suggest metal-rich (or highly ionized) gas that originated from a galactic outflow, and that unbound absorbers were usually associated with low $\log{N_{\rm H~I}}$ \citep{adelberger+2003,simcoe+2006,turner+2014,turner+2015,pratt+2018,rudie+2019}. \textnormal{Indeed, enhanced \ion{O}{6} absorption (and \ion{C}{4}) have been found to favor a scenario where the ions arise from metal-rich ($Z>-0.1$) hot ($\rm T>10^5~K$) gas \citep[][]{simcoe+2004,simcoe+2006,turner+2015}.}

We now place limits on the timescales associated with the ejection of the absorbers using their measured velocities as a sanity check. Starburst driven galactic superwinds are known to have velocities of $v_{winds}\sim$200 - $>$1000 \kms from $z=0-1$ \citep[][]{heckman+1990,heckman+2015,prusinski+2021}. Indeed, the average velocities of the two complexes are consistent with this range. Using the lower and upper limit for galactic superwind velocity over 20.8 kpc gives an ejection time between $\lesssim$21-104 Myr ago. That is ample time for the proposed galactic outflow to have launched and made its way to \textnormal{$b$ given the stellar age} of G1 ($\sim$600 Myr).

Altogether, these complexes appear to provide direct evidence of metal enrichment of the CGM (and/or IGM) from galactic outflows based on their velocities, \ion{H}{1} and metal column densities, multiphase nature of the absorbers, and limits on the timescales of ejected absorbers.


\section{Discussion and Caveats} \label{sec:discussion}

\subsection{Are We Actually Probing CGM Gas?}

One can never prove that an absorber is associated with a particular galaxy because there are multiple explanations for the presence of gas at the same redshift as the galaxy: ejected/stripped gas from another galaxy, chance alignment of a satellite galaxy intersecting the QSO LOS, intragroup medium of an unresolved galaxy group, etc. Therefore, we must rely on the ensemble of overlapping physical properties of the galaxy and absorber to argue whether it is \textit{reasonable} to assume that they are physically associated with one another. 

The velocity offset between the systemic redshift acquired from the rest-optical non-resonant nebular emission and the best-fit DLA HI component for both G1 and N1 are $\rm|\Delta v| <10~\kms$. 
One would expect a larger velocity offset if the objects were truly not physically in the same vicinity.

The impact parameters of the QSO sightlines are well within the virial radius of the galaxies ($b \rm \ll 100~kpc$) where one would expect large column densities from CGM gas absorption detectable within a single sightline. The occurrence rate of high \ion{H}{1} column gas decreases steeply with impact parameter \citep[]{rudie+2012,werk+2014,krogager+2017}. 

The metallicity difference seen between the ISM (12+$\log{(\rm O/H)}$) and CGM ([O/H]) fits well with the picture of the DLA arising from CGM gas that has been enriched by metal products from the ISM with non-instantaneous mixing for G1. However, N1's flat metallicity difference is puzzling but could be explained by a more complex interplay between N1 and its satellite galaxies e.g., the origin of the DLA could be the stripped ISM of HU6 or the satellite galaxies are driving outflows that are enriching the DLA.

An interesting argument against a galactic origin of the DLA gas towards G1 is related to its metal abundance pattern. After accounting for dust depletion, we found that the DLA has moderate-metallicity $\rm [Fe/H]_d=-0.76$ but has no $\alpha$-enhancement ($\rm [\alpha / Fe]\sim0$), suggesting that it has not seen recent enrichment from galactic outflows dominated by CCSNe like that seen in VMP DLAs ($\rm [\alpha / Fe]\sim0.5$). These two points seem to contradict G1's apparent ability to drive galaxy-scale outflows, deduced from the instantaneous SFR (and SFR$_{\rm SED}$), ionization, and optical line ratios. After all, galaxy-scale outflows are ubiquitous at $z\sim2-3$. Nonetheless, we know that the CGM is patchy, so the sightline we are probing might have been ``missed'' by the most recent outflow events \textit{or} G1s abundance pattern deviates from solar values. 

Finally, it is possible that the DLA arises in the ISM of a (physically) small satellite galaxy that has not yet been identified;  this possibility is difficult to rule out, but the level of enrichment attained by the gas would make it implausible to associate the absorption with a very faint host. 
We have leveraged very deep IFU cubes and the natural masking of the \textnormal{QSO} by the DLAs to show that there is no faint/diffuse Ly$\alpha$ emission on top of the QSO other than emission directly associated with the galaxies.  Our 3$\sigma$ detection limit corresponds to a \textnormal{surface brightness} of $\rm SB \geq 1.5 \times 10^{-19}~\sbunits$. Unfortunately, our QSO subtraction technique only allows us to probe $b\geq$1.25'' from the QSO center, so future improvements to our technique may result in discoveries of new galaxies. The HST images contain no detectable continuum sources within 0.7\arcs of the QSOs, which places a limit on the hypothetical projected radius $r\lesssim$2.8 kpc at both galaxies' redshifts. With more sensitive IFUs on larger telescopes with longer integration times, we will be able to probe lower limits of flux (stellar mass) until either a galaxy is found or the limiting \textnormal{stellar mass} is less than that corresponding to $\log{N_{\rm H~I}}$.

Altogether, it is reasonable to interpret the DLAs as the CGM of the galaxies due to the small CGM-ISM velocity offset ($\rm <10~km~s^{-1}$), small impact parameters ($b<$21 kpc, well within $R_{\rm vir}$), and the positive ISM-CGM metallicity gradient (for G1).  Some puzzles remain: the CGM abundance pattern (for G1) and the flat CGM-ISM metallicity gradient (for N1). 

\subsection{\textnormal{Comparison with other Absorber-Galaxy Systems}}
\textnormal{
In this section we compare G1 and N1 to several works that analyzed absorber-host galaxies at z$\geq$2. One caveat to this comparison is that the two galaxy-absorber pairs presented  in this paper do not represent a homogeneous selection and thus should not be used to infer properties of some underlying population.}

\textnormal{\citet{krogager+2017} analyzed a sample drawn from two decades of searches for $z>2$ DLA host galaxies. Their new search focused on metal-rich DLAs under the assumption that their hosts would follow the mass-metallicity relation, i.e., the hosts would be more likely to be detected \citep[e.g.,][]{fynbo+2008,fynbo+2011}. Similarly, a very recent survey was conducted by \citet{oyarzun+2024} who searched for \lya\ emission of metal-rich DLA hosts using KCWI. This sample was unique because of the inclusion of galaxies with CO detections associated with DLAs. Both studies report a galaxy detection rate (\lya\ and/or CO detection) close to $\sim 50 \%$, well above the typical blind detection rate of $\sim10-15\%$ (i.e., no metal selection), adding more evidence that metal-rich DLAs are more likely to have detectable host galaxies \citep[][]{krogager+2017,fumagalli+2015}. This is consistent with our findings that neither G1 nor N1 is associated with a metal-poor DLA. Interestingly though, the DLA metallicities of the detected galaxies ranged from low--moderate to metal--rich ([M/H]=-1.39 -- -0.27) putting G1 in the middle of the distribution.}

\textnormal{Comparing further, we see that the \ion{H}{1} column density of the DLAs were comparable to that of G1 ($\log{(N_{\rm H~I})} \sim 20.6$); the impact parameters spanned $b$=0 -- 70 kpc, the majority (4/6) of which being $b<12$ kpc, which is about 1.5$\times$ smaller than G1 and N1; and the \lya\ fluxes ranged from 0.5 -- 17 $\times 10^{-17} \fluxunits$, comparable with the fluxes we measured for N1 and G1 (3, 13 $\times 10^{-17} $\fluxunits).}

\textnormal{The MUSE Analysis of Gas Around Galaxies \citep[MAGG;][]{lofthouse+2020} survey and the MUSE Quasar-field Blind Emitters Survey \citep[MUSEQuBES;][]{muzahid+2020} have recently searched for $z\gtrsim 3$ LAEs at the same redshift as \ion{H}{1} absorbers. Both samples are galaxy-selected (via \lya\ emission detection) and MAGG additionally focuses on QSOs with strong \ion{H}{1}-absorbers (LLS, and higher). Both samples discovered a large number ($N\gtrsim100$) of LAEs within $b\sim 10-300$ kpc of QSOs and $\pm1000~\kms$ of a mix of LLS, sub-DLAs, and DLAs ($\log{(N_{\rm H~I}/cm^{-2})}>16.5$) that have moderate to low metallicities ([M/H]$\sim$-3.5 -- -1.5). The novelty of N1 compared to these samples is threefold: First, its impact parameter is almost an order of magnitude smaller than a typical detection ($b_{\rm N1}\sim 20~kpc$ vs. $b_{\rm MAGG,median}\sim 165-200$ kpc).  Second, it is associated with a (sub-)DLA whereas the majority of the MAGG LAEs are associated with LLSs.  Third, N1 was a continuum-selected galaxy as opposed to the \lya\ selection from MUSEQuBES and MAGG. }

\textnormal{MAGG has found more than 120 LAEs but only six (confident) detections are found at $b<50$ kpc, whereas MUSQuBES discovered 96 LAEs, 5 of which have $b<50~kpc$. Interestingly, the LAEs with the smallest impact parameter in their sample are not found to be associated with DLAs but instead LLS (or lower), suggesting the CGM of N1 may not be typical, although larger samples at small impact parameters would be required for this statement to be made with high confidence.  The closest (confident) LAE-DLA hosts are typically found at $b>50$ kpc while the brightest (confident) LAEs are found at $b>100$ kpc. Though seemingly uncommon, there have been discoveries of $z\sim3$ DLA host galaxies \citep[see e.g.,][]{fumagalli+2017}.}
    
\textnormal{Finally, we compare N1 to dwarf-galaxy CGM studies given its lower mass $M_* \sim 10^9 M_\odot$. MUSEQuBES probes a lower stellar mass ($M_* \sim 10^{8.6} M_\odot$) and lower \lya\ luminosity ($L_{\lya}\sim 10^{42}~erg~s^{-1}$) than N1. Among some of their findings is an excess of \ion{H}{1} and \ion{C}{4} at line of sight (radial) velocities $v<500~\kms$; they measure a range of $\log{N(\rm{CIV}/cm^{-2}})\sim$11.4--14.3, placing N1 at the high end of their distribution ($\log{N(\rm{CIV}/cm^{-2}})_{N1}=$13.9); about 45\% of the \ion{C}{4} absorbers were at velocities that were unbound to the halo, which is not the case for N1 which showed no unambiguously unbound absorbers (see Section \ref{sec:baryoncycle_unboundgas}).}

\textnormal{At low-redshift there have been multiple dwarf-galaxy CGM studies that have found a low frequency of CGM metal absorption from low- (e.g., \ion{Si}{2}) and intermediate-ions (e.g., \ion{C}{4}) but a high frequency of high-ionization gas \ion{O}{6}, all associated with low \ion{H}{1} gas \citep[][]{johnson+2017,qu+2022,zheng+2024} \citep[but see][]{bordoloi+2014}. Therefore, N1 appears to support this trend that at high-$z$ there is a large fraction of neutral--intermediate ionization metal absorption, though a larger sample of $z\sim3$ lower mass galaxies will be required to more convincingly support this.}

\textnormal{The full KBSS-InCLOSE sample will provide significantly improved statistics on the bulk properties of the inner CGM surrounding galaxies at $z\sim2$, including the frequency of high-N(HI) absorption such as that seen in N1 and G1, herein.} 


\section{Summary and Conclusions} \label{sec:summary}
We have presented the design and first results from the new KBSS-InCLOSE survey which focuses on the Inner CGM of QSO Line Of Sight Emitting (InCLOSE) galaxies at $z\sim$2-3. \textnormal{KBSS-}InCLOSE allows us to connect galaxy properties (e.g., stellar mass, interstellar medium ISM metallicity) with the physical conditions (e.g., kinematics, metallicity) of the inner circumgalactic medium to directly observe the galaxy-scale baryon cycle of  
$z\sim2-3$ star-forming galaxies.  

In this paper we focused on two QSO fields, Q2343+1232 and Q2233+1310, for which we have added extended optical IFU coverage with deep KCWI pointings. \textnormal{We leveraged the fact that these fields have $z\sim2-3$ star-forming galaxies near the QSO line of sight (nLOS galaxies), Q2343-G1 (G1, z=2.4312; a typical $z\sim2$ \textnormal{star-forming} galaxy) and Q2233-N1 (N1, z=3.1509; a lower mass LBG with strong, extended nebular emission that makes it analagous to $z\sim2$ EELGs),} to develop observation strategies, QSO subtraction techniques, and coupled ISM-CGM analyses that will be applied to the remaining KBSS-InCLOSE fields.

We summarize our main findings below:
\begin{enumerate}
    \item KCWI is an efficient and effective instrument for finding new nLOS galaxies. We discovered more than \textnormal{15} new galaxies across the two QSO fields. The QSO subtraction techniques that we develop, use, and repurpose were key to this discovery process since the QSO point spread function (PSF) dominates emission in the wings, where new galaxies are likely to be found.

    \item For the first time at this $z$, we explicitly \textnormal{compare the gas-phase N/O (vs O/H) in the CGM and ISM finding that \textnormal{\textit{G1's}} CGM is metal-poor and less chemically evolved than its ISM suggesting that it was enriched by a previous starburst, perhaps when its ISM was in the low metallicity regime.} 
    
    \textnormal{\textit{N1's}} CGM has a comparable metallicity to its ISM \textnormal{($\rm \Delta [O/H]_{CGM-ISM}\sim$0)} which may be a result of stripped ISM gas from a previous interaction with one of its satellite, or simultaneous enrichment \textnormal{from one} or more of its satellite galaxies.

    \item \textnormal{The inferred CGM line of sight dust extinction is an an order of magnitude less than the ISM for both galaxies suggesting there is little dust in their CGM.}

    \item G1's CGM \textnormal{appears to have unbound, metal-rich, hot gas} that may be the product of undiluted CCSNe \textnormal{ejecta} driven by an energetic galactic outflow between $\rm \sim20-100~Myr$ ago

    \item Both galaxies' CGM absorption (HIRES) revealed: a high incidence of metal absorption showing detections of e.g., C, N, O, Al, Si, S, Fe, etc.; multiphase gas is common from neutral to quintuply ionized transitions with similar kinematic structure; kinematicatically complex absorbers, requiring at least 10 components (and up to 31 components!) to fit most metal transitions and spread out to $\rm \sim-800~km~s^{-1}$; there are low to moderate levels of dust depletion ($\delta_d=0.6$ dex for G1; $\delta_d=0.05$ dex for N1) ratios; and \textnormal{\textit{G1}} has a puzzling abundance pattern (chemically evolved by some ratios [$\alpha$/Fe]$\sim$0 but chemically young in some ratios [N/Fe]$\sim$-0.8) that may deviate significantly from solar, and \textnormal{\textit{N1}} has an abundance pattern typical of chemically young absorbers.

    \item \lya\ surface brightness \textnormal{and} velocity maps \textnormal{(KCWI)} revealed that 
    \textnormal{\textit{G1}} has a single-peaked \lya\ profile ($v_{red}=+$271 \kms), an extended \lya\ halo (28 kpc), and has two \lya\ satellite galaxies in proximity to it. 
    \textnormal{\textit{N1}} has a double-peaked \lya\ profile ($v_b\sim-420~\kms,v_r\sim 171~\kms$), strong \lya\ emission \textnormal{($EW_{\lya}\sim-40~\AA$)}, a very extended \lya\ halo ($s\sim$100 kpc), and has \lya\ emitting satellite galaxies near it
    
    \item Rest-optical spectra (MOSFIRE) and SED modeling of the galaxies revealed that 
    \textnormal{\textit{G1}} is a typical $z=2-3$ star-forming galaxy in terms of its stellar mass ($\rm \log{M_*/M_\odot}_{\rm G1}= 9.9 \pm 0.1,$), star formation rate (SFR=6-15 \msun), dust extinction \textnormal{($A_{\rm V}=0.21$)}, ionization parameter \textnormal{($\log{U}\sim \rm -2.7$)}, and \lya\ escape fraction ($f_{\rm esc,\lya,G1}=6\%$). 
    \textnormal{\textit{N1}} is a \textnormal{lower} mass ($\rm \log{M_*/M_\odot}_{\rm G1}= 8.7-9.2 \pm 0.1,$), young ($t_*$=30-50 Myr), relatively \textnormal{low dust ($A_{\rm V}\sim0.1$) star-forming galaxy with strong nebular emission (\lya\, [\ion{O}{3}]), complex and extended \lya\ halo morphology ($s\sim100~\rm kpc$), and high \lya\ escape fraction ($f_{\rm esc,\lya,N1}=30\%$) putting in the tail end of the $z\sim3$ LBG population, and reminiscent of $z\sim2$ EELGs.}

\end{enumerate}

The diversity of the two galaxies' CGM-ISM properties highlights the need to build a large sample of nLOS galaxy-QSO pairs to place global observational constraints on the $z\sim2-3$ CGM. The complete KBSS-InCLOSE sample will accomplish this by analyzing e.g., the abundance of unbound gas, how CGM measurables vary as a function of galaxy property (e.g., $M_*$, SFR), metallicity distribution as a function of ion and impact parameter, the fraction of thermally vs. non-thermally supported gas a function of galaxy properties,  further constrain the gas and metal mass of the CGM, etc. 

Fundamentally, these first results show that the observational strategies, QSO subtraction techniques, and analyses that we have adopted/developed for KBSS-InCLOSE are well suited to expand our understanding of the $z\sim2-3$ baryon cycle.

\section*{Acknowledgements}
\textnormal{We thank Sebastiano Cantalupo for assistance, guidance, and troubleshooting with CubExtractor. We thank Holly Christenson and Regina Jorgenson for providing insights and sharing data of Q2233-N1. We thank Hsiao-Wen Chen, Sean Johnson, and Mandy Chen for useful discussions regarding datacube QSO subtraction.}

This material is based upon work supported by the National Science Foundation Graduate Research Fellowship under Grant No. 2019286522. CCS has been supported, in part, by NSF grant AST-2009278 and by the Caltech/JPL President's and Director's program.  E.N.K.\ acknowledges support from NSF CAREER grant AST-2233781. Z.Z.\ acknowledges the financial support from NASA FINESST grant 80NSSC22K1755. \textnormal{D.K.E. is supported by the NSF through Astronomy \& Astrophysics grant AST1909198. RJC is funded by a Royal Society University Research Fellowship, and acknowledges support from STFC (ST/T000244/1, ST/X001075/1).}

The authors wish to recognize and acknowledge the very significant cultural role and reverence that the summit of Maunakea has always had within the Native Hawaiian community. We are most fortunate to have the opportunity to conduct observations from this mountain.

This research has made use of the Keck Observatory Archive (KOA), which is operated by the W. M. Keck Observatory and the NASA Exoplanet Science Institute (NExScI), under contract with the National Aeronautics and Space Administration

Some of the data presented in this work were obtained from the Keck Observatory Database of Ionized Absorbers toward QSOs (KODIAQ), which was funded through NASA ADAP grant NNX10AE84G

\facility{Keck:II (KCWI)}
\facility{Keck:I (MOSFIRE)}
\facility{Keck:I (LRIS)}

\software{
NUMPY \citep{numpy_harris2020array},
MATPLOTLIB \citep{MATPLOTLIB_Hunter:2007},
Astropy \citep{astropy_2013,astropy_2018},
DS9 \citep{DS9_2000ascl.soft03002S},
IMEXAM \citep{IMEXAM_2022ascl.soft03004S},
PHOTUTILS \citep{bradley_2022},
VoigtFit \citep{krogager+2017},
ALIS \citep{ALIS},
MOSPEC \citep{strom+2017},
CubExtractor \citep{cantalupo+2019}
}

\bibliography{citations}{}
\bibliographystyle{aasjournal}

\appendix
\section{\textnormal{Impact of Quasar Spectral Cube Subtraction on the Individual Galaxy Spectra}} \label{sec:ifsfit_sampling}
\textnormal{In the section we wanted to show effects on the QSO subtraction from the KCWI cube using IFSFIT discussed in Section \ref{sec:ifsfit}.}

We wanted to ensure that the continuum and spectral lines detected in the extracted galaxy spectra from the QSO-spectrall subtracted cubes were not strongly dependent on the sampling of the input QSO spectrum to IFSFIT. To test this, we varied the radius of the circular aperture used to extract the \textnormal{QSO} spectrum 
then ran it through IFSFIT, then extracted the spectra of G1 and N1.

\begin{figure}[hb]
    \centering
    \includegraphics[scale=0.53]{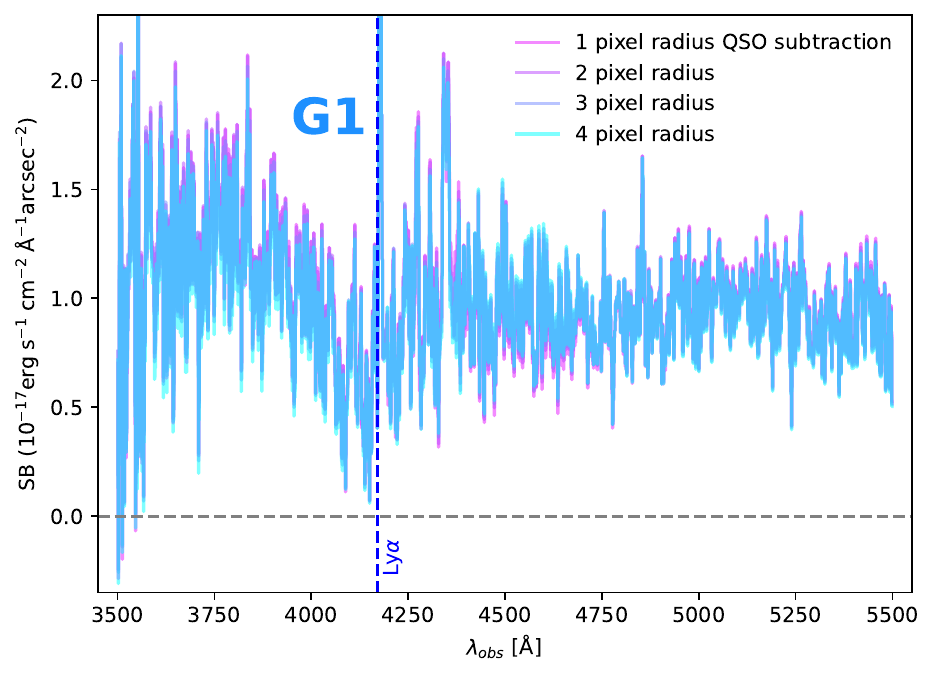}
    \includegraphics[scale=0.53]{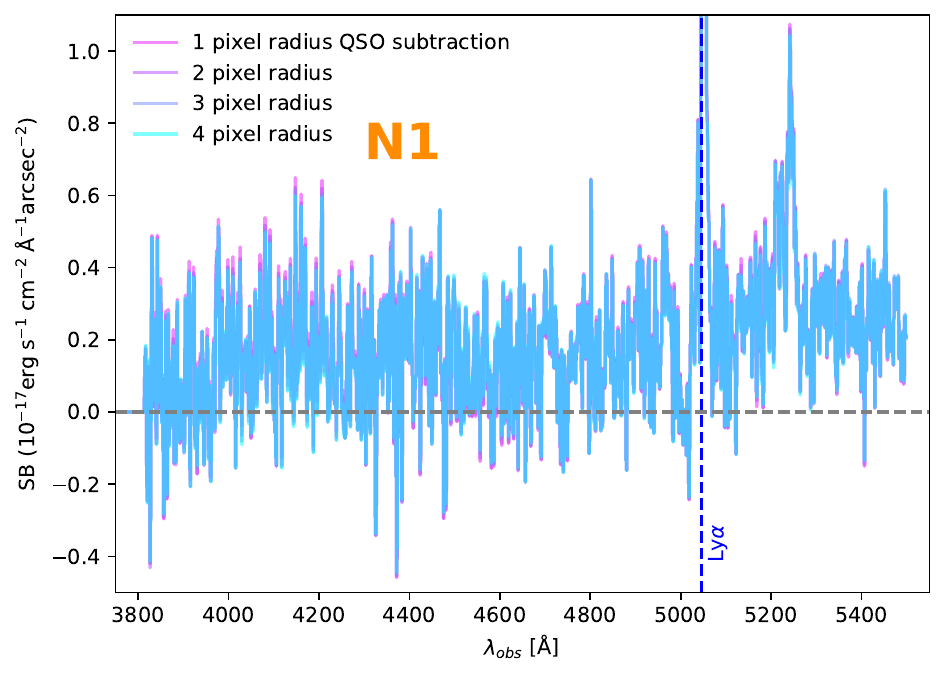}
    \caption{Extracted rest-FUV Keck/KCWI spectra as a function of extraction aperture size (used in the \textnormal{QSO} subtraction input). Ly$\alpha$ emission peak is marked as a blue dashed line in each spectrum. \textit{\textnormal{Left}:} Sum of spaxels that contain galaxy G1. The blue end of the spectrum shows the largest difference (in flux) between the apertures size but are still $\lesssim10\%$ of one another while the shape stays identical. \textit{\textnormal{Right}:} Sum of spaxels that contain galaxy N1. Similar behavior is seen except the percent differences between the aperture sizes \textnormal{are $\lesssim5\%$.}}
    \label{fig_kcwi_psfsub_aperture}
\end{figure}

We can see from both the plots in Figure \ref{fig_kcwi_psfsub_aperture} that the radius of the extraction aperture does not affect absorption line strength or centroid, emission line strength or centroid, or continuum shape. 
Even the smallest aperture size is sufficient to recover the main galaxy spectral features. Adding more pixels just removes flux from the galaxy continuum at the 5-10$\%$ level depending on the portion of the spectrum.

\begin{figure}
    \centering
    \includegraphics[scale=0.46]{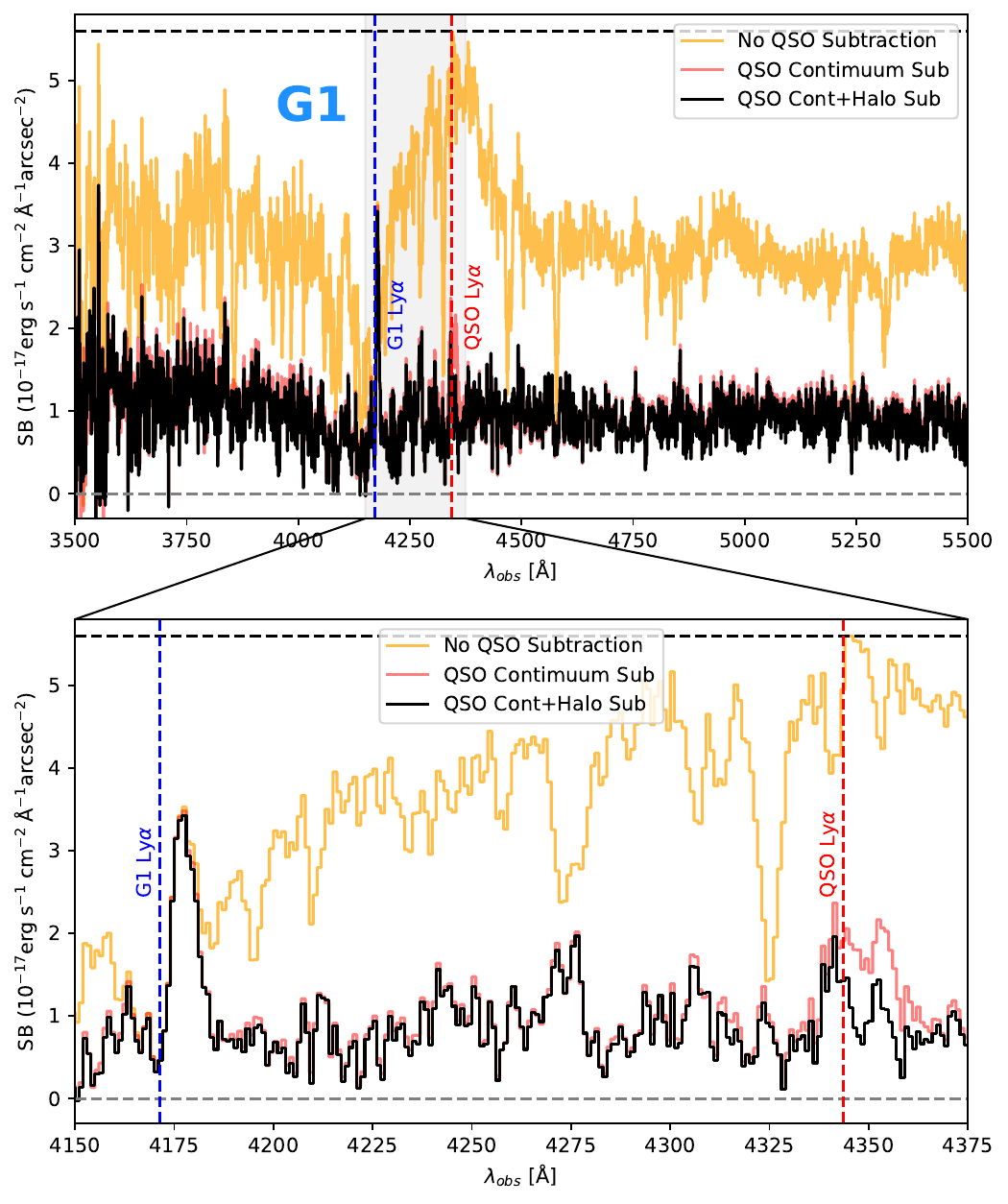}
    \includegraphics[scale=0.46]{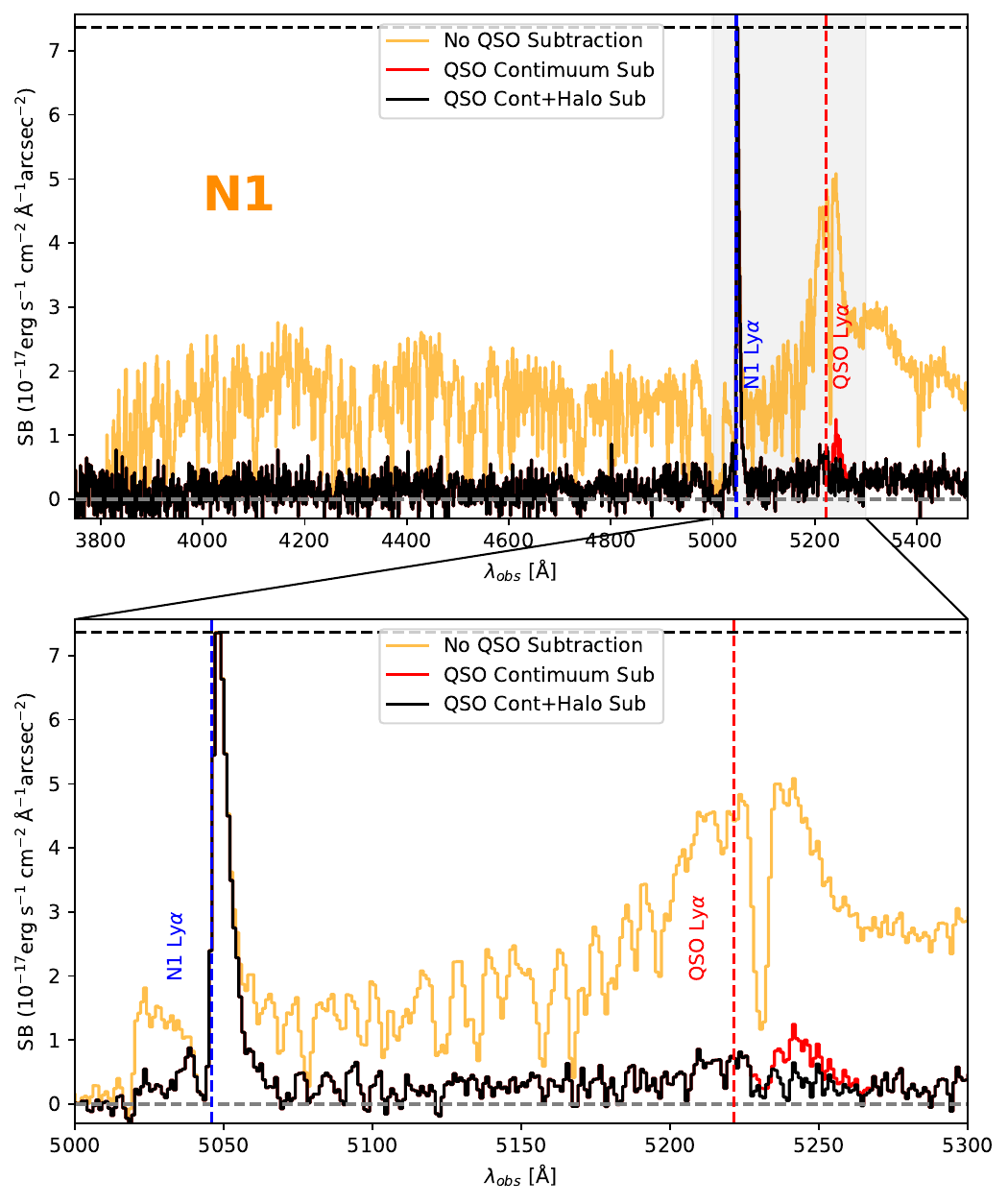}
    \caption{\textnormal{Extracted rest-FUV KCWI spectra during all stages of QSO subtraction for G1 (left two panels) and N1 (right two panels). \textit{Top panels:} Orange shows the spectrum without subtraction; red shows the first subtraction (i.e., QSO continuum subtraction); and black shows the second/final subtraction (removal of QSO \lya\ halo+QSO continuum removal). \textit{Bottom panels:} Zoom-in showing the QSO \lya\ halo removal.}}
    \label{fig_NBsub}
\end{figure}

\textnormal{In Figure \ref{fig_NBsub} we show the effects of steps 3 and 4 from the KCWI cube spectral subtraction (i.e., QSO \lya\ halo subtraction; Section \ref{sec:ifsfit}) on the extracted spectra of galaxies Q2343-G1 and Q2233-N1. The red spectrum show the results from the QSO continuum subtraction (steps 1 and 2), while the black lines show the results from the QSO continuum+halo subtraction (Steps 3 and 4). 
In both galaxies the QSO Ly$\alpha$ peak is reduced by more than 50\% while the rest of the spectrum is unaffected.
}

\section{Complete Voigt Profile Fitting Results} \label{sec:full_voigt}
\begin{figure*}
    \centering
    \includegraphics[scale=0.78]{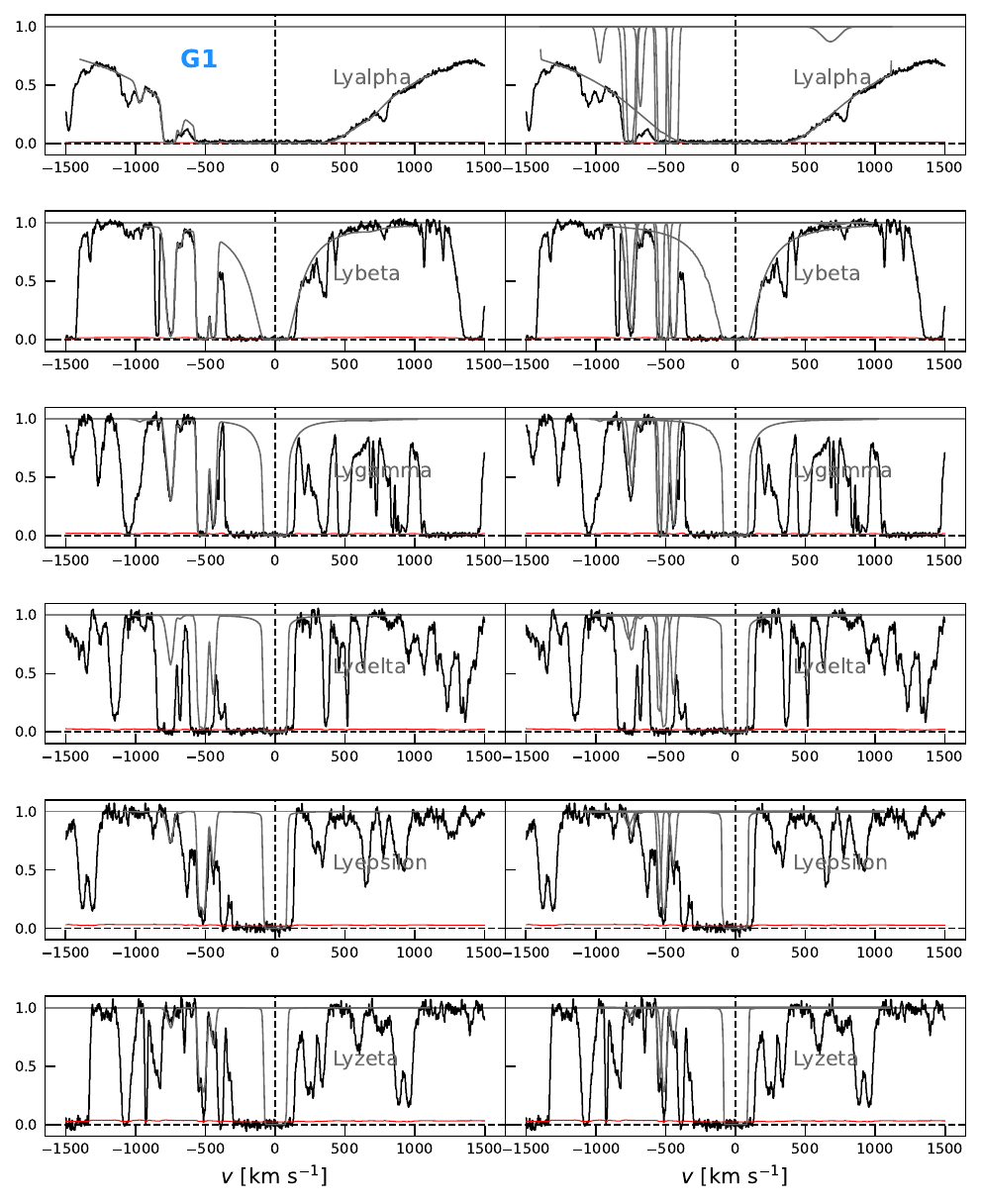}
    \caption{G1's \ion{H}{1} CGM absorption fits from \citetalias{rudie+2012} centered at $z_{\rm G1}$=2.4312. The lines and colors are \textnormal{the} same as Figure \ref{fig_cgmfit_q2343rep}. 
    }
    \label{fig_cgmfit_q2343hydrogen}
\end{figure*}

\begin{figure*}
    \centering
    \includegraphics[scale=0.78]{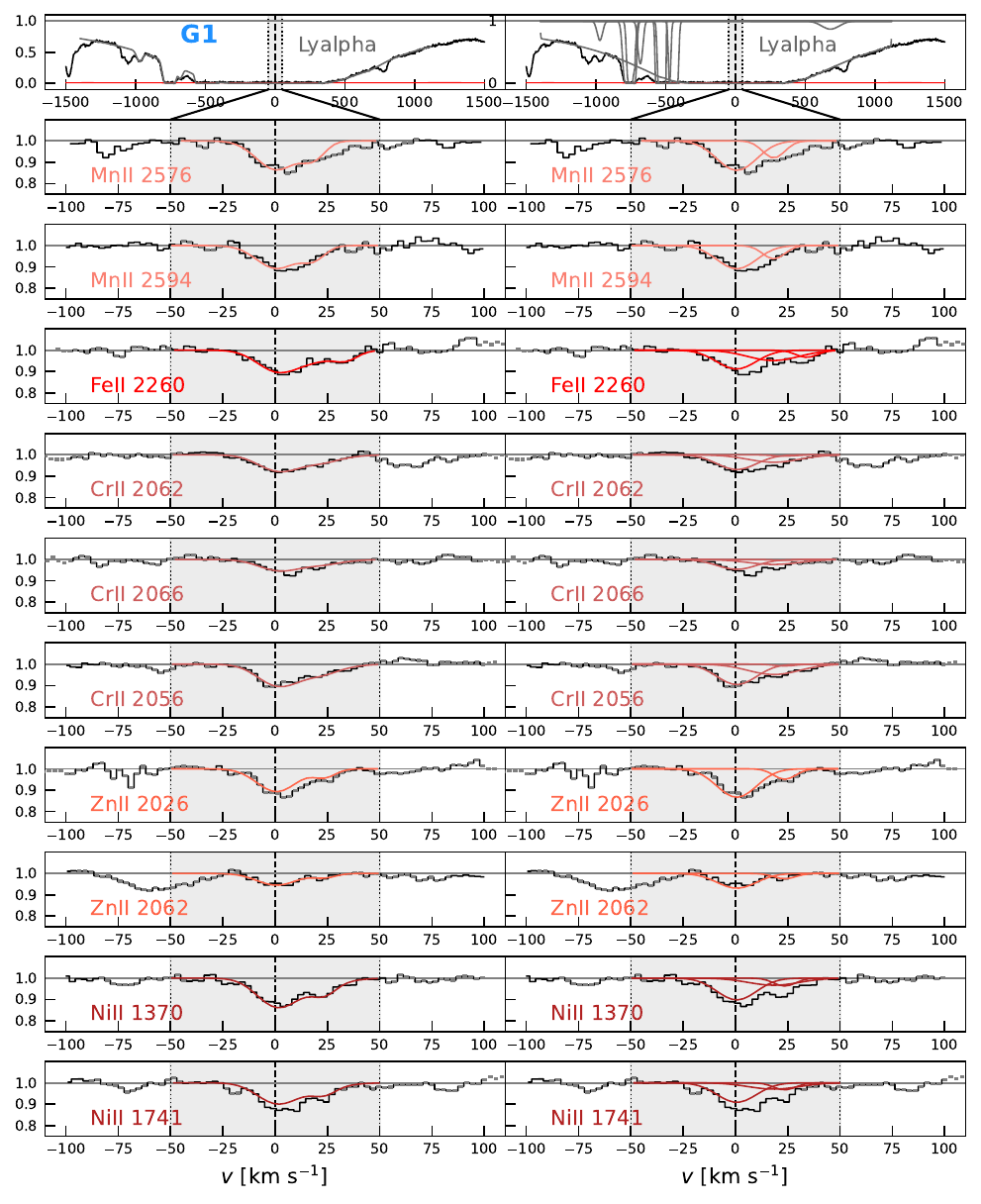}
    \caption{G1's Fe-peak element CGM absorption and best-fit Voigt profiles centered at $z_{\rm G1}$=2.4312. The lines and colors are \textnormal{the} same as Figure \ref{fig_cgmfit_q2343rep}. 
    }
    \label{fig_cgmfit_q2343fepeak}
\end{figure*}

\begin{figure*}
    \centering
    \includegraphics[scale=0.78]{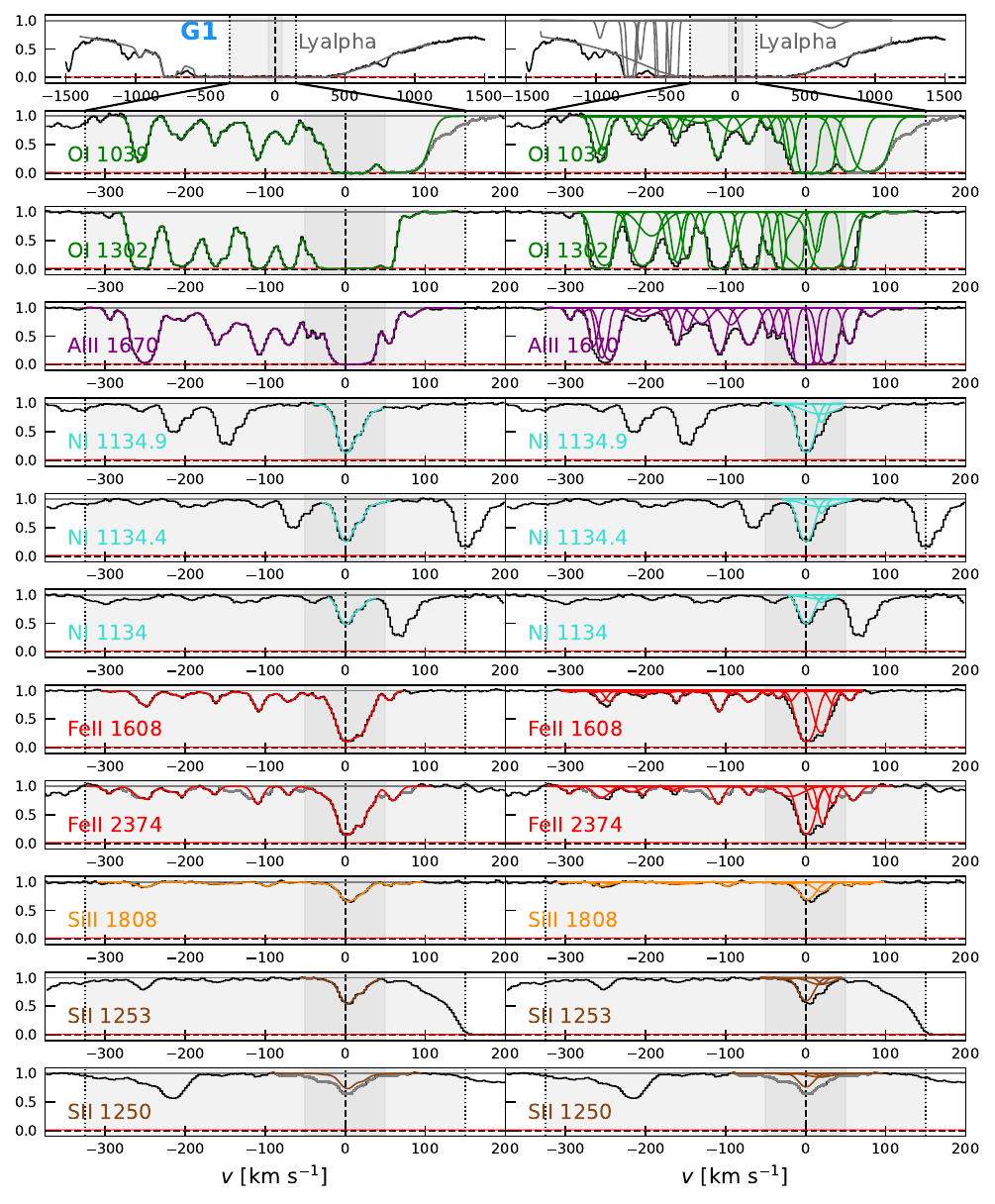}
    \caption{G1's neutral- and low-ionization metal CGM absorption and best-fit Voigt profiles centered at $z_{\rm G1}$=2.4312. The lines and colors are \textnormal{the} same as Figure \ref{fig_cgmfit_q2343rep}. 
    }
    \label{fig_cgmfit_q2343low}
\end{figure*}

\begin{figure*}
    \centering
    \includegraphics[scale=0.78]{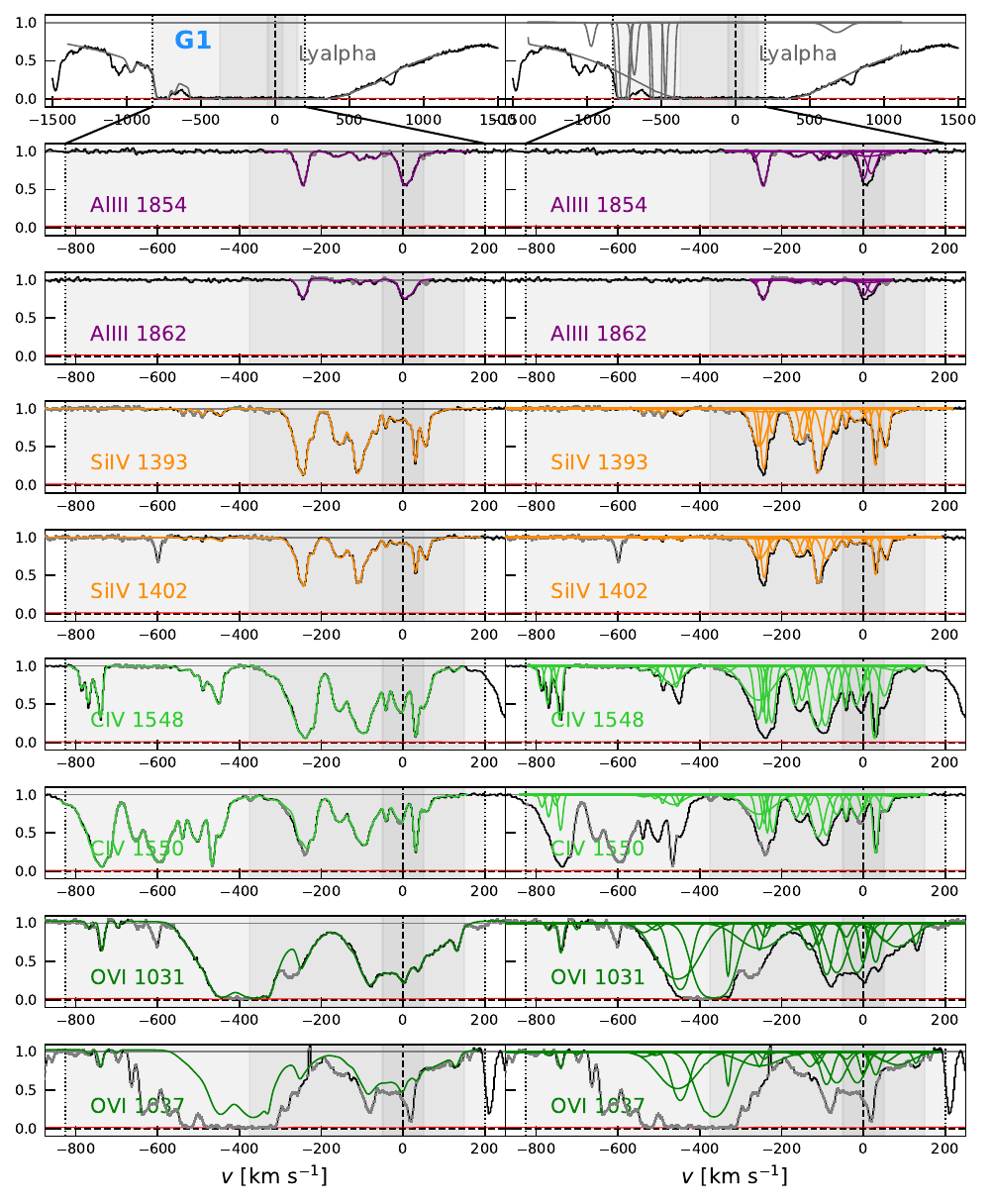}
    \caption{G1's intermediate- and high-ionization metal CGM absorption and best-fit Voigt profiles centered at its systemic redshift z$_{\rm G1}$=2.4312. The lines and colors are \textnormal{the} same as Figure \ref{fig_cgmfit_q2343rep}. 
    }
    \label{fig_cgmfit_q2343medhigh}
\end{figure*}

\textnormal{In Figures \ref{fig_cgmfit_q2343hydrogen}-\ref{fig_cgmfit_q2343medhigh} we show the all of the best-fit Voigt profiles to G1s CGM absorption.}

\end{document}